\documentclass[a4paper]{JHEP3}  
\usepackage[centertags]{amsmath}
\usepackage{amssymb}
\usepackage{graphicx}

\usepackage{psfrag}

\title{Small Hairy Black Holes in $AdS_5 \times S^5$}

\author{
Sayantani Bhattacharyya$^a$\footnote{sayanta@theory.tifr.res.in}, \
Shiraz Minwalla$^a$\footnote{minwalla@theory.tifr.res.in} and 
Kyriakos Papadodimas$^b$\footnote{k.papadodimas@uva.nl}\\
\small{\emph{$^{a}$ Tata Institute of Fundamental Research,}}\\
\small{\emph{~~~Homi Bhabha Rd, Mumbai 400005.}} \\ 
\small{\emph{$^{b}$ Institute for Theoretical Physics,}}\\
\small{\emph{~ Valckenierstraat 65, 1018 XE Amsterdam,}}\\
\small{\emph{~ The Netherlands}}\\
}

\abstract{We study small hairy black holes in a consistent truncation
  of ${\cal N}=8$ gauged supergravity that consists of a single
  charged scalar field interacting with the metric and a $U(1)$ gauge
  field. Small very near extremal RNAdS black holes in this system
  are unstable to decay by superradiant emission. The end point of
  this instability is a small hairy black hole that we construct
  analytically in a perturbative expansion in its charge. Unlike their
  RNAdS counterparts, hairy black hole solutions exist all the way down
  to the BPS bound, demonstrating that ${\cal N}=4$ Yang Mills theory
  has an ${\cal O}(N^2)$ entropy at all energies above
  supersymmetry. At the BPS bound these black holes reduce to
  previously discussed regular, supersymmetric horizon free
  solitons. We use numerical methods to continue the construction of
  these solitons to large charges and find that the line of soliton
  solutions terminates at a singular solution $S$ at a finite charge. We
  conjecture that a one parameter family of singular supersymmetric
  solutions, which emerges out of $S$, constitutes the BPS 
  limit of hairy black holes at larger values of the charge. We analytically 
  determine the near singularity behaviour of $S$, demonstrate that both the 
  regular and singular solutions exhibit an infinite set of damped `self 
  similar' oscillations around $S$ and analytically compute the frequency of 
  these oscillations. At leading order in their charge, the 
  thermodynamics of the small hairy black holes constructed in this paper turns 
  out to be correctly reproduced by modeling these objects as a non 
  interacting mix of an RNAdS black hole and the supersymmetric soliton in 
  thermal equilibrium. Assuming that a
  similar non interacting model continues to apply upon turning on
  angular momentum, we also predict a rich family of rotating hairy black
  holes, including new hairy supersymmetric black holes. This
  analysis suggests interesting structure for the space of (yet to be
  constructed) hairy charged rotating black holes in $AdS_5\times S^5$, 
  particularly in the near BPS limit.}

\preprint{TIFR/TH/\\
ITFA10-13}

\begin{document}

\section{Introduction}\label{sec:intro}

Black hole solutions of IIB theory on $AdS_5 \times S^5$ constitute
the thermodynamic saddle points of ${\cal N}=4$ Yang Mills theory on
$S^3$ via the AdS/CFT correspondence. A complete understanding of the
space of stationary black hole solutions in $AdS_5\times S^5$ is
consequently essential for a satisfactory understanding of the state
space of ${\cal N}=4$ Yang Mills theory at energies of order
$N^2$. While the Kerr RNAdS black hole solutions are well known
\cite{Gibbons:2004uw, Gibbons:2004ai, Cvetic:2004ny, Chong:2005hr,
  Chong:2005da, Chong:2006zx, Cvetic:2005zi}, it seems likely that
several additional yet to be determined families of black hole
solutions will play an important role in the dynamics and
thermodynamics of ${\cal N}=4$ Yang Mills theory.

In this paper we will construct a new class of asymptotically
$AdS_5\times S^5$ black hole solutions. The black holes we construct
are small, charged, and have parametrically low temperatures; our
construction is perturbative in the black hole charge. Our solutions
are hairy, in the sense that they include condensates of charged
scalar fields\footnote{See \cite{Sachdev:2010ch,
    Horowitz:2010gk,Hartnoll:2009qx, Hartnoll:2009sz,Herzog:2009xv}
  and references therein for reviews of recent work - sparked by an
  observation by Gubser \cite{Gubser:2008px} -on hairy black branes in
  $AdS$ spaces.}. 
In the BPS limit these hairy black holes reduce to regular horizon free 
solitons. 
We also use numerical techniques to continue our perturbative construction 
of these solitons to charges of order unity, and uncover an intricate self 
similar behaviour in the space of solitons in the neighborhood of 
a finite critical value of the charge. The solutions presented in this paper 
suggest a qualitatively new picture of the near BPS spectrum of ${\cal N}=4$ 
Yang Mills.

Our perturbative construction of hairy black holes is close in spirit and 
technique to the constructions presented in the recent paper \cite{our}, which 
may be regarded as an immediate precursor to the current work.  For
this reason we first present a brief review of \cite{our} before
turning to a description and discussions of the new black hole
solutions constructed in this paper.

It was demonstrated in \cite{our} that small charged black holes in
global $AdS$ spaces are sometimes unstable to the condensation of
charged matter fields. More precisely any system governed by the
Lagrangian
\begin{equation}\label{form}\begin{split} 
 &\int d^5 x \sqrt{g}\left[ \frac{1}{2}\left(R + 12\right)  - \frac{3}{8}F_{\mu\nu}F^{\mu\nu} - 
\frac{3}{16}\left(|D_\mu \phi|^2 +\Delta(\Delta-4)\phi\phi^*\right) +
{\rm Interactions} \right]\\
&D_\mu\phi  = \partial_\mu\phi - e i A_\mu \phi 
\end{split}
\end{equation}
possesses small RNAdS black holes that are unstable to decay by
superradiant discharge of the scalar field $\phi$ whenever $e >
\Delta$. The end point of this superradiant tachyon condensation
process is a hairy black hole. The authors of \cite{our} constructed
these hairy black hole solutions (working with a particular toy model
Lagrangian of the form \eqref{form}) in a perturbation expansion in
their mass and charge. At leading order in this expansion, the hairy
black holes of \cite{our} are well approximated by a non interacting
mix of a small RNAdS black hole and a weak static solitonic scalar
condensate. In particular, it was shown in \cite{our} that the leading
order thermodynamics of small hairy black holes could be reproduced
simply by modeling them as a non interacting mix of an RNAdS black
hole and a regular charged scalar soliton.
 
The results of \cite{our} suggest that the density of states of 
certain field theories with a gravity dual description might be dominated in
certain regimes by previously unexplored phases consisting of an
approximately non interacting mix of a normal charged phase and a Bose
condensate. In order to make definitive statements about the actual
behaviour of ${\cal N}=4$ Yang Mills theory, however, it is necessary
to perform the relevant calculations in IIB supergravity on $AdS_5
\times S^5$ rather than a simple toy model Lagrangian; this is the
subject of the current paper. As the 
IIB theory on $AdS_5 \times S^5$ is a very special system, the reader might 
anticipate that hairy black holes in this theory have some distinctive 
special properties not shared by equivalent objects in the toy model studied 
in \cite{our} at least at generic values of parameters. As we will see 
below, this indeed turns out to be the case.

In order to avoid having to deal with the full complexity of IIB
SUGRA, in this paper we identify\footnote{In unpublished work, 
S. Gubser, C. Herzog and S. Pufu have independently identified this 
consistent truncation, and have numerically investigated hairy black branes 
in this set up. We thank C. Herzog for informing us of this.} work with 
a consistent truncation
of gauged ${\cal N}=8$ supergravity (itself a consistent truncation of
IIB SUGRA on $AdS_5 \times S^5$).  Of the complicated spectrum of
${\cal N}=8$ supergravity, our truncation retains only a single
charged scalar field $\phi$, a gauge field $A_\mu$ and the
metric.\footnote{Under the $AdS/CFT$ correspondence, $\phi$ is dual to
  the operator $Tr X^2+Tr Y^2 +Tr Z^2$ while $A_\mu$ is dual to the
  conserved current $J^{X {\bar X}}_\mu+J^{Y {\bar Y}}_\mu + J^{Z
    {\bar Z}}_\mu$.  Here $X$, $Y$ and $Z$ denote the three complex
  chiral scalars in the ${\cal N}=4$ Lagrangian.} The Lagrangian for
our system is given by
\begin{equation}\label{spclag}\begin{split}
S=& \frac{N^2}{4 \pi^2} \int \sqrt{g}\left[ \frac{1}{2}\left(R + 12\right)  - 
\frac{3}{8}F_{\mu\nu}F^{\mu\nu} - \frac{3}{16}\left(|D_\mu \phi|^2- 
\frac{\partial_\mu(\phi\phi^*)\partial^\mu(\phi\phi^*)}{4\left(4 + \phi\phi^*\right)} - 4 \phi\phi^*\right)\right]\\
D_\mu\phi & = \partial_\mu\phi - 2 i A_\mu \phi \\
F_{\mu\nu}& = \partial_\mu A_\nu - \partial_\nu A_\mu 
\end{split}
\end{equation}
and is of the general structure \eqref{form} with $e=2$ and
$\Delta=2$. As $e=\Delta$, small very near extremal black holes in
this system lie at the precipice of the  super radiant instability
discussed in \cite{our}. The question of whether they fall over this
precipice requires a detailed calculation. We perform the necessary
computation in this paper. Our results demonstrate that very near
extremal small black holes {\it do} suffer from the super radiant
instability; we proceed to construct the hairy black hole that
constitutes the end point of this instability.

The calculations presented in this paper employ techniques that are
similar to those used in \cite{our}. In particular we work in a
perturbation expansion in the scalar amplitude using very near
extremal vacuum RNAdS black holes as the starting point of our
expansion. This expansion is justified by the smallness of the charge
of our solutions. We implement our perturbative procedure by matching
solutions in a near (horizon) range, intermediate range and far field
range.  This matching procedure is justified by the parametrically large 
separation of scales between the horizon radius of the black holes 
we construct and AdS curvature radius.  We refer the reader to the 
introduction of
\cite{our} for a more detailed explanation of physical motivation for
this perturbative expansion and its formal structure. Actual details
of our calculations may be found in sections \ref{sec:soliton} and
\ref{ssec:ordreps} below. In the rest of this introduction we simply
present our results and comment on their significance.

We study black holes of charge $Q$ \footnote{$Q$ is the charge of the black 
hole solutions under each of the three diagonal U(1) Cartan's of SO(6). 
The consistent truncation of this paper forces these three U(1) charges to
be equal.} (normalized so that each of the
three complex Yang Mills scalars has unit charge) and mass $M$
(normalized to match the scaling dimensions of dual operators). We find 
it convenient to deal with the `intensive' mass and
charge, $m$ and $q$, given by
$$m=\frac{M}{N^2} ~~~q=\frac{Q}{N^2}.$$ 
As in \cite{our}, in this
paper we are primarily interested in small black holes for which $q
\ll 1$ and $m \ll 1$. We now recall some facts about RNAdS black holes
in this system. First, the masses of such black holes obey the
inequality
$$m \geq m_{ext}(q)=3q+3q^2 -6q^3 +{\cal O}(q^4)$$ 
Black holes that
saturate this inequality are extremal, regular and have finite entropy
(see subsection \ref{ssec:RNAdSBH} for more details). The chemical
potential $\mu_{ext}(q)$ of extremal black holes is given by 
$$ \mu_{ext}(q)=1+2q -6q^2+{\cal O}(q^3)$$ 
and approaches unity in the limit of small charge. Note also that extremal 
black holes lie above the BPS bound
$$m_{BPS}(q)=3q$$
and in particular that 
$m_{ext}(q)-m_{BPS}(q)=3q^2 +{\cal O}(q^3)$

We will now investigate potential superradiant instabilities (see the
introduction of \cite{our} for an explanation of this term) of these
black holes. Recall that a mode of charge $e$ and energy $\omega$
scatters off a black hole of chemical potential $\mu$ in a
superradiant manner whenever $ \mu e > \omega$. The various modes of
the scalar field $\phi$ in \eqref{spclag} have energies $\omega=2, 3,
\ldots$ and all carry charge $e=2$. As the chemical potential of a
small near extremal black hole is approximately unity, it follows that
only the ground state of $\phi$ (with $\omega=\Delta=2$) could
possibly scatter of a near extremal RNAdS black hole in a superradiant
manner.  This mode barely satisfies the condition for superradiant
scattering; as a consequence we will show in this paper that small
RNAdS black holes in \eqref{spclag} suffer from a superradiant
instability into this ground state mode only very near to extremality,
i.e. when
$$m -m_{ext}(q) \leq 6q^3+{\cal O}(q^4).$$ Unstable black holes
eventually settle down into a new branch of stable hairy black hole
solutions.  We have  constructed these hairy black holes 
in a perturbative expansion in their charge in Section \ref{ssec:ordreps}; 
we now proceed to present a qualitative description of these solutions and 
their thermodynamics. 

Recall that the zero mode of the scalar field $\phi$ obeys the BPS
bound (and so is supersymmetric) at linear order in an expansion about
global $AdS_5$.  It has been demonstrated in
\cite{Chong:2004ce,Liu:2007xj, Chen:2007du} (and we reconfirm in
section \ref{sec:soliton} below) that this linearized BPS solution
continues into a nonlinear BPS solution upon increasing its
amplitude. In this paper we will refer to this regular solution as the
supersymmetric soliton. The hairy black holes of this paper may
approximately be thought of as a small, very near extremal RNAdS black
hole located in the center of one of these solitons. Although the
soliton is supersymmetric, the black hole at its center is not, and so
hairy black holes are not BPS in general. These solutions exist in the
mass range
\begin{equation}\label{rangeintro} 
3q \leq m \leq 3q+3q^2+{\cal O}(q^4)
\end{equation}
At the lower bound of this range \eqref{range} hairy black holes
reduce to the supersymmetric soliton. At the upper bound (which is
also the instability curve for RNAdS black holes) they reduce to RNAdS
black holes.

\begin{figure}[h]
\centering
\psfrag{a}[t]{ $h_0$}
\psfrag{b}[r]{$Q$}
\includegraphics[totalheight=0.22\textheight]{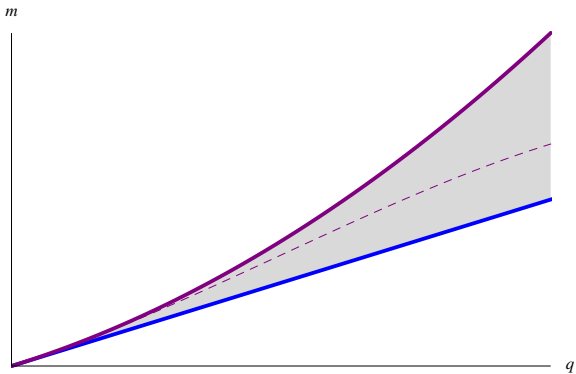}
\caption{Phase diagram as a function of charge $q$ (x axis) and mass
  $m$ (y axis) at small $q$. The solid blue line at the bottom is the
  BPS bound along which the soliton lives. Hairy black holes exist -
  and are the dominant phase - in the shaded region. RNAdS black holes
  are the only known solutions (so in particular the dominant phase)
  in the unshaded region above the solid red curve at the top. RNAdS
  black holes also exist (but are dynamically unstable and
  thermodynamically sub dominant) between the solid red curve and the
  dashed curve.  The solid red curve is described by $m=3 q + 3 q^2 +{\cal O}(q^4)$, while
the blue curve by $m=3q$. The dashed curve corresponds to extremal RNAdS
black holes and is given by $m=3q+3q^2 -6q^3+{\cal O}(q^4)$. The curves have not been drawn
to scale to make the diagram more readable.}
\label{one}
\end{figure}

In Fig. \ref{one} below we have plotted the near extremal micro
canonical `phase diagram' for our system.  As is apparent from
Fig. \ref{one} our system undergoes a phase transition from an RNAdS
phase to a hairy black hole phase upon lowering the energy at fixed
charge. This phase transition occurs at the upper end of the range
\eqref{rangeintro}.  Note that the phase diagram of Fig. \ref{one} has
several similarities with the phase diagram depicted in Fig. 1 in
\cite{our}; however there is also one important difference. The
temperature of the hairy black holes of this paper decreases with
decreasing mass at fixed charge, and reaches the value zero at the BPS
bound. In contrast the temperature of the hairy black holes of
\cite{our} increases with decreasing mass (at fixed charge),
approaching infinity in the vicinity of the lower bound.
 
As we have emphasized, the phase diagram depicted in Fig. \ref{one}
applies only in the limit of small charges and masses. We would now
like to inquire as to how this phase diagram continues to large
charges and masses. In order to address this question we first focus
on solitonic solutions.  These solutions may be determined much more
simply than the generic hairy solution, as they obey the constraints
of supersymmetry rather than simply the equations of motion. It turns
out that spherically symmetric supersymmetric solutions are given as
solutions to a single nonlinear, second order ordinary differential
equation \cite{Chong:2004ce,Liu:2007xj, Chen:2007du}.  The solitons
constitute the unique one parameter set of regular solutions to this
equation.  It is easy to continue our perturbative construction of the
solitonic solutions to large charges by solving this equation
numerically: in fact this exercise was already carried out in
\cite{Liu:2007xj}. This numerical solution reveals that the
solitonic branch of solutions terminates at a finite charge $q_c=
0.2613$. For $q>q_{c}$ there are no supersymmetric spherically
symmetric solutions to the equations of motion of
\eqref{spclag}\footnote{To be more precise, there are smooth solitonic
  solutions up to a slightly higher value $q_m=0.2643$, but in a sense
  that will be explained in section \ref{sec:nsoliton}, the point
  $q_c$ marks the boundary between regular solitonic solutions and
  singular ones.}.

Recall that solitons constitute the lower edge of the space of hairy
black hole solutions of Fig. \ref{one}.  The non existence of regular
supersymmetric solutions for $q>q_c$ might, at first, suggest that at
these charges the space of hairy black hole solutions terminates at a
mass greater than $3q$ (i.e. does not extend all the way down to
supersymmetry). While this is a logical possibility, we think it is likely 
that the truth lies elsewhere. As we will explain in section \ref{sec:nsoliton}
the solitonic branch of supersymmetric solutions terminates in a
distinguished singular solution $S$. It turns out that $S$ is also the
end point (or origin) of a one parameter set of supersymmetric
solutions that are all singular at the origin. The charges of these
solutions increase without bound (indeed we have found an explicit
analytic solution for the singular supersymmetric solution in the
limit of arbitrarily large charge). The two one parameter families of
solutions, regular and singular ones, are joined at the special
solution $S$. We conjecture that smooth hairy black hole solutions exist
in our system at every $q$ and for $m>3q$. Upon taking the limit $m
\rightarrow 3 q$, these smooth solutions reduce to the smooth soliton
for $q<q_c$ but reduce to the singular supersymmetric solutions
described above when $q>q_c$.  In summary, we conjecture that the
phase diagram of our system takes the form displayed in Fig. \ref{two}
below.

\begin{figure}[h]
\centering
\psfrag{a}[t]{ $q$}
\psfrag{b}[r]{$m$}
\includegraphics[totalheight=0.22\textheight]{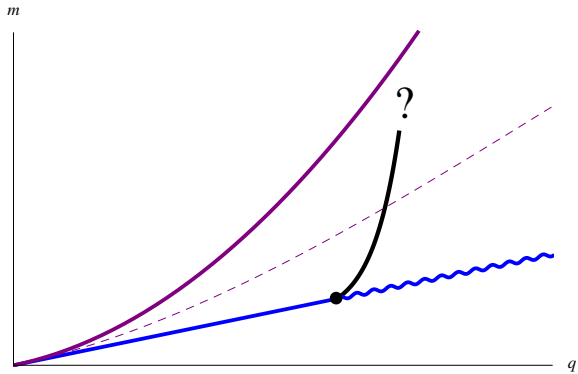}
\caption{Conjectured phase diagram as a function of charge $q$ (x
  axis) and mass $m$ (y axis) for all values of $q$. The blue line at
  the bottom is the BPS bound along which the regular soliton lives
  (straight part) and the singular supersymmetric solutions (wiggly
  part). The solid red curve at the top marks the phase transition
  between the regime of RNAdS black holes (above the line) and that of
  smooth hairy black holes (below). The black curve indicates a
  phase transition between different types of hairy black holes. This
  curve originates from the BPS line at the black dot which is close
  to the point $q=q_c$ and could end either in the bulk of the hairy
  black hole region or could extend all the way up to the red line.}
\label{two}
\end{figure}

The distinguished solution $S$ clearly plays a special role in the space 
of spherically symmetric supersymmetric solutions. In subsection \ref{class}
we analytically determine the near singularity behaviour of this solution. 
Viewing the 2nd order differential equation that determines supersymmetric 
solutions as a dynamical system in the `time' variable $\ln r$, we demonstrate
that the solution $S$ is a stable fixed point of this system, and analytically 
compute the eigenvalues that characterize the approach to this fixed point. 
This eigenvalue has an imaginary part (which damps fluctuations) and a real
part (that results in oscillations). Solitonic - and singular - 
solutions in the neighborhood of $S$ may be thought of as configurations 
that that flow to $S$ at large $\ln r$. The oscillations\footnote{We are
extremely grateful to M. Rangamani for suggesting that we look for this 
`self similar' structure in the space of solitons in the neighborhood of 
$q=q_c$. The results reported in this paragraph are the outcome of 
investigations that were spurred directly by this suggestion.}
referred to above result in the following 
phenomenon: the system develops a multiplicity of supersymmetric solitonic 
(or singular) at charges $q$ when $q$ comes near enough to $q_c$. The  
number of solutions diverges as $q \to q_c$. The space of solitonic
and singular supersymmetric solutions are usefully plotted as a curve 
on a plane parametrized by the charge $q$ and the expectation value of the 
operator dual to the scalar $\phi$. On this plane supersymmetric solitons 
spiral into the point $S$, while the singular solutions spiral out of 
the same point (see Fig. \ref{par2}); the two spirals are non self 
intersecting \footnote{We are very grateful to V. Hubeny for suggesting the 
possibility of a spiral structure for these curves.}. We find this extremely
intricate structure of supersymmetric solutions quite fascinating, and 
feel that its implications for ${\cal N}=4$ Yang Mills physics certainly 
merits further investigation.

In this paper we have so far only considered charged black holes with
vanishing angular momentum. Such solutions are spherically symmetric;
i.e. they preserve the $SO(4) =SU(2) \times SU(2)$ rotational isometry
group.  As all known supersymmetric black holes in $AdS_5\times S^5$
possess angular momentum
\cite{Gutowski:2004yv,Gutowski:2004ez,Kunduri:2006ek}, it is of
interest to generalize the study of this paper to black holes with
angular momentum. Let us first consider a spinning Kerr RNAdS black
hole that preserves only $U(1)\times U(1) \in SU(2)\times
SU(2)$. Perturbations about such a solution are functions of an angle and
a radius and are given by solutions to partial rather than ordinary
differential equations. However there exist RNAdS black holes with
self dual angular momentum.  The angular momentum of such a black hole
lie entirely within one of the two $SU(2)$ above, and so preserve a
$U(1) \times SU(2)$ subgroup of the rotation group. Perturbations
around these solutions may be organized in representations of $SU(2)$
and so obey ordinary rather than partial differential
equations.\footnote{We thank A. Strominger for pointing this out to
  us.} Consequently, the generalization of the hairy black hole
solutions determined in this paper to solutions with self dual spin
appears to be a plausibly tractable project; which we however leave to
future work.

Even though we do not embark on a serious analysis of charged rotating
black holes in this paper, we do present a speculative appetizer for
this problem.  In more detail we present a guess (or a prediction) for
the leading order thermodynamics of these spinning hairy
solutions. Our guess is based on the observation that the
thermodynamics of the hairy solutions constructed in this paper can be
reproduced, at leading order, by modeling the hairy black hole as a
non interacting mix of a RNAdS black hole and the soliton. In section
\ref{rot} we simply assume that a similar model works for charged
spinning black holes, and use this model to compute the thermodynamics
of a certain class of spinning hairy black holes. The most interesting 
aspect of our
results concern the BPS limit. Our non interacting model predicts that
extremal hairy black holes are BPS at every value of the angular
momentum and charge.  This is in stark contrast with Kerr RNAdS black
holes that are BPS only on a co dimension one surface of the space of
extremal black holes.  According to our non interacting model, BPS
hairy black holes are a non interacting mix of Gutowski Reall black
holes \cite{Gutowski:2004yv,Gutowski:2004ez,Kunduri:2006ek} and
supersymmetric solitons.  Such a mix is thermally equlibrated at all
values of charge and angular momentum because of an important property
of Gutowski Reall black holes; their chemical potential is exactly
unity\footnote{We are very grateful to S. Kim for explaining this to us.}. 
We find this result both intriguing and puzzling (see
e.g. \cite{FernandezGracia:2009em}). We emphasize that our prediction
is based on the non interacting superposition model, which may or may
apply to actual black hole solutions. We leave further investigation of 
this extremely interesting issue to future work.

This paper has been devoted largely to the study of small very near
extremal charged black holes in $AdS_5$ that are smeared over
$S^5$. As uncharged small smeared black holes are well known to suffer
from Gregory-Laflamme type instabilities \cite{Hubeny:2002xn}, the
reader may wonder whether the black holes studied in this paper might
suffer from similar instabilities.  We believe that this is not the
case. Recall that the likely end point of a Gregory Laflamme type
instability is a small black hole of proper horizon radius $r_H$ localized 
on the 
$S^5$. In order
that this black hole be near extremal, it has to zip around the $S^5$
at near the speed of light, i.e. at $v=1-\delta$ with $\delta \ll
1$. The $AdS_5$ charge of such a black hole is given by $q \propto
\frac{r_H^7}{\sqrt{\delta}}$ while its energy above extremality of such
a black hole is given by $m-3q \propto r_H^7 \sqrt{\delta}$.\footnote{To
  see this let the sphere be given by equations
  $|z_1|^2+|z_2|^2+|z_3|^2=1$ where $z_i$ are the three complex
  embedding coordinates. A black hole we study is located at
  $|z_1|=|z_2|=|z_3|= \frac{1}{\sqrt{3}}$, and moves with speed
  $\frac{(1-\delta)}{\sqrt{3}}$ in each of the three orthogonal
  planes. Let the proper mass of the black hole be $m_p \propto
  r_H^7$. Its angular momentum in each plane, $q=r \times p$, is given by
  $\frac{1}{\sqrt{3}}\times \frac{m_p \gamma
    (1-\delta)}{\sqrt{3}}=\frac{ m_p \gamma (1-\delta)}{3}$ where
  $\gamma=\frac{1}{\sqrt{1-(1-\delta)^2}}$. The energy of the black
  hole is $m_p \gamma$.} We may now solve for $r_H$ and $\delta$ as a
function of $q$ and $m-3q$. In the near BPS limit of interest to this
paper $m-3q \sim q^2$ and we find $r_H^7 \propto q^\frac{3}{2}$ and
$\sqrt{\delta} \propto \sqrt{q}$. It follows that the entropy of such
a localized black hole $\propto r_H^8 \propto q^{\frac{3}{2} \times
  \frac{8}{7}}$, and so is smaller than the entropy, $(\propto
q^{\frac{3}{2}})$, of the black holes studied in this paper. As the
the black holes studied in this paper have higher entropy than $S^5$
localized black holes with the same charge, there seems no reason to
expect them to suffer from Gregory- Laflamme type
instabilities\footnote{Were we interested in black holes with small
  $q$ and $m-3q \sim {\cal O}(q)$ then we would have found $\delta
  \sim 1$ and $r_H^7 \sim q$. The entropy of the localized black hole
  would then have been $\propto q^{\frac{8}{7}}>q^{\frac{3}{2}}$. As
  such black holes can increase their entropy by condensing, they
  presumably do suffer from Gregory Laflamme type instabilities.}.
Another pointer to the same conclusion is the fact it was very important 
for the analysis of \cite{Hubeny:2002xn} that the black holes they studied
had negative specific heat. The charged black holes at the center of 
the hairy solutions here all have positive specific heat \footnote{We thank
V. Hubeny and M. Rangamani for discussions on this point.}.

We also note that the Gubser Mitra instability
\cite{Gubser:2000ec,Gubser:2000mm} afflicts three equal charge black
holes only when the black holes in question have large enough charge.
It follows that the small black holes primarily studied in this paper
do not suffer from Gubser Mitra type instabilities\footnote{We thank
  M. Rangamani for a discussion on this point.}.

It is conceivable that the solutions presented in this paper might
suffer from further superradiant instabilities, once embedded in IIB
theory on $AdS_5 \times S^5$. In order to see why this might be the
case, let us recall once again why the field $\phi$ - dual to the
chiral Yang Mills operator $Tr X^2+Tr Y^2 + Tr Z^2$ condensed in the
presence of very near extremal charged RNAdS black hole. The reason is
simply that the energy $\Delta=2$ of this field is equal to its charge
$e=2$. As a consequence the Boltzmann suppression factor,
$e^{-\beta(\Delta-e)}$ of this mode exceeds unity when $\mu >1$
causing this mode to Bose condense.  However exactly the same
reasoning applies to, for instance, the field $\phi_n$ dual to the chiral 
operator $Tr X^n + Tr Y^n+ Tr Z^n$ all of which have $\Delta=e$ \footnote{
This statement is true more generally of every operator in the ${\cal N}=1$ 
chiral ring of the theory.}.  
It seems likely that there exist other
hairy solutions in which some linear combination of $\phi_n$ (rather
than simply $\phi_2$) condense \footnote{In the BPS limit any linear
  combination of $\rho_ns$ can condense and we have an infinite
  dimensional moduli space of solutions (see
  \cite{Gava:2006pu,Chen:2007du}). We expect the introduction of a
  black hole to lift this moduli space, to a discrete set of
  solutions.}. It is important to know whether any solution of this
form has higher entropy than the black holes with pure $\phi_2$
condensate presented in this paper. If this is the case then the hairy
black holes of our paper would likely suffer from superradiant
instabilities towards the condensation to the entropically dominant
black hole. On the other hand the black holes of this paper, with
$\phi_2$, the lightest chiral scalar operator that preserves all 
discrete symmetries of the problem, as the only condensate,
are quite special. It seems quite plausible to us that the solution
presented in this paper has the largest entropy of all the hairy
solutions with $\rho_n$ condensates.  If this is indeed the case then
the hairy black hole solutions presented in this paper constitute the
thermodynamically dominant saddle point of ${\cal N}=4$ Yang Mills
very near to supersymmetry; and the entropy of ${\cal N}=4$ Yang Mills
very near to the BPS bound is given the formula \eqref{enthair} below.

To end this introduction we would like to emphasize that the black
hole solutions of this paper give a qualitatively different picture of
the density of states of ${\cal N}=4$ Yang Mills theory at finite
charge compared to a picture suggested by RNAdS black holes. As we
have seen above, there exist no RNAdS black holes with masses between
$m_{BPS}(q)$ and $m_{ext}(q)$, a fact had previously been taken to
suggest that, for some mysterious reason, there are less than ${\cal
  O}(N^2)$ states in Yang Mills theory between $m_{BPS}(q)$ and
$m_{ext}(q)$. The new black hole solutions of this paper establish, on
the other hand, that ${\cal N}=4$ Yang Mills theory has ${\cal
  O}(N^2)$ states all the way down to the BPS bound at least at small
charge, and plausibly at all values of the charge (see
Fig. \ref{two}).\footnote{This difference is starkest in the limit of large charge,
  i.e. in the Poincare Patch limit. The energy density, $\rho_E$, of
  RNAdS black branes is bounded from below by $c \rho_Q^{\frac{4}{3}}$
  where $\rho_Q$ is the charge density. Fig. \ref{two} , on the other
  hand predicts that the energy density of a charged black brane can
  be arbitrarily small at any given value of the charge density.} The
saddle point that governs near BPS behaviour is a
mix of a charged Bose condensate and a normal charged
fluid. It would be fascinating to find some (even qualitative)
confirmation of this picture from a direct field theory analysis.

\section{A Consistent Truncation and its Equations of Motion}\label{sec:cst}

\subsection{A Consistent Truncation of Gauged Supergravity}

${\cal N}=8$ gauged supergravity constitutes a consistent truncation of 
IIB theory on $AdS_5 \times S^5$. In addition to the metric, the 
bosonic spectrum of this theory consists of 42 scalar fields, 15 gauge 
fields and 12 two form fields. 
The scalars transform in the ${\bf 20}$ +${\bf 10_c}$ +${\bf 1}$ +$ {\bf 1}$ 
of SO(6), the gauge fields transform in the 15 dimensional adjoint 
representation, while the two form fields transform in the 
${\bf 6}_c$ representation of $SO(6)$. 

It has been shown \cite{Cvetic:2000nc} that ${\cal N}=8$ gauged
supergravity admits a further consistent truncation that retains only
the scalars in the {\bf 20} and the vector fields in the ${\bf 15}$
together with the metric, setting all other fields to zero. The action
for this consistent truncation is given by \cite{Cvetic:2000nc}
\begin{equation}\label{cvetlag}
\begin{split}
S =&\frac{1}{16 \pi G_5}
\int \sqrt{g} \bigg[ R - \frac{1}{4}T_{ij}^{-1}\left(D_\mu T_{jk}\right)T_{kl}^{-1}\left(D^\mu T_{li}\right)-\frac{1}{8}T_{ik}^{-1}T_{jl}^{-1}F^{ij}_{\mu\nu}(F^{kl})^{\mu\nu} - V\\
&-\frac{1}{48}\epsilon_{i_1 \cdots i_6}\left(F^{i_1 i_2}F^{i_3 i_4} A^{i_5 i_6}- F^{i_1 i_2}A^{i_3 i_4} A^{i_5 j}A^{j i_6} + \frac{2}{5}A^{i_1 i_2}A^{i_3 j} A^{j i_4 }A^{ i_5 k}A^{ k i_6}\right) \bigg]
\end{split}
\end{equation}
where
\begin{equation}\label{not}
 \begin{split}
  V =& \frac{1}{2}\left(2 T_{ij} T_{ij} - (T_{ii})^2\right)\\
F^{ij} =& dA^{ij} + A^{ik}\wedge A^{kj}\\
D_\mu T_{ij} = & \partial_\mu T_{ij} + A_\mu^{ik}T_{kj} + A_\mu^{jk}T_{ik}\\
G_5=& \frac{\pi}{2 N^2}
\end{split}
\end{equation}
Here $(i,j,\cdots)$ denote the $SO(6)$ vector indices and $(\mu,\nu,\cdots)$ 
are the space time indices. $T_{ij}$ are symmetric unimodular (i.e. $T_{ij}$ 
is a matrix of unit determinant) $SO(6)$ tensors. Further $N$ is the rank of 
the gauge group of the dual ${\cal N}=4$ Yang Mills theory, and we work 
in units in which the $AdS_5$ with unit radius solves \eqref{cvetlag}.

We will now describe a further consistent truncation of  
\eqref{cvetlag}. For this purpose we find it useful to move to a complex 
basis for the $SO(6)$ vector indices that appear summed in \eqref{cvetlag}.
Let $(x_j~~~j = 1,\cdots,6)$ denote $SO(6)$ Cartesian directions. We define 
the complex coordinates 
$$x_{2j - 1} + i x_{2j}= z_j,~~x_{2j - 1} - i x_{2j} = {\bar z_j}~~j=1,\cdots,3$$
We will now argue that the restriction 
\begin{equation}\label{fildcont}
 \begin{split}
  T_{z_1z_1} &=T_{z_2z_2} =T_{z_3z_3} =\frac{\phi}{4}\\
T_{{\bar z_1}{\bar z_1}} &=T_{{\bar z_2}{\bar z_2}} =T_{{\bar z_3}{\bar z_3}} =\frac{\phi^*}{4}\\
 T_{z_1{\bar z_1}} &=T_{z_2{\bar z_2}} =T_{z_3{\bar z_3}} =\frac{\sqrt{4 + \phi\phi^*}}{4}\\
A_\mu^{{z_1{\bar z_1}}} & = A_\mu^{{z_2{\bar z_2}}} =A_\mu^{{z_3{\bar z_3}}} = 2 i A_\mu\\
&{\rm All~~ Others =}0
 \end{split}
\end{equation}
 constitutes a consistent truncation of \eqref{cvetlag}. To see this is the 
case note that the permutations of labels 
$(1,2,3)$, as also separate rotations by $\pi$ in the 
$z_1$, $z_2$ and $z_3$ planes, can each be generated by separate $SO(6)$ 
gauge transformations. It follows that these discrete transformations are 
symmetries of  \eqref{cvetlag}. Now it is easy to convince oneself that  
\eqref{fildcont} is the most general field configuration of 
\eqref{cvetlag} that is invariant separately under each of these four 
discrete symmetries. It follows that \eqref{fildcont} is a consistent
truncation of the system \eqref{cvetlag}.

The consistent truncation \eqref{fildcont} is governed by the Lagrangian
\begin{equation}\label{restlag}\begin{split}
S =& \frac{1}{8 \pi G_5} \int \sqrt{g}\left[ \frac{1}{2}\left(R + 12\right)  - 
\frac{3}{8}F_{\mu\nu}F^{\mu\nu} - \frac{3}{16}\left(|D_\mu \phi|^2- \frac{\partial_\mu(\phi\phi^*)\partial^\mu(\phi\phi^*)}{4\left(4 + \phi\phi^*\right)} - 4 \phi\phi^*\right)\right]\\
=& \frac{N^2}{4 \pi^2} \int \sqrt{g}\left[ \frac{1}{2}\left(R + 12\right)  - 
\frac{3}{8}F_{\mu\nu}F^{\mu\nu} - \frac{3}{16}\left(|D_\mu \phi|^2- \frac{\partial_\mu(\phi\phi^*)\partial^\mu(\phi\phi^*)}{4\left(4 + \phi\phi^*\right)} - 4 \phi\phi^*\right)\right]\\
D_\mu\phi & = \partial_\mu\phi - 2 i A_\mu \phi \\
F_{\mu\nu}& = \partial_\mu A_\nu - \partial_\nu A_\mu 
\end{split}
\end{equation}
Note that $\phi$ has charge 2 and  $m^2=-4$. 
Under the AdS/CFT dictionary this field maps to an operator
of dimension $\Delta=2.$ Note also that the kinetic term of the gauge 
field has the factor prefactor $\frac{3}{8}$ rather than the (more usual)
$\frac{1}{4}$ as employed, for instance, in \cite{our}. 
\footnote{Consequently gauge fields and chemical potentials 
in this paper and \cite{our} are related by 
$$ A_{here}= \sqrt{\frac{2}{3}} A_{there}, ~~~ \mu_{here}=
\sqrt{\frac{2}{3}} \mu_{there}$$
Note also that $G_5$ was set to to unity in \cite{our}, while 
$G_5=\frac{\pi}{2 N^2}$ in this paper. It follows that 
$$ \frac{M_{here}}{N^2}= \frac{2}{\pi} M_{there}, 
~~ \frac{S_{here}}{N^2}= \frac{2}{\pi} S_{there}, ~~
\frac{Q_{here}}{N^2}= \frac{2}{\pi} \sqrt{\frac{1}{6}} Q_{there}$$
The factor of $\sqrt{\frac{1}{6}}$ above ensures that 
$T dS_{there}= d M_{there}-\mu_{there} d Q_{there}$
implies 
$T dS_{here}= d M_{here}-3 \mu_{here} dQ_{here}$
as is required on on physical grounds.}

\subsection{Equations of Motion}

We now list the equations of motion that follow from varying \eqref{restlag}.  
We find the Einstein equation
\begin{equation}\label{meteq}
R_{\mu\nu} - \frac{1}{2}g_{\mu\nu}R -6 g_{\mu \nu}= -\frac{3}{2}T^{EM}_{\mu\nu} + \frac{3}{8}T^{mat}_{\mu\nu}
\end{equation}
where  
\begin{equation}\label{strtn}
\begin{split}
 T^{EM}_{\mu\nu} =& {F_\mu}^\sigma F_{\sigma\nu} - \frac{1}{4}g_{\mu\nu}F^{\alpha\sigma}
 F_{\sigma\alpha}\\
T^{mat}_{\mu\nu} =& \frac{1}{2}\left[D_\mu\phi\left( D_\nu\phi\right)^* + D_\nu\phi \left(D_\mu\phi\right)^*\right] -\frac{1}{2}g_{\mu\nu} |D_\sigma\phi|^2 + 2 \phi\phi^* g_{\mu\nu}\\
&-\frac{1}{4(4 + \phi\phi^*)}\left[\partial_\mu (\phi\phi^*)\partial_\nu(\phi\phi^*) - \frac{1}{2}g_{\mu\nu}[\partial_\sigma(\phi\phi^*)]^2\right] 
\end{split}
\end{equation}
the Maxwell equation
\begin{equation}\label{gaugeeq}
\nabla_\sigma {F_\mu}^\sigma = \frac{i}{4}\left[\phi(D_\mu\phi)^* - \phi^*D_\mu\phi\right]
\end{equation}
and the scalar equation
\begin{equation}\label{scaleq}
\begin{split}
 &D_\mu D^\mu\phi + \phi\left[\frac{[\partial_\sigma(\phi\phi^*)]^2}{4 
(4 + \phi\phi^*)^2} - \frac{\nabla^2(\phi\phi^*)}{2(4 + \phi\phi^*)} +4\right]=0.\\
\end{split}
\end{equation}

In this paper we study static spherically symmetric configurations 
of the system \eqref{restlag}
We adopt a Schwarzschild like gauge and set 
\begin{equation} \label{ansatz}
\begin{split}
ds^2&=-f(r) dt^2+ g(r) dr^2+ r^2 d \Omega_3^2\\
A_t&=A(r)\\
A_r&=A_i=0\\
\phi&=\phi^*=\phi(r)\\
\end{split} 
\end{equation}

The four unknown functions $f(r)$, $g(r)$, $A(r)$ and $\phi(r)$ are
 constrained by Einstein's equations, the Maxwell equations and the 
scalar equations. It is possible to demonstrate that $f, g, A , \phi$ are 
solutions to the equations of motion if and only if
\begin{equation}\label{maineqs}
 \begin{split}
 E_1 = & g(r) \left(-\frac{3 \left[A(r)^2+f(r)\right] \phi (r)^2}{4
   f(r)}-\frac{3}{r^2}-6\right)\\
&+\frac{3}{4} \left[-\frac{\phi '(r)^2}{\phi
   (r)^2+4}+\frac{A'(r)^2+\frac{2 f'(r)}{r}}{f(r)}+\frac{4}{r^2}\right]=0\\
E_2 = & \frac{g(r)^2}{f(r)}\left(-A(r)^2 \phi (r)^2-\frac{A'(r)^2}{g(r)}\right)
+  \left(\frac{2 r g'(r)-4
   g(r)}{r^2}\right)\\
&- \frac{g(r) \phi '(r)^2}{\phi
   (r)^2+4}+\left(\frac{4}{r^2}+8+\phi (r)^2\right)g(r)^2=0\\
E_3 = &2 A(r) g(r) \phi (r)^2 
+ A'(r)
   \left(\frac{f'(r)}{f(r)}+\frac{g'(r)}{g(r)}-\frac{6}{r}\right)
-2 A''(r)=0\\
E_4 =&\left(\frac{1}{4 + \phi(r)^2}\right) \nabla^2\phi(r) +\left(1 + \frac{A(r)^2}{f(r)} 
-\frac{\phi '(r)^2}{g(r)\left[\phi(r)^2+4\right]^2}\right)\phi(r)=0\\
\end{split}
\end{equation}
where 
$$\nabla^2\phi(r) = \frac{g(r) \left[\left(\frac{f'(r)}{f(r)}+\frac{6}{r}\right) \phi '(r)+2
   \phi ''(r)\right]-g'(r) \phi '(r)}{2 g(r)^2}$$

The equations $E_1$ and $E_2$ are derived from the $rr$ and $tt$ components 
of the Einstein equations, $E_3$ is the $t$ component of the Maxwell equation 
and $E_4$ is the equation of the scalar field.

As in \cite{our} the equations \eqref{maineqs} contain only 
first derivatives of $f$ and $g$, 
but depend on derivatives up to the second order for 
$\phi$ and $A$. It follows  that \eqref{maineqs} admit a 6 parameter set 
of solutions. One of these solutions is empty AdS$_5$  space, given by $f(r)=r^2+1$, 
$g(r)=\frac{1}{1+r^2}$, $A(r)=\phi(r)=0$. We are interested in those solutions to \eqref{maineqs} that asymptote to AdS space time, i.e. solutions whose large $r$ behaviour 
is given by  
\begin{equation}\label{gCondns} \begin{split}
f(r)&=r^2+1+{\cal O}(1/r^2)\\
g(r)&=\frac{1}{1+r^2} + {\cal O}(1/r^6)\\
A(r)&={\cal O}(1) + {\cal O}(1/r^2) \\
\phi(r)&={\cal O}(1/r^2)\\
\end{split} 
\end{equation}
As in \cite{our} it turns out that these conditions effectively impose 
two conditions on the 
solutions of \eqref{maineqs}, so that the system of equations admits a 
four parameter set of asymptotically AdS  solutions.
We usually also be interested only in solutions that are regular 
(in a suitable sense) in the interior. This requirement will usually 
cut down solution space to distinct classes of two parameter space of 
solutions; the parameters may be thought of as the mass and charge of the 
solutions.

\subsection{RNAdS Black Holes}\label{subsec:RNBH}

The AdS-Reissner-Nordstrom black holes constitute a very well known 
two parameter set of solutions to the equations \eqref{maineqs}.
These solutions are given by 
\begin{equation}\label{bhsol}
 \begin{split}
  f(r) =& \frac{\mu ^2 R^4}{r^4}-\frac{\left(R^2+\mu ^2+1\right) R^2}{r^2}+r^2+1\\
=& \frac{1}{r^4}\left[r^2 - R^2\right]\left[r^4 +r^2(R^2 + 1) - \mu^2 R^2\right]\\
g(r) =& \frac{1}{f(r)}\\
A(r) =& \mu\left(1 - \frac{R^2}{r^2}\right)\\
\phi(r)&=0 \\
\end{split}
\end{equation}
where $\mu$ is the chemical potential of the RNAdS black hole. The
function $V(r)$ in \eqref{bhsol} vanishes at $r=R$ and 
consequently this solution has a horizon at $r=R$. In fact, it 
can be shown that $R$ is the outer event horizon provided 
\begin{equation}\label{extbound}
\mu^2 \leq (1+2 R^2).
\end{equation}

As explained in \cite{our} and in the introduction,  
\eqref{bhsol} is unstable to 
superradiant decay provided in the presence of field of charge $e$ and 
minimum energy $\Delta$ provided $e\mu > \Delta$. Now our field 
$\phi$ has $\Delta=2$ and $e=2$. Moreover, in the limit $R \to 0$, 
RNAdS black holes have $\mu \leq 1$ (this inequality is saturated 
at extremality). It follows that small extremal black holes lie at the 
edge of instability, as mentioned in the introduction. We show below that
very near extremal RNAdS black holes do in fact suffer from super radiant 
instabilities.

\section{The Supersymmetric Soliton in Perturbation Theory}
\label{sec:soliton}

In this section we will construct the analogue of the ground state 
soliton in \cite{our}. The new feature in here is that the soliton 
turns out to be supersymmetric (this is obvious at linearized order). 

In this section we generate the solitonic solution in perturbation theory. 
We use only the equations of motion without imposing the constraints of 
supersymmetry, but check that our final solution is supersymmetric 
(by verifying the BPS bound order by order in perturbation theory). 
This method has the advantage that it generalizes in a straightforward 
manner to the construction of non supersymmetric hairy black holes in 
subsequent sections.

In Section \ref{sec:nsoliton}  
we will revisit this solitonic solution; we will rederive 
it by imposing the constraints of supersymmetry from the start. 
That method has the advantage that it permits a relatively
simple extrapolation of supersymmetric solutions to large charge. 

\subsection{Setting up the perturbative expansion}

We now turn to the description of our perturbative construction. 
To initiate the perturbative construction of the supersymmetric soliton we set
\begin{equation}
 \begin{split}
f(r)&=1+r^2 + \sum_{n} \epsilon^{2n} f_{2n}(r)\\
g(r)&=\frac{1}{1+r^2}+ \sum_{n=1}^\infty \epsilon^{2n} g_{2n}(r)\\
A(r)&=1 + \sum_{n=1}^\infty \epsilon^{2n} A_{2n}(r)\\
\phi(r)&=\frac{\epsilon}{1+r^2} + \sum_{n=1}^\infty
\phi_{2n+1}(r) \epsilon^{2n+1}\\
\end{split}
\end{equation}
and plug these expansions into \eqref{maineqs}. We then expand out 
and solve these equations order by order in $\epsilon$. All equations are 
automatically solved up to ${\cal O}(\epsilon)$. At order 
$\epsilon^{2n}$ the last equation in \eqref{maineqs} is trivial while the first 
three take the form 
\begin{equation}\label{bahirs}
 \begin{split}
  \frac{d}{dr}\bigg(r^2(1 + r^2)^2 g_{2n}(r)\bigg) &= P^{(g)}_{2n}(r)\\
\frac{d}{dr}\left(\frac{f_{2n}(r)}{1+r^2}\right) &= \frac{2(1 + 2 r^2)}{r}
g_{2n}(r) + P^{(f)}_{2n}(r)\\
\frac{d}{dr}\left(r^3\frac{dA_{2n}(r)}{dr}\right) &= P^{(A)}_{2n}(r).\\
 \end{split}
\end{equation}
On the other hand, at order $\epsilon^{2n+1}$ the first three equations in 
\eqref{maineqs}
is trivial while the last equation reduces to 
\begin{equation} \label{phs}
\frac{d}{dr}\bigg(\frac{r^3}{1 + r^2}\frac{d}{dr}
\left[(1 + r^2)\phi_{2n +1}(r)\right]\bigg) = P^{(\phi)}_{2n+1}(r)
\end{equation}
Here the source terms $P^{(g)}_{2n}(r),~P^{(f)}_{2n}(r),~P^{(A)}_{2n}(r)~~ \text{and}~~P^{(\phi)}_{2n+1}(r)$ are completely 
determined by the solution to lower orders in perturbation theory, and 
so should be thought of as known functions, in terms of which we wish to 
determine the unknowns $f_{2n}$, $g_{2n}$, $A_{2n}$ and $\phi_{2n+1}$. 

The equations \eqref{bahirs} are all easily integrated. It also turns out that 
all the integration constants in these equations are uniquely determined by 
the requirements of regularity, normalisability and our definition of 
$\epsilon$, exactly as in \cite{our}. The solution is given by 
\begin{equation}\label{inteq}
\begin{split}
g_{2n}(r) =& \frac{1}{r^2(1 + r^2)^2}\left[\int_0^r P^{(g)}_{2n}(s)ds\right]\\
 f_{2n}(r) =& -(1 + r^2)\int_r^\infty \left[ \frac{2(1 + 2 s^2)}{s}
g_{2n}(s) + P^{(f)}_{2n}(s)\right]ds\\
A_{2n}(r) =&-\int_r^\infty \frac{ds}{s^3} \left(\int_0^{s}P^{(A)}_{2n}(s')ds'\right)\\
\phi_{2n + 1}(r) =& -\frac{1}{1 + r^2}\left[\int_r^\infty ds\left( \frac{1 + s^2}{s^3}\right) \left(\int_0^{s}P^{(\phi)}_{2n+1}(s')ds'\right)\right]
\end{split}
\end{equation}

\subsection{The Soliton up to ${\cal O}(\epsilon^9)$}

The perturbative procedure outlined in this subsection is very easily implemented
to arbitrary order in perturbation theory. In fact, by automating the 
procedure described above, we have implemented this perturbative series to 17th order in a Mathematica programme. In the rest of this subsection we content 
ourselves with a presentation of our 
results to ${\cal O}(\epsilon^9)$.
\begin{equation}\label{sol23}
\begin{split}
 g_2(r) &= 0\\
f_2(r) &= -\frac{1}{4(1 + r^2)}\\
A_2(r) &= -\frac{1}{8(1 + r^2)}\\
\phi_3(r) &= \frac{1}{8(1 + r^2)^3}
\end{split}
\end{equation}

\begin{equation}\label{sol45}
\begin{split}
 g_4(r) &= \frac{r^4}{192(1 + r^2)^5}\\
f_4(r) &= -\frac{r^4}{192(1 + r^2)^5}\\
A_4(r) &= -\frac{r^4+3 r^2+3}{384 \left(r^2+1\right)^3}\\
\phi_5(r) &= \frac{6 r^4+4 r^2+55}{2304 \left(r^2+1\right)^5}
\end{split}
\end{equation}

\begin{equation}\label{sol67}
\begin{split}
 g_6(r) &= \frac{r^4 \left(4 r^4+15 r^2+20\right)}{7680 \left(r^2+1\right)^7}\\
f_6(r) &= -\frac{12 r^8+45 r^6+60 r^4+20 r^2+5}{23040 \left(r^2+1\right)^5}\\
A_6(r) &= -\frac{6 r^8+30 r^6+60 r^4+55 r^2+25}{23040 \left(r^2+1\right)^5}\\
\phi_7(r) &= \frac{120 r^8+460 r^6+1095 r^4+558 r^2+2368}{460800 \left(r^2+1\right)^7}
\end{split}
\end{equation}

\begin{equation}\label{sol89}
\begin{split}
 g_8(r) &= \frac{r^4 \left(169 r^8+1024 r^6+2640 r^4+3320 r^2+2180\right)}{2211840
   \left(r^2+1\right)^9}\\
f_8(r) &= -\frac{\left(5 \left(169 r^8+1024 r^6+2580 r^4+3344 r^2+2288\right)
   r^2+3096\right) r^2+516}{11059200 \left(r^2+1\right)^7}\\
A_8(r) &= -\frac{\left(5 \left(\left(169 \left(r^4+7 r^2+21\right) r^2+5819\right)
   r^2+5543\right) r^2+14721\right) r^2+4191}{22118400
   \left(r^2+1\right)^7}\\
\phi_9(r) &= \frac{\left(5 \left(1014 r^8+6124 r^6+18257 r^4+30484 r^2+36676\right)
   r^2+75784\right) r^2+155759}{132710400 \left(r^2+1\right)^9}
\end{split}
\end{equation}

The soliton obeys the BPS relation $m=3q$ to the order to which we have
carried out our computation (we present more details of the thermodynamics
in Section \ref{sec:thermoMicro}).

\section{The Hairy Black Hole in Perturbation Theory }\label{ssec:ordreps}

\subsection{Basic Perturbative strategy}

We will now present our perturbative construction of hairy black hole 
solutions. In order to set up 
the perturbative expansion we expand the metric gauge field and the 
scalar fields as 
\begin{equation}\label{outexp}
 \begin{split}
 f(r) = & \sum_{n = 0}^\infty \epsilon^{2 n } f_{2n}(r)\\
  g(r) = & \sum_{n = 0}^\infty \epsilon^{2 n } g_{2n}(r)\\
 A(r) = & \sum_{n = 0}^\infty \epsilon^{2 n } A_{2n}(r)\\
  \phi(r) = & \sum_{n = 0}^\infty \epsilon^{2 n + 1} \phi_{2n + 1}(r)\\
 \end{split}
\end{equation}
where the unperturbed solution is taken to be the RNAdS black hole
\begin{equation}\label{bhsolm}
\begin{split}
f_0(r,R) &= V(r),~~g_0(r,R) = \frac{1}{V(r)}\\
A_0(r,R)&=\mu_{0} (1-\frac{R^2}{r^2} )\\
V(r)&=1+r^2\left(1-\frac{R^2 \mu_{0}^2+R^2+R^4}{r^4 } + 
\frac{R^4 \mu_0^2}{r^6}\right)\\
\end{split}
\end{equation}
The chemical potential of our final solution will be given by 
an expression of the form
\begin{equation} \label{muexp} \begin{split}
\mu &= \mu(\epsilon, R)=\sum_{n=0} \epsilon^{2n} \mu_{2n}(R)\\
\mu_{2n}(R)&= \sum_{k=0}^\infty \mu_{(2n,2k)} R^{2k}\\
\mu_{(0,0)}&=1
\end{split}
\end{equation}  
Note that, at the leading order in the perturbative 
expansion, $\mu=1$.

Our basic strategy is to plug the expansion \eqref{outexp} into 
the equations of motion and then to recursively solve 
the later in a power series in $\epsilon$.   We expand our equations in a power series in $\epsilon$. At each order in $\epsilon$ we have a set of linear 
differential 
equations (see below for the explicit form of the equations),
 which we solve subject to the requirements of the normalisability 
of $\phi(r)$ and $f(r)$ at infinity together with the regularity of 
$\phi(r)$ and the metric at the horizon. These four physical requirements
turn out to automatically imply that $A(r=R)=0$ i.e. the gauge field 
vanishes at the horizon, as we would expect of a stationary solution.
These four physical requirements determine 4 of the six integration 
constants in the differential equation, yielding a two parameter 
set of solutions. We fix the remaining two integration constants by adopting 
the following conventions to label our solutions: we require that $\phi(r)$ 
fall off at infinity like $\frac{\epsilon}{r^2}$ (definition of $\epsilon$) 
and  
that the horizon area of our solution is $2 \pi^2 R^3$ (definition of $R$). 
This procedure completely determines our solution as a function of $R$ and $\epsilon$.
We can then read of the value of $\mu$ in \eqref{muexp} on our solution from the 
value of the gauge field at infinity.

As in \cite{our}, the linear differential equations 
that arise in perturbation theory are difficult to solve exactly, but 
are easily solved in a power series expansion in $R$, by matching near field, 
intermediate field and far field solutions. At every order in $\epsilon$ 
we thus have a solution as an expansion in $R$. Our final solutions are, 
then presented in a double power series expansion in $\epsilon$ and $R$. 

\subsection{Perturbation Theory at ${\cal O}(\epsilon)$}

In this section we present a 
detailed description of the implementation of our 
perturbative expansion at ${\cal O}(\epsilon)$. The procedure described 
in this subsection applies, with minor modifications, to 
the perturbative construction at ${\cal O}(\epsilon^{2m+1})$ for all 
$m$. 

Of course all equations are automatically obeyed at 
${\cal O}(\epsilon^0)$. 
The only nontrivial equation at ${\cal O}(\epsilon)$ is 
$D^2 \phi=0$  where $D_\mu = \nabla_\mu - 2 i A_\mu$ is  the linearized 
gauge covariantised derivative about the background \eqref{bhsolm}. 
We will now solve this equation subject to the 
constraints of normalisability at infinity, regularity at the horizon, 
and the requirement that 
$$\phi(r) \sim \frac{\epsilon}{r^2}+{\cal O}(1/r^4)$$
at large $r$.

\subsubsection{Far Field Region ($r \gg R$)}\label{sssec:farfield}

Let us first focus on the region $r \gg R$. In this region the black hole 
\eqref{bhsolm} 
\begin{equation}\label{bsol}
\begin{split}
  f_0(r) =& \frac{\mu_0 ^2 R^4}{r^4}-\frac{\left(R^2+\mu ^2+1\right) R^2}{r^2}+r^2+1\\
=& \frac{1}{r^4}\left[r^2 - R^2\right]\left[r^4 +r^2(R^2 + 1) - \mu_0^2 R^2\right]\\
g_0(r) =& \frac{1}{f(r)}\\
A_0(r) =& \mu_0\left(1 - \frac{R^2}{r^2}\right)\\
\mu_0 =& \sum_{k = 0}^\infty R^{2k}\mu_{(0,2k)}\\
\mu_{(0,0)}=& 1
\end{split}
\end{equation}
is a small perturbation about global AdS space. 
For this reason we expand
\begin{equation}\label{phiexpout}
\phi_1^{out}(r)=\sum_{k=0}^\infty R^{2k} \phi^{out}_{(1,2k)}(r) ,
\end{equation}
where the superscript {\it out} emphasises that this expansion is good at large $r$.
In the limit $R \to 0$, \eqref{bsol} reduces to global AdS space time 
with  $A_t=1$. A stationary linearised fluctuation about this background 
is gauge equivalent to a linearised fluctuation with time dependence 
$e^{-i t}$ about global AdS space with $A_t=0$ ($A_t$ is the temporal 
component of the gauge field). The required solution is 
simply the ground state excitation of an $m^2=-4$ minimally coupled scalar 
field about global AdS and is given by 
\begin{equation}\label{normalisable}
\phi^{out}_{(1,0)}(r) = \frac{1}{1 + r^2}
\end{equation} 
The overall normalisation of the mode is set by our definition of 
$\epsilon$ which implies  
$$\phi^{out}_{(1,0)}(r)=\frac{1}{r^2} +{\cal O}(1/r^4).$$

We now plug \eqref{phiexpout} into the equations of motion $D^2 \phi=0$ and 
expand to ${\cal O}(R^2)$ to solve for $\phi^{out}_{(1,2)}$. 
Here $D^2$ is the gauge covariant Laplacian about the background 
\eqref{bsol}. Now  
$$(D^2)^{out}= (D_0^2)^{out}+R^2 (D_2^2)^{out} + \ldots$$
where $(D_0^2)^{out}$ is the gauge covariant Laplacian about global AdS 
space time with background gauge field $A_t=1$. It follows that, at 
${\cal O}(R^2)$, 
$$(D_0^2)^{out}\phi^{out}_{(1,2)}= -(D_2^2)^{out} \phi_{(1,0)}^{out}=-(D_2^2)^{out} \left[\frac{1}{1+r^2} \right]$$
This equation is easily integrated and we find 
\begin{equation}\label{phout}
\phi^{out}_{(1,2)}(r) =-\left(\frac{1}{1 + r^2}\right)
\left(\frac{\mu_{(0,2)}}{r^2}-2 \left[\mu_{(0,2)}-2\right] 
\log (r)+\left[\mu_{(0,2)}-2\right] \log \left(r^2+1\right)+
\frac{2}{r^2+1}\right)
\end{equation}

We could iterate this process to generate $\phi^{out}_{(1,2k)}$ upto any 
desired order $k$. The equations we need to solve, at order ${\cal O} 
(R^{2k})$, 
takes the form
\begin{equation}\label{R2kph}
\frac{d}{dr}\bigg(\frac{r^3}{1 + r^2}\frac{d}{dr}
\left[(1 + r^2)\phi^{out}_{(1,2k)}(r)\right]\bigg) = P^{out}_{(1,2k)}(r)
\end{equation}
where $P^{out}_{(1,2k)}(r)$ is a source function, whose form is determined 
by the results of the expansion at lower orders in perturbation theory.

As in \eqref{phout}, it turns out that the expressions 
$\phi^{out}_{(1,2k)}$ are increasingly singular as $r \to 0$.  In fact it may 
be shown that the most singular piece of $\phi^{out}_{(1,2k)}$ scales like 
$\frac{1}{r^{2k}}$, upto logarithmic corrections. In other words the 
expansion of $\phi^{out}$ in powers of $R^2$ is really an expansion
in $\frac{R^2}{r^2}$ (upto log corrections)  and breaks down at $r \sim R$.

To end this subsection we summarize our results to ${\cal O}(R^2)$. 
We have 
\begin{equation}\label{2sol}
 \begin{split}
\phi^{out}_1 =& \frac{1}{1 + r^2} +\\
& + R^2 \left[ -\left(\frac{1}{1 + r^2}\right)\left(\frac{\mu_{(0,2)}}{r^2}-2 \left[\mu_{(0,2)}-2\right] \log (r)+\left[\mu_{(0,2)}-2\right] \log \left(r^2+1\right)+\frac{2}{r^2+1}\right) \right] \\
&+{\cal O}(\frac{R^4}{r^4})\\
\end{split}
\end{equation}
Expanding $\phi_1^{out}(r)$ in a Taylor series about $r = 0$ we find
\begin{equation}\label{phmexp}
 \begin{split}
 \phi_1^{out}(r)&= \left[1 - r^2 + {\cal O}(r^4)\right]\\
& +R^2\left[\frac{\mu_{(0,2)}}{r^2}  +(\mu_{(0,2)}-2)
   (2 \log (r)+1)  + {\cal O}(r^2)\right]\\
& + {\cal O}(\frac{R^4}{r^4})(1 + {\cal O}(r^2))
+ \ldots  
 \end{split}
\end{equation} 

Note that this result depends on the as yet unknown parameter $\mu_{(0,2)}$. 
This quantity will be determined below by matching with the intermediate
field solution of the next subsection.

\subsubsection{Intermediate Field Region  $r\ll 1$ and $(r -R)\gg R^3$}
\label{sssec:nearfield}

Let us now turn to intermediate region $ R^3 \ll r -R\ll 1$. 
Over these length scales the small black hole is far from a small 
perturbation about AdS$_5$  space. 
Instead the simplification in this region stems 
first from the fact that we focus on radial distances 
of order $R$ ( $r \sim R \ll 1$). Over these small length scales 
the background gauge field, which is 
of order unity, is negligible compared to the mass scale set by the horizon 
radius $\frac{1}{R}$. 

A second simplification results from the fact that we insist that 
$(r -R)\gg R^3$, i.e. we do not let our length scales become too small. 
 At these distances the back hole that we perturb around are effectively 
extremal (rather than slightly non extremal) at leading order. Moreover 
the black hole may also be thought of (at leading order) as a small black hole
in flat rather than global $AdS$ space. \footnote{As we will see below, deviations of 
the black hole from extremality (and deviations of the form of its metric 
from the metric of a flat space black hole) are crucial to dynamics at 
$r-R \sim R^3$, but are small perturbations on dynamics when $(r -R)\gg R^3$.}

In this region it is convenient to work in 
a rescaled radial coordinate $y=\frac{r}{R}$ and a rescaled time coordinate 
$\tau=\frac{t}{R}$. Note that the near field region consists of space time 
points with $y$ of order unity (but not too near to unity). Points with 
$y$ of order $\frac{1}{R}$ (or larger) and $y-1$ of order ${\cal O}(R^2)$ 
(or smaller) are excluded from the considerations of this subsection.

The metric and the gauge field of the background black hole take the form
\begin{equation}\label{bhsolmy}
 \begin{split}
  ds^2 =& R^2 \left[-V(y) d\tau^2 + \frac{dy^2}{V(y)} + y^2 d\Omega_3^2 \right]\\
V(y) =& \left[1 - \frac{1}{y^2}\right]\left[1 - \frac{\mu_0^2}{y^2} + R^2(1 + y^2)\right]\\
A_\tau(y) =& R \mu_0\left( 1 - \frac{1}{y^2}\right)\\
\mu_0 =& \sum_{k = 0}^\infty R^{2k}\mu_{(0,2k)}\\
\mu_{(0,0)}=& 1
\end{split}
\end{equation}

As in the previous subsection we expand
\begin{equation}\label{pertphmid}
 \begin{split}
 \phi^{mid}_1(y) = \sum_{k = 0}^\infty R^{2k}\phi^{mid}_{(1,2k)}(y)
 \end{split}
\end{equation}

To determine the unknown functions in this expansion, we must solve the 
equation $D^2 \phi^{mid}=0$, where $D^2$ is the gauge covariant Laplacian 
about the background \eqref{bhsolmy}. Our solutions must match with 
the far field expansion of the previous subsection, and the near field 
expansion of the next subsection, but are subject to no intrinsic 
boundary regularity requirements.

At ${\cal O}(R^{2k})$ our equations take the form
\begin{equation}\label{R2kphm}
\frac{1}{y^3}\frac{d}{dy}\bigg(y^3 V_0(y)\frac{d}{dy}
\bigg)\phi^{mid}_{(1,2k)}(y) = P^{mid}_{(1,2k)}(y)
\end{equation}
where $$V_0(y) = \left(1 - \frac{1}{y^2}\right)^2$$
and  $P^{mid}_{(1,2k)}(y)$ is a source term determined (recursively) 
by the perturbative procedure. Ignoring the requirements of matching, 
for a moment, the solution to this equation is determined only upto two 
integration constants at every order. It turns out that 
$\phi^{mid}_{(1,2k)}(y)$ grows like $y^{2k}$ (upto possible logarithmic 
corrections) at large $y$ and grows like $\frac{1}{(y-1)^k}$ as $y$  
approaches unity. It follows that the expansion \eqref{R2kphm} is good 
only when 
$$R^2 \ll (y-1) \ll \frac{1}{R}$$ 

We now work out the explicit solutions at low orders. $P^{mid}_{(1,0)}(y)=0$
vanishes, so the solution for $ \phi^{mid}_{(1,0)}(y)$ is particularly 
simple, and takes the form
$$ \phi^{mid}_{(1,0)}(y) = c_1 + \frac{c_2}{y^2-1}$$
$c_1$ and $c_2$ are the two constants.
It is easy to check that the matching of $\phi^{mid}_{(1,0)}(y)$ with 
$\phi^{out}_{(1,0)}(r)$ sets $c_1 = 1$. It follows on general grounds that 
matching with 
the (as yet undetermined) near field solution forces $c_2$ to vanish. This 
is because, were $c_2$ to be nonzero, it would match onto a near field solution
of order ${\cal O}\left(\frac{1}{R^2}\right)$ in the near field region  (see 
the next subsection for details), violating the requirement that that 
our solution has a smooth $R \to 0$ limit. 

We can now iterate the procedure of this subsection to solve to order in $R^2$
in the intermediate field region. We find
\begin{equation}\label{2insol}
 \begin{split}
\phi^{mid}_{(1,0)}(y) =& 1\\
\phi^{mid}_{(1,2)}(y) =&-y^2 - 2\log(y^2 - 1) + c_3 + \frac{c_4}{y^2-1}\\
\end{split}
\end{equation}
so that 

Here $c_3$ and $c_4$ are the two integration constants. $c_3$ may immediately
be determined by matching with the far field solution; it turns out that 
this procedure also determines $\mu_{(0,2)}=0$, the chemical potential that 
was left undetermined in the previous subsection. In order to perform this 
matching we expand  $\phi^{mid}_1(y)$ about large $y$
\begin{equation}\label{midphexp}
 \phi^{mid}_1(y) = 1 + R^2\left[ - y^2 + c_3 - 4 \log(y) + {\cal O}\left(\frac{1}{y^2}\right)\right] +{\cal O}(R^4 y^4)
\end{equation}

The strategy is now to substitute $y = \frac{r}{R}$ in \eqref{midphexp} 
and then to compare with \eqref{phmexp}. Of course one should only compare 
those terms that are reliable in both expansions. Terms of order $R^{2m} r^{2n}$
are reliably computed from \eqref{midphexp} only when $m+n \leq 1$. Terms 
of the same form are reliably computed from \eqref{phmexp} only when 
$m \leq 1$. Consequently, the only terms that one may reliably compare 
are those of the form ${\cal O}(R^0 r^0)$, ${\cal O}(R^0 r^2)$, ${\cal O}(R^2 r^0)$  together with logarithmic corrections. The difference between the sum 
of the corresponding terms (in \eqref{midphexp} and \eqref{phmexp}) is given
by  
$$ \text{Difference} = R^2\left( \frac{\mu_{(0,2)}}{r^2} + 2\mu_{(0,2)} \log(r) + \mu_{(0,2)} - 2 -4\log(R) - c_3\right)$$
and vanishes provided $\mu_{(0,2)}=0$ and $c_3 = -2(1 + 2\log R)$
so that
\begin{equation}\label{2insolf}
\phi^{mid}_{1}(y)= 1+ R^2 
\left( -y^2 - 2\log(y^2 - 1) -2(1 + 2\log R) + \frac{c_4}{y^2-1} \right)
+{\cal O}(R^4)
\end{equation}
 $c_4$ will be determined below by 
matching to the near field region. To facilitate this matching in the next 
subsection, we present the expansion of $\phi^{mid}_1(y)$ expanded around 
$y = 1$.
\begin{equation}\label{midinexp} \begin{split}
 \phi^{mid}_1(y) &=1 + R^2\left[\frac{c_4}{2(y - 1)} -\left(\frac{c_4}{4} + 3 + 2\log 2 + 4\log R\right) - 2 \log(y-1) + {\cal O}(y-1)\right]\\
+& {\cal O}(\frac{R^4}{(y-1)^2)}
\end{split}
\end{equation}

\subsubsection{Near Field Region  $(r -R) \ll R$ or $(y-1)\ll 1$}

In this subsection we will determine the scalar field in the near field 
region $r-R \ll R$. More particularly, we will work in terms of a further 
rescaled radial coordinate $ z = \frac{y - 1}{R^2} $. Note that the black hole
horizon occurs at $z=0$. Note points at finite $z$ are located at $r-R \sim R^3$
or $y-1 \sim R^2$. It is also convenient to work with the new time coordinate 
$T= R t = R^2 \tau$ As in the previous subsection, the background 
gauge field makes a small direct contribution to dynamics in this region. 
However deviation of the black hole metric from extremality (and the 
difference between an $AdS$ and flat space black hole metric) are all important
in this region, and have to be dealt with exactly rather than perturbatively.

In the new coordinates, the metric and gauge field take the form 
\begin{equation}\label{inback}
 \begin{split}
  \frac{ds^2}{R^2} =& V(z) dT^2 + \frac{ dz^2}{V(z)} + (1 + R^2 z)^2 d\Omega_3^2\\
A_T(y) =& \frac{\mu_0}{R}\left( 1 - \frac{1}{\left(1 + R^2 z\right)^2}\right)
= 2\mu_0 R z^2 +{\cal O}(R^3 z^4) \\
\mu_0 =& \sum_{k = 0}^\infty R^{2k}\mu_{(0,2k)}\\
\mu_{(0,0)}=& 1,~~\mu_{(0,2)}= 0\\
V(z) =&  \frac{1}{R^4} \times \left(1 +\frac{\mu_0 ^2}{\left(R^2 z+1\right)^4} 
- \frac{\mu_0 ^2+1 + R^2}{\left(R^2 z+1\right)^2}
+ R^2\left(1 + R^2 z\right)^2 \right)
\\
=&= 4 z(1 + z ) 
+ {\cal O}(R^2) \\
\end{split}
\end{equation}

As in previous subsections we expand the field $\phi(z)$ as
\begin{equation}\label{pertphin}
 \begin{split}
 \phi^{in}_1(z) = \sum_{k = 0}^\infty R^{2k}\phi^{in}_{(1,2k)}(z)
 \end{split}
\end{equation}
and plug this expansion into the equations of motion. At ${\cal O} (R^{2k})$ 
the equations take the form 
\begin{equation}\label{R2kphin}
\frac{d}{dz}\bigg(4 z(1 + z ) \frac{d}{dz}
\bigg)\phi^{in}_{(1,2k)}(z) = P^{in}_{(1,2k)}(z)
\end{equation}
where, as usual, $P^{in}_{(1,2k)}(z)$ is a source term whose form is determined
from the results of perturbation theory at lower orders. We solve the 
equation \eqref{R2kphin} subject to the requirement of regularity at $z=0$. 
It is possible to argue that the solution to $\phi^{in}_{(1,2k)}(z)$ behaves
like $z^{k-1}$ (upto logarithmic corrections) at large $z$. 

At lowest order $(k=0)$ $P^{in}_{(1,0)}(y)=0$ vanishes, and the unique 
regular solution for  $ \phi^{in}_{(1,0)}(y)$ is  the constant. Matching 
determines the value of the constant to be unity. 
 
At next order, (i.e. ${\cal O}(R^2)$) the solution - after imposing the 
requirement of regularity - is given by 
\begin{equation}\label{p12bh}
\begin{split}
\phi^{in}_{(1,2)}(z) &= 1\\
 \phi^{in}_{(1,2)}(z) &= \alpha-\frac{1}{2} \log ^2(z+1)-2 \log (z+1) -\text{Li}_2(-z)
\end{split}
\end{equation}
where $\alpha$ is the constant which we will now determine by matching with 
the intermediate field solution.
Expanding $\phi^{in}_1(z)$ around $z = \infty$ we find 
\begin{equation}\label{inphexp}
 \phi^{in}(z) = 1 + R^2\left[\alpha + \frac{\pi^2}{6} - 2\log z + 
{\cal O}\left(\frac{1}{z}\right)\right]
\end{equation}
We now substitute $z = \frac{y-1}{R^2}$   in \eqref{inphexp} and then compare
with \eqref{midinexp}. We find a perfect match provided 
$\alpha$ and $c_4$ are chosen to be the following
$$c_4 = 0~~\text{and}~~\alpha = - \left(\frac{\pi^2}{6} + 3 + 2\log 2 + 8 \log R\right)$$

\subsection{Perturbation theory at ${\cal O}(\epsilon^2)$}\label{ssec:ordepssq}

We now briefly outline the procedure used to evaluate the solution at 
${\cal O}(\epsilon^2)$. We proceed in close imitation to the previous 
subsection. The main difference is that at this (and all even orders) in 
the $\epsilon$ expansion, perturbation theory serves to determine the 
corrections to the functions $f$, $g$ and $A$ rather than the function $\phi$. 
The procedure described here applies, with minor modifications, to 
the perturbative construction at ${\cal O}(\epsilon^{2m})$ for all 
$m$. 

\subsubsection{Far Field Region, $r \gg R$}

When $r \gg R$ we expand 
\begin{equation}\label{pertph}
 \begin{split}
 f^{out}_2(r) &= \sum_{k = 0}^\infty R^{2k}f^{out}_{(2,2k)}(r)\\
g^{out}_2(r) &= \sum_{k = 0}^\infty R^{2k}g^{out}_{(2,2k)}(r)\\
A^{out}_2(r) &= \sum_{k = 0}^\infty R^{2k}A^{out}_{(2,2k)}(r)
 \end{split}
\end{equation}
 where 
\begin{equation}\label{zerout}
 \begin{split}
  f^{out}_0(r) =& V(r),~~ g^{out}_0(r) = \frac{1}{V(r)},~~A^{out}_0(r) = \mu_0\left(1 - \frac{R^2}{r^2}\right)\\
\mu_0 =& \sum_{k = 0}^\infty R^{2k}\mu_{(0,2k)}\\
\mu_{(0,0)}=& 1,~~~\mu_{(0,2)} = 0\\
 \end{split}
\end{equation}
As in the previous subsection, we plug this expansion into the equations of 
motion and solve the resultant equations recursively. The equations take 
the form 
\begin{equation}\label{outgmeq}
 \begin{split}
  \frac{d}{dr}\bigg(r^2(1 + r^2)^2 g^{out}_{(2,2k)}(r)\bigg) &= \text{Source}\\
\frac{d}{dr}\left(\frac{f^{out}_{(2,2k)}(r)}{1+r^2}\right) - \frac{2(1 + 2 r^2)}{r}
g^{out}_{(2,2k)}(r) &= \text{Source}\\
\frac{d}{dr}\left(r^3\frac{dA^{out}_{(2,2k)}(r)}{dr}\right) &= \text{Source}.\\
 \end{split}
\end{equation}
and may be thought of as the equations governing sourced linearized 
fluctuations about empty global $AdS$ space with $A_t=1$. 

The equations \eqref{outgmeq} are easily solved by integration. One of the 
integration 
constant in the first equation is fixed by the requirement that 
$f^{out}_{(2,2k)}(r)$ is normalizable (see \eqref{gCondns}). 
The remaining three integration constants (one in the first equation and 
two in the last) will be fixed by matching with the intermediate field 
solution below.

The constraints of matching are particularly simple at ${\cal O}(R^0)$; they
require that the solutions for $g^{out}_{(2,0)}$, $f^{out}_{(2,0)}$ and 
$A^{out}_{(2,0)}$ are all regular at $r = 0$. This is because a far field 
solution of the form $\frac{1}{r^k}$ would match onto an intermediate 
solution of the form $\frac{1}{y^k R^k}$. But this contradicts our basic
assumption that our solutions have a smooth $R \to 0$ limit. It follows that 
at ${\cal O}(R^0)$ \footnote{At higher orders the same reasoning does not
forbid singularities, but determines them in terms of the known intermediate
field behaviour at one order lower.} all our functions obey the same 
equations - and boundary conditions - for the 
2nd order fluctuations about the supersymmetric soliton and we obtain the 
same (unique) solution
\begin{equation}\label{zrootsol}
 \begin{split}
  g^{out}_{(2,0)}(r) &= 0\\
 f^{out}_{(2,0)}(r) &= -\frac{1}{4(1 + r^2)}\\
 A^{out}_{(2,0)}(r) &= -\frac{1}{8(1 + r^2)}\\
\end{split}
\end{equation}

At order ${\cal O}(R^2)$ we find 
\begin{equation}\label{twootsol}
 \begin{split}
  g^{out}_{(2,2)}(r) &= -\frac{1}{4 r^2 \left(r^2+1\right)^3} - \frac{k}{r^2(1 + r^2)^2} \\
 f^{out}_{(2,2)}(r) &= \frac{3 + 5 r^2}{4 r^2(1 + r^2)^2} + \frac{1}{ 1 + r^2}\log\left(\frac{1 + r^2}{r^2}\right) + \frac{k}{r^2}\\
 A^{out}_{(2,2)}(r) &= \frac{1 + 2 r^2}{r^2(1 + r^2)^2} + \frac{2}{ 1 + r^2}\log\left(\frac{1 + r^2}{r^2}\right) + h_1 + \frac{h_2}{r^2}\\
\end{split}
\end{equation}
Here $k,~h_1~ \text{and}~h_2$ are the three undetermined constants, which 
will be determined by matching with the intermediate field solution. 
To facilitate this determination below we end this subsection by presenting 
an expansion of  \eqref{twootsol} 
about $r = 0$
\begin{equation}\label{gmoutx}
 \begin{split}
g^{out}_2(r) &= R^2 \left[-\frac{k+\frac{1}{4}}{r^2}+{\cal O}(r^0)\right] + {\cal O}(R^4)\\
  f^{out}_2(r) &= \left[-\frac{1}{4}+{\cal O}(r^2)\right] + R^2\left[\frac{k+\frac{3}{4}}{r^2}+{\cal O}(r^0)\right] + {\cal O}(R^4)\\
A^{out}_2(r) &= \left[-\frac{1}{8}  + {\cal O}(r^2)\right] + R^2\left[\frac{4h_2 + 1}{4r^2}  + {\cal O}(r^0)\right] + {\cal O}(R^4)
 \end{split}
\end{equation}

\subsubsection{Intermediate field region,  $r \ll 1$ and $(r -R)\gg R^3$}

As in the previous section, we find it convenient to work with 
the variables $ y = \frac{r}{R} $ and $\tau=\frac{t}{R}$ in the intermediate 
field region. Recall also that, in these coordinates, the leading order 
metric has an overall factor of $R^2$. The metric variables that obey 
simple equations have this factor of $R^2$ stripped from them. For that 
reason we define  
Here $$f^{mid}(y) = \frac{g_{\tau \tau}}{R^2}= g_{tt} ~~\text{and}~~ 
g^{mid}(y) = \frac{g_{yy}}{R^2} =g_{r r}$$
(here $g_{\mu\nu}$ are metric components. In a similar fashion we define 
$A^{mid}= \frac{A_\tau}{R}=A_t$. With these definitions we expand
\begin{equation}\label{phmidrt}
 \begin{split}
 f^{mid}_2(y) &= \sum_{k = 0}^\infty R^{2k}f^{mid}_{(2,2k)}(y)\\
g^{mid}_2(y) &= \sum_{k = 0}^\infty R^{2k}g^{mid}_{(2,2k)}(y)\\
A^{mid}_2(y) &= \sum_{k = 0}^\infty R^{2k}A^{mid}_{(2,2k)}(y)
 \end{split}
\end{equation}
 where 
\begin{equation}\label{zermid}
 \begin{split}
  f^{mid}_0(y) =& V(y),~~ g^{mid}_0(y) = \frac{1}{V(y)},~~A^{mid}_0(y) = \mu_0\left(1 - \frac{1}{y^2}\right)\\
\mu_0 =& \sum_{k = 0}^\infty R^{2k}\mu_{(0,2k)}\\
\mu_{(0,0)}=& 1,~~~\mu_{(0,2)} = 0\\
 \end{split}
\end{equation}

The equations are slightly simpler when rewritten in terms of a new function 
$$K_{(2,2k)}(y) = V_0(y) g^{mid}_{(2,2k)}(y) + \frac{f^{mid}_{(2,2k)}(y)}{V_0(y)}$$
 where $$V_0(y) = \left(1 - \frac{1}{y^2}\right)^2$$

 In terms of this function the final set of equations take the form
\begin{equation}\label{samikaran}
 \begin{split}
\frac{dK_{(2,2k)}(y)}{dy} &= \text{Source}\\
\frac{d}{dy}\left(y^3\frac{dA^{mid}_{(2,2k)}(y)}{dy}\right)-\left(\frac{d K_{(2,2k)}(y)}{dy}\right) &= \text{Source} \\
\frac{d}{dy}\bigg(y^2f^{mid}_{(2,2k)}(y)\bigg)-2 y K_{(2,2k)}(y) +2\left( \frac{dA^{mid}_{(2,2k)}(y)}{dy}\right) &= \text{Source} \\
\end{split}
\end{equation}
These equations are all easily solved by integration, upto four undetermined
integration constants (one each from the first and third equation, and 
two for the second). It will turn out that two of these constants are 
determined by matching with the far field solution while the other two 
are determined by matching with the near field solution. As in the previous 
section we will find that $k^{th}$ order solutions scale like $y^{2k}$ at 
large $y$, but scale like $\frac{1}{(z-1)^k}$ at small $z$.

The solution at leading order,  $R^0$, is given by
\begin{equation}\label{gmmid}
 \begin{split}
  f^{mid}_{(2,0)}(y) = & \alpha_1\left(1 - \frac{1}{y^2}\right) -\frac{2 \alpha _2 }{y^2} + \frac{2\alpha_3}{y^4} + \frac{\alpha_4}{y^2}\\
g^{mid}_{(2,0)}(y) =&-\frac{\alpha_1 y^4}{(y^2 - 1)^3} - \frac{2\alpha_3 y^4}{(y^2 - 1)^4}+ \frac{(2\alpha_2 -\alpha_4) y^6}{(y^2 - 1)^4}\\
A^{mid}_{(2,0)}(y) =& \alpha_2 - \frac{\alpha_3}{y^2}
 \end{split}
\end{equation}
Here $\alpha_1,~~\alpha_2,~~\alpha_3~~\text{and} ~~\alpha_4$ are the 
four integration constants to be determined by matching.

Expanding \eqref{gmmid} around $y = \infty$ one finds
\begin{equation}\label{midgmexp}
\begin{split}
 f^{mid}_2(y) =& \alpha_1 +\frac{\left(\alpha_4 - \alpha_1 - 2\alpha_2\right)}{y^2} + {\cal O}\left(\frac{1}{y^4}\right) + {\cal O}(R^2)\\
g^{mid}_2(y) =& \frac{\left(2\alpha_2 -\alpha_1-\alpha_4\right)}{y^2}+ {\cal O}\left(\frac{1}{y^4}\right) + {\cal O}(R^2)\\
A^{mid}_2(y) =& \alpha_2 - \frac{\alpha_3}{y^2} + {\cal O}(R^2)\\
\end{split}
\end{equation}
As usual, we substitute as $y = \frac{r}{R}$ and then match relevant 
terms of \eqref{midgmexp} and \eqref{gmoutx}. This determines 
\begin{equation}\label{constdet}
 \begin{split}
  \alpha_1 =& -\frac{1}{4}\\
\alpha_2 =&-\frac{1}{8}\\
h_2 =&-\left(\alpha_3 + \frac{1}{4}\right)\\
k =& \alpha_4 - \frac{1}{4}\\
 \end{split}
\end{equation}
To facilitate matching with the near field region in the next subsection 
we expand 
$f^{mid}_2(y),~~g^{mid}_2(y) ~~ \text{and}~~A^{mid}_2(y)$ about $y = 1$
\begin{equation}\label{ingmexp}
\begin{split}
 f^{mid}_2(y) =&\bigg[ \left(\frac{1}{4} + 2 \alpha_3 + \alpha_4\right)-\left(1 + 8 \alpha_3 + 2\alpha_4\right)(y - 1) + \left(\frac{3}{2} + 20\alpha_3 + 3\alpha_4\right)(y - 1)^2\\
 &+{\cal O}\left((y-1)^3\right) \bigg]+ {\cal O}(R^2)\\
g^{mid}_2(y) =&\bigg[-\frac{1 + 8\alpha_3 + 4\alpha_4}{64(y-1)^4} -\frac{1 + 8\alpha_3 + 8\alpha_4}{32(y-1)^3}-\frac{1 + 8\alpha_3 + 44\alpha_4}{128(y-1)^2}\\
&+{\cal O}\left(\frac{1}{y-1}\right) \bigg]+ {\cal O}(R^2)\\
A^{mid}_2(y) =& -\left(\frac{1}{8} + \alpha_3\right)+ 2\alpha_3 (y -1) +{\cal O}\left((y-1)^2\right)+ {\cal O}(R^2)\\
\end{split}
\end{equation}

\subsubsection{Near Field Region $(r -R) \ll R$ or $(y-1)\ll 1$} 

As in the previous section we work with the shifted and rescaled radial 
coordinate $ z = \frac{y - 1}{R^2} $. In this coordinate the black hole 
horizon is at $z = 0$. As we have seen in the previous section, $g_{TT}$ 
and $g_{rr}$ have an overall factor of $R^2$ even at leading order. For 
this reason the natural dynamical variables in the problem are 
$$\frac{g_{TT}}{R^2} = \frac{g_{tt}}{R^4} 
~~\text{and}~~  \frac{g_{zz}}{R^2}=R^4 g_{rr}$$
(here $g_{\mu\nu}$ is the metric).
For easy of matching with the intermediate field solution however, 
we will continue to use the notation 
$$f^{in}=g_{tt}= R^4 \times \frac{g_{TT}}{R^2}$$ 
$$g^{in}=g_{rr}= \frac{1}{R^4} \times \frac{g_{zz}}{R^2}$$
$$A^{in}=A_t=R^2 \times \frac{A_T}{R}$$

And so our perturbative expansion takes the form (note the lower limits of 
the sums)
\begin{equation}\label{gminrt}
 \begin{split}
 f^{in}_2(z) &= \sum_{k = 2}^\infty R^{2k} f^{in}_{(2,2k)}(z)\\
g^{in}_2(z) &= \sum_{k = -2}^\infty R^{2k}g^{in}_{(2,2k)}(z)\\
A^{in}_2(z) &= \sum_{k = 1}^\infty R^{2k}A^{in}_{(2,2k)}(z)
 \end{split}
\end{equation}

We now come to an important subtlety of our expansion procedure. First recall 
that the radial coordinate $r$ employed in this paper has geometrical 
significance; it parametrizes the volume of the $S^3$ at that point.
For this reason reparametrizations of $r$ do not form a symmetry of 
the equations in this paper, in general. At leading order in the near 
field region, however, the  metric metric takes the form 
\begin{equation}\label{lidmet}
\begin{split}
 \frac{ds_0^2}{R^2} = -4  z(1 + z) dT^2 + \frac{ dz^2}{4 z (1 +z)} + d\Omega_3^2
\end{split}
\end{equation}
Note in particular that the size of three sphere (at leading order) is a 
constant independent of $z$. For this reason the leading order metric 
equations in the near field region admit a whole functions worth (instead
of 4 numbers worth) of solutions, parametrized by any 
${\cal O}(\epsilon^2R^0)$ redefinition of $z$ coordinate. So without even doing
any calculations, we have deduced that one linear combination of the three 
functions is undetermined at leading order. 

Now let us move to the next order, ${\cal O}(R^2)$. As the homogeneous part
of the equations are same at every order, the same linear combination of 
second order fluctuations disappears from (i.e. is undetermined by) the 
second order equations. However the 0 order `gauge transformation' is now
not a symmetry of the ${\cal O}(R^2)$ equations (because, at this order, we 
see the fact that the size of the sphere is not really constant). So the 
zero order `gauge transformation function' shows up in the second order 
equations. As this term comes with an explicit $R^2$ (without this factor 
the equations cannot
 distinguish it from pure gauge) it cannot multiply any 
of the 2nd order unknowns, and so appears as a genuine unknown all by itself. 
The net upshot of all this is that at every order other than the leading, 
we actually do have as many equations as variables. The variables, however, 
consist of two unknown functions at that order coupled with the one unknown
`gauge transformation' at the previous order!

We will now say all of this more precisely. Our equations can be simplified
 by performing the following redefinition of the functions ($W$ is essentially
the `gauge transformation')
\begin{equation}\label{redefv}
 \begin{split}
f^{in}_{(2,2k)}(z) =&  \frac{d}{dz}\left[4z(1 + z)\right]W_{(2,2k)}(z)\\
g^{in}_{(2,2k)}(z) =& \zeta_{(2,2k)}(z)+\frac{d}{dz}\left[\frac{1}{4z(1 + z)}\right] W_{(2,2k)}(z)  +\left[\frac{1}{2 z(1 + z)}\right]\frac{d}{dz}\left[ W_{(2,2k)}(z)\right] \\
A^{in}_{(2,2k)}(z) = & \chi_{(2,2k)}(z) + 2 W_{(2,2k)}(z)\\
\end{split}
\end{equation}
 In terms of these functions the equations at order $R^{2k}$ take the form
\begin{equation}\label{homoeq}
 \begin{split}
 \frac{d}{dz}\left[z(1 + z)(1 + 2z)^2\zeta_{(2,2k)}(z)\right] =&~ \text{Source}\\
\frac{d}{dz}\left[ \chi_{(2,2k)}(z)\right] - 4 z(1 + z)\zeta_{(2, 2k)}(z) 
=&~ \text{Source} \\
\frac{d^2}{dz^2}\left[ W_{(2,2k-2)}(z)\right] = &~\text{Source}
 \end{split}
\end{equation}

As we anticipated above, $W_{(2,2k)}(z)$ does not appear in the homogeneous 
equations at ${\cal O}(R^{2k})$ as at this order it is pure gauge. But 
it appears in the homogeneous equations of ${\cal O}(R^{2k+2})$. Therefore 
to completely determine the metric and gauge field (upto integration 
constants) at any given order $R^{2k}$ , one has to solve one more equation 
at the order $R^{2k + 2}$ along with all the equations at order $R^{2k}$.

The equations \eqref{homoeq} are completely well posed, and may easily be 
integrated to solve for $\zeta_{(2,2k)}(z)$, $\chi_{(2,2k)}(z)$ and 
$W_{(2,2k-2)}(z)$ in terms of four integration constants.
Two of these constants are determined by the requirement that 
$f^{in}_{(2,2k)}(z)$ and  $A^{in}_{(2,2k)}(z)$ vanish at the horizon $z = 0$. 
The remaining  two constants  are determined by matching with the intermediate
range solution.

Solving the first two equations at ${\cal O}(R^0)$ and the third equation at 
${\cal O}(R^2)$ one can find $\zeta_{(2,0)}(z),~\chi_{(2,0)}(z)~\text{and}~W_{(2,0)}(z)$ respectively.
The solution is the following.
\begin{equation}\label{intsol}
 \begin{split}
  \zeta_{(2,0)}(z) =& \frac{\Lambda_1}{z(1 + z)(1 + 2z)^2}\\
\chi_{(2,0)}(z) = &\frac{4\Lambda_1z}{1 + 2z} + \Lambda_2\\
W_{(2,0)}(z) = & -\frac{\log[8(1 + z)]}{8} + \frac{\Lambda_1}{1 + 2z} + z \beta_1 + \beta_2
 \end{split}
\end{equation}
Regularity at the horizon  implies that
$$\Lambda_1 = \beta_2 + \frac{3\log2}{8},~~\Lambda_2 = 0$$
After imposing the regularity at $z = 0$ the solution at the leading order
\begin{equation}\label{ingmsol}
 \begin{split}
 f^{in}_{(2,4)}(z)=& -\frac{1}{2}(1 + 2z)\log(1 + z) -3(\log2)z + 4\beta_1 z(1 + 2z) + 8 \beta_2z\\
g^{in}_{(2,-4)}(z) =& \frac{1}{32 z^2 (1 + z)^2}\left[ (1 + 2z)\log(1 + z) +2z (3\log 2 - 1)\right] + \frac{\beta_1 - 2\beta_2}{4z(1 + z)^2}\\
A^{in}_{(2,2)}(z) = & -\frac{1}{4}\log(1 + z) + 2 z \beta_1
 \end{split}
\end{equation}
Here $\beta_1~\text{and}~\beta_2$ are the two constants which are to be determined by matching.
Expanding around $z=\infty$ one finds
\begin{equation}\label{ingminexp}
 \begin{split}
 f^{in}_2(z)=& R^4\left[8 \beta_1 z^2 + {\cal O}(z) \right]+ {\cal O}(R^6)\\
g^{in}_2(z) =& {\cal O}(R^{-2}) \\
A^{in}_2(z) = & R^2\left[2 \beta_1 z +{\cal O}{(z^0)}\right] + {\cal O}(R^4)\\
\end{split}
\end{equation}
After substituting $z = \frac{y - 1}{R^2}$ this expansion will match with \eqref{ingmexp} provided one chooses the constants in the following way
\begin{equation}\label{consteqd}
 \begin{split}
  \alpha_3 =& -\frac{1}{8}\\
\alpha_4 =& 0\\
\beta_1 =& -\frac{1}{8}
 \end{split}
\end{equation}
In the whole solution at this order there are two constants left undetermined.
The first is $\beta_2$ in the near field solution and the second is $h_1$ 
in the far field solution. In the expansion of $f^{in}_2(z)$ the 
constant $\beta_2$ appears at ${\cal O}( R^4z)$ which is equivalent to 
a term of ${\cal O}\left[ R^2(y-1)\right]$ in expansion of 
$f^{mid}_2(y)$. Therefore to compute this constant one needs the solution 
upto ${\cal O}(R^2)$ in the intermediate region. Upon determining this solution
to ${\cal O}(R^2)$ one can solve for $\beta_2$ (as well as all the new 
constants appearing in the ${\cal O}(R^2)$ intermediate solution) in terms 
of $h_1$. It turns out that 
 $$\beta_2 = \frac{1}{16} + \frac{3}{8}\log2 -\frac{h_1}{2}$$
So at the end the full solution to ${\cal O}(R^2)$ is determined in terms 
of a single constant,
$h_1$, which in turn is determined only by the $\epsilon^3$ 
order scalar field analysis. It turns out that 
$$h_1  = 0$$

\section{All Spherically Symmetric Supersymmetric Configurations}
\label{sec:nsoliton}
In this section we will analyze the set of spherically symmetric
supersymmetric solutions of \eqref{spclag}. The configurations we will
find will include the solitons of section \ref{sec:soliton} (determined
more simply than in that section), but will also include
several solutions that are singular at the origin. In particular (as
we have explained in the introduction) we will identify a one
parameter set of singular supersymmetric solutions which we will conjecture
to be physical; we will conjecture that these singular configurations
may be obtained as the limit of nonsingular nonextremal solutions.

\subsection{The Equations of Supersymmetry}
\label{ssec:susysolitoneq}

The action \eqref{spclag} is a consistent truncation of ${\cal N}=8$
gauged supergravity. Hence any solution of the equations of motion
\eqref{maineqs}, which saturates the BPS bound $m=3q$, corresponds to
a supersymmetric solution of ${\cal N}=8$ gauged supergravity and
consequently of IIB SUGRA on $AdS_5\times S^5$. Here we present a more
direct analysis of the supersymmetry equations for the consistent
truncation.  \footnote{Let us first briefly describe how one could 
supersymmetrize
the action \eqref{spclag}. The bosonic field content of our theory is
that of minimal gauged supergravity (i.e. the graviton and the $U(1)$
gauge field) coupled to matter (the charged scalar $\phi$). The scalar
field can be thought of as a member of a hypermultiplet. A complete
hypermultiplet would contain 2 complex scalar fields. Therefore, to
supersymmetrize the action \eqref{spclag} we have to add one more
scalar field, besides the fermions, and the resulting theory is
minimal gauged supergravity coupled to a hypermultiplet. However, in
the set of solutions that we are interested in, the additional scalar
field can be consistently set to zero so we can ignore it in what
follows.}

The supersymmetry conditions in theories of this type have been
analyzed in \cite{Chong:2004ce, Liu:2007xj,Chen:2007du,Liu:2007rv}.
In these works supersymmetric solutions were found for a more general
truncation of ${\cal N}=8$ supergravity to $U(1)^3$ gauged
supergravity coupled to 3 hypermultiplets.  Our theory is a special
case of theirs where all three U(1) charges and three hyperscalars are
taken to be equal. Specializing their results to our theory we find
that spherically symmetric supersymmetric configurations can be
written as follows. The metric, gauge field and scalar
are\footnote{Let us explain our notation in relation to the notation
  of \cite{Liu:2007xj}. We have: $ r_{there}= \rho_{here},\quad
  (H_1)_{there}= (H_2)_{there}=(H_3)_{there}= h_{here},\quad
  A_{there}=-A_{here},\quad 2\sinh(\phi_{there})=\phi_{here}$.}
\begin{equation}
\label{metricbps}
\begin{split}
& ds^2 = - \, {1+\rho^2 h^3 \over h^2}\,\, dt^2 + {h\over 1+\rho^2 h^3} 
\,\, d\rho^2 +  \rho^2\,h\, d\Omega_3^2\cr\cr
& A = h^{-1} dt,\qquad  \phi =
2\sqrt{(h + \rho h'/2)^2-1}
\end{split}
\end{equation}
The entire solution is then determined by the single function
$h(\rho)$ which is constrained to satisfy the following ordinary differential
equation
\begin{equation}\label{solode}
(1 +\rho^2\, h ^3) \left(3\,h' + \rho\, h''\right)=\rho \left[4- \left(2\,h +
 \rho\, h'
\right)^2\right] h^2
\end{equation}
Notice that prime denotes differentiation with respect to the variable
$\rho$.

This parametrization of the metric is somewhat different from the one
that we used in the previous section, so we explain how the two are
related. Comparing the coefficient of $d\Omega_3^2$ in the metric
\eqref{metricbps} to that of \eqref{ansatz} we see that the two radial
coordinates are related by
\begin{equation}\label{radiuschange}
r^2 = \rho^2\,h(\rho) 
\end{equation}
Comparing the other coefficients of the two metrics we find
\begin{equation}
\label{identmetric}
f(r)  = {1 + \rho^2 h^3\over h^2} \qquad ,\qquad g(r) = {4 \rho^2 h^2 \over (2\rho h + \rho^2 h')^2(1+\rho^2 h^3)}
\end{equation}
With these identifications it is a matter of algebra to verify that
equation \eqref{solode} is sufficient for the equations of motion
\eqref{maineqs} to be satisfied.

In summary the most general spherically symmetric supersymmetric solutions 
to the equations of motion \eqref{maineqs} is given by the configuration
\begin{equation}\label{susyconst}\begin{split}
g(r) &= {4 \rho^2 h^2 \over (2\rho h + \rho^2 h')^2(1+\rho^2 h^3)}\\
f(r) & = {1 + \rho^2 h^3\over h^2}\\
A(r)&=\frac{1}{h(r)}\\
\phi(r)&=2\sqrt{(h + \rho h'/2)^2-1}\\
\end{split}\end{equation}
with 
\begin{equation}\label{defsus}
r^2=\rho^2 h(\rho)
\end{equation}
and $h(\rho)$ any function that obeys \eqref{solode}.

\subsection{Classification of Supersymmetric Solutions}\label{class}

As we have explained in the previous subsection, supersymmetric solutions 
to the equations of motion are given by solutions to the second order 
differential equation
\begin{equation}\label{solode}
(1 +\rho^2\, h ^3) \left(3\,h' + \rho\, h''\right)=\rho \left[4- \left(2\,h +
 \rho\, h'
\right)^2\right] h^2
\end{equation}
In this paper we are only interested in regular normalizable solutions to 
these equations. It is of crucial importance to this section that the 
condition of normalizability is automatically met; 
an analysis of \eqref{solode} at large $\rho$ 
immediately reveals that ${\it all}$ solutions to this equation behave at 
large $\rho$ like $$h(\rho) = 1+ {2q\over \rho^2}+...$$
\footnote{Using \eqref{radiuschange} and \eqref{metricbps} we find
that this implies the large $r$ behavior of the gauge field 
$$
A(r) =  1 -{2 q \over  r^2} +...
$$
so the constant $q$ may be identified with the electric charge of the 
solution, in the conventions of previous sections. }
ensuring normalizability for all physical fields\footnote{This fact 
has a natural explanation from the viewpoint of the dual ${\cal N}=4$ 
Yang Mills field theory; a deformation of the Lagrangian of that theory
by only mass term $Tr X^2+ Tr Y^2 +Tr Z^2$ preserves no supersymmetry.}. 
The importance of this observation is the following; one may study 
the small $\rho$ behaviour of supersymmetric solutions in a purely local 
manner, without having to worry about when the solutions we study have 
acceptable large $\rho$ behaviour, as that is always guaranteed. This fact 
allows us, in this section,  to use local analysis to present a simple 
classification of normalizable supersymmetric solutions. Relatedly, 
supersymmetric solutions may be obtained by solving \eqref{solode} as an 
initial value problem with initial conditions set at small $\rho$. This is 
numerically and conceptually simpler than the boundary value problem we would 
have to solve by shooting methods off supersymmetry. 

It remains to impose the condition of `regularity'. Let us first explain 
what we mean by this term. We call a supersymmetric configuration 'regular' 
if it can be regarded as the limit of a one parameter set of smooth 
nonsupersymmetric (and so non extremal) solutions to the equations of motion 
\eqref{maineqs}. While every smooth supersymmetric solution is automatically 
`regular', a singular susy solution may also be `regular', if its singularity 
can be removed upon heating the solution up infinitesimally. 

Solutions to \eqref{solode} can develop singularities only at $\rho=0$. 
In this subsection we will classify all possible behaviours of solutions to 
\eqref{solode} near $\rho=0$. In a later subsection we will then go on 
to present conjectures about which of these solutions are `regular'. 

In order to investigate possible behaviours of solutions to
\eqref{solode} at small $\rho$ we plug in the ansatz $h(\rho)
=\frac{A}{\rho^{\alpha}}$ into the equation. It is easy to check that
the only values of $\alpha$ that solve the equation near $\rho=0$ are
$\alpha=0, \frac{2}{3}, 1, 2$.  It is also possible to demonstrate
(see below) that there is a unique solution with
$\alpha=\frac{2}{3}$. On the other hand solutions with $\alpha=0$ and
$\alpha=1$ both appear in a one parameter family. Finally, the generic
solution to the differential equation has $\alpha=2$; solutions with
$\alpha=2$ appear in a 2 parameter family.\footnote{ One special
  exact solution of \eqref{solode} is $h(\rho) = 1+ {q\over \rho^2}$,
  the so-called ``superstar'' \cite{Behrndt:1998ns}. This solution has
  $\alpha=2$. For this solution we notice that the scalar field is not
  turned on ($\phi=0$). So in a sense it is qualitatively different
  from the ``hairy'' configurations of interest to us in this paper.}

\subsubsection{ $h(\rho) \approx \rho^{-\frac{2}{3}}$}

As we have mentioned above, there is a unique solution with 
$\alpha=\frac{2}{3}$. This solution may be expanded at small $\rho$ 
as follows
\begin{equation}
\label{twothirds}
h(\rho) = \rho^{-2/3} + {9 \over 26} \,\rho^{2/3} -{243\over 20956}\, \rho^2 +{\cal O}(\rho^{8/3})
\end{equation}
We now present a crude estimate for validity domain of the expansion 
\eqref{twothirds}.
Note that the formal procedure that generates the series expansion
\eqref{twothirds} treats the term proportional to $4$ (in the RHS of
\eqref{solode} ) as subleading to the term proportional to $h$. This procedure
is valid whenever $\rho \ll 1$; as a consequence we expect the
expansion \eqref{twothirds} to be valid whenever $\rho \ll 1$ but to
break down at larger values of $\rho$.  

For $\rho\ll 1$ the metric, gauge field and scalar corresponding to this
solution take the following form 
\begin{equation}
\label{metricS}
\begin{split}
& ds^2 \approx -2 r^2 dt^2 + {9 \over 8} dr^2 + r^2 d\Omega_3^2 \cr
& A(r) \approx r dt\cr
&\phi(r) \approx {4\over 3r}
\end{split}
\end{equation}
where we used the relations \eqref{susyconst} and \eqref{defsus} to bring
the solution in the form of \eqref{ansatz}. 

We will denote the distinguished singular solution of this subsection
by $S$. We will now explain that there is a sense in which $S$ is a
fixed point of the equation \eqref{solode} viewed as a flow equation
in the variable $\log \rho =x$ (see
\cite{Heinzle:2003ud, Vaganov:2007at, Hammersley:2007rp} for similar
discussions in distinct but similar contexts). For this purpose we redefine
$$h(\rho) = e^{-{2\over 3} x} f(x)
$$ 
The differential equation \eqref{solode} becomes
$$
9f''(1+f^3)+3f'(2+3 f^2 f'+10 f^3)+8(f^4-f) - 36 e^{4 x/3} f^2=0
$$ For very small $\rho$ (that is for $x\rightarrow- \infty$) the last
term in the equation can be ignored, and the equation becomes
approximately time translation invariant (or an autonomous equation,
in the language of dynamical systems).  With this approximation the
system has an exact solution $f(x)=1$, which is precisely the leading
small $\rho$ approximation to the solution $S$. We will restrict
attention to large negative values of $x$ in the rest of this
subsubsection, and so study the truncated equation
\begin{equation}\label{trunc}
9f''(1+f^3)+3f'(2+3 f^2 f'+10 f^3)+8(f^4-f) =0
\end{equation}
Let us consider a small perturbation about $f=1$, i.e. we set
\begin{equation}\label{sp}
f(x) = 1 + \varepsilon g(x) . 
\end{equation}
To linear order in $\varepsilon$ the \eqref{trunc} turns into the linear ODE
\begin{equation}\label{trunclin}
3g'' + 6g' + 4g =0
\end{equation}
The two linearly independent solutions to this equation are 
\begin{equation} \begin{split}\label{linsol}
g(x) &= e^{\lambda x}\\ 
\lambda &= -1 \pm {i \over \sqrt{3}}\\
\end{split}
\end{equation}
Note that the real part of the each of these eigenvalues is negative, which 
demonstrates that $f=1$ is a stable fixed point of the dynamical system 
\eqref{trunc}. 
Moving back to the variable $\rho$, it follows that arbitrary small perturbations around the solution $S$ behaves like
$$
h(\rho) \approx \rho^{-2/3} \left( 1 + \varepsilon\, \frac{1}{\rho} 
\cos\left({1\over \sqrt{3}} \log\rho + a\right) \right)
$$ 
Note that all perturbations die out for $\rho \gg \epsilon$ (this is a 
restatement of the fact that $f=1$ is a stable fixed point).

\subsubsection{ $h(\rho) \approx h_o +{\cal O}(\rho^2)$}
\label{sssec:matcha}

We now turn to regular solutions to \eqref{solode}, i.e. solutions with 
$\alpha=0$. Such solutions appear in a one parameter set, labeled by
$h_0=h(0)$.  The small $\rho$ expansion of \eqref{solode} is given by  
\begin{equation}
\label{hexpand}
h(\rho) = h_0 + {1\over 2}(h_0^2 - h_0^4) \rho^2 + {1\over 6}(h_0^3
-5 h_0^5 + 4 h_0^7)\rho^4 + {\cal O}(\rho^6)
\end{equation}
The solutions of this subsection are simply the solitons studied in 
section \ref{sec:soliton}. These solutions were generated perturbatively 
(i.e. at $h_0-1$ small) in section \ref{sec:soliton}. 

Let us now turn to the opposite limit of large $h_0$. Let us first inquire 
as to the validity domain of the expansion \eqref{hexpand}. 
As the term $h^3 \rho^2$ on the LHS of \eqref{solode} is of order
$\rho^2$, the formal process that generates the power series expansion
\eqref{hexpand} treats this term as subleading compared to unity. This
is actually correct only when $h_0^3 \rho^2 \sim 1$.  It follows that
when $h_0$ is large the series expansion \eqref{hexpand} will break
down at the small value $\rho \sim \rho_{rb}\sim h_o^{-\frac{3}{2}}$.

\begin{figure}[h]
\begin{minipage}[t]{0.5\linewidth} 
\centering
\psfrag{a}[t]{$\rho$}
\psfrag{b}[r]{$h(\rho)$}
\includegraphics[totalheight=0.18\textheight]{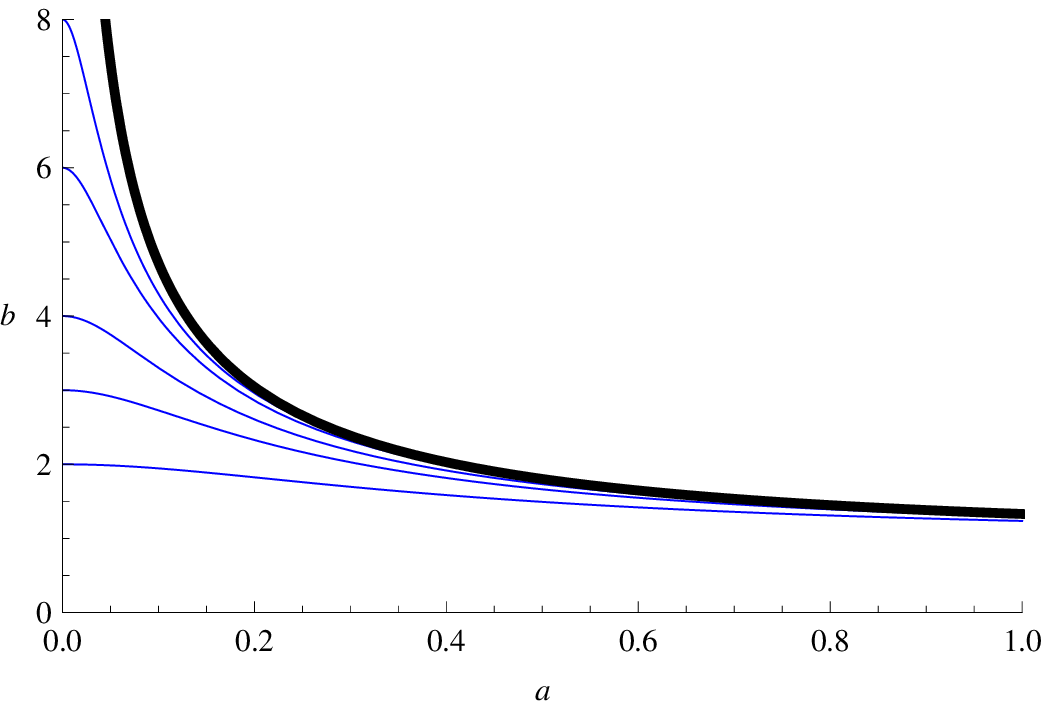}
\caption{Convergence of the numerical solutions for the regular
solitons to the special singular solution $S$ as we increase $h_0 = h(0)$.
The black line corresponds to the solution $S$ with $\rho^{-2/3}$ behavior
near $\rho=0$. The blue lines correspond to regular solitons of $h_0 = 2,3,4,6,8$, starting from the lowest blue curve and going up.}
\label{nc1}
\end{minipage}
\hspace{0.5cm} 
\begin{minipage}[t]{0.5\linewidth}
\centering
\psfrag{a}[t]{$\log h_0$}
\psfrag{b}[r]{${q-q_c\over e^{-{3\over 2} \log h_0}}$}
\includegraphics[totalheight=0.18\textheight]{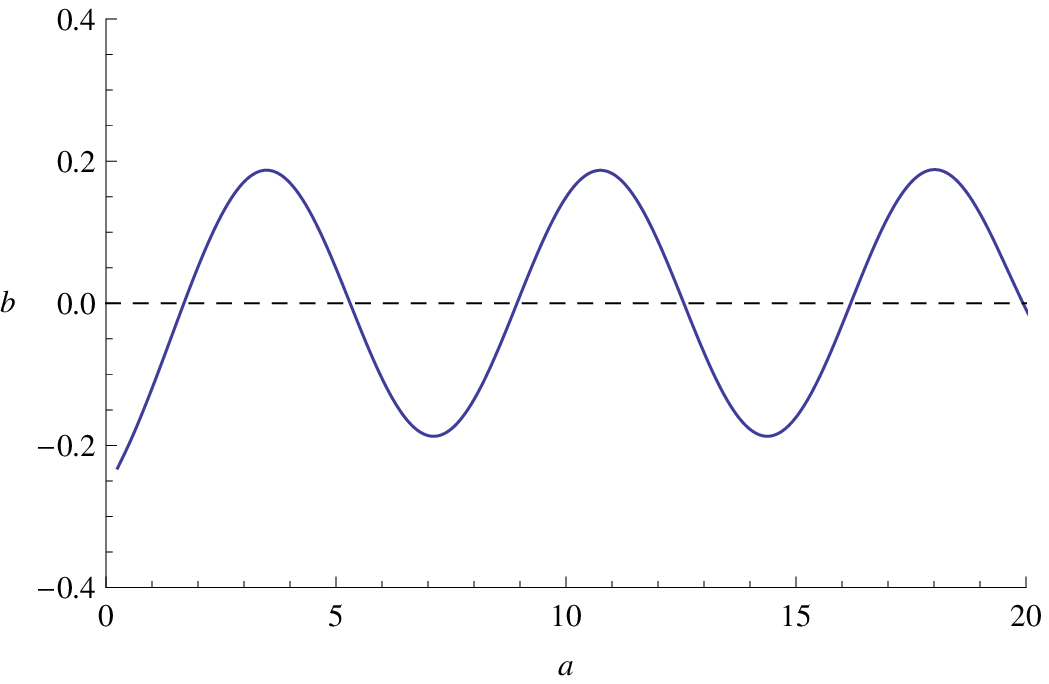}
\caption{The damped oscillations of $q$ around the critical value 
$q_c$ for large $h_0$.}
\label{roscq}
\end{minipage}
\end{figure}

When $h_0$ is large,  the expansion \eqref{hexpand} does not apply in the range 
$\rho \gg h_o^{-\frac{3}{2}}$  . If $\rho \ll 1$ in this range, however, 
the general arguments presented above guarantee that our solution 
behaves like $\rho^{-\alpha}$ for one of the allowed values of $\alpha$ 
described above. What solution does the expansion \eqref{hexpand} match 
onto in this range? A clue to the answer to this question is given by 
noting that the value of $h(\rho)$,  at the point of break down 
of \eqref{hexpand} is approximately given by $h_0\sim \rho_{rb}^{-\frac{2}{3}}$. 
Thus the solution \eqref{hexpand} could smoothly match onto 
the special solution $S$ of previous subsubsection, at $\rho \sim \rho_{rb}$. 
This suggests that  the special solution $S$ of the 
previous subsubsection is the limit as $h_0 \to \infty$ of the solutions
of  this subsubsection. We now present numerical evidence that strongly 
supports this guess. In Fig. \eqref{nc1} we present numerically generated
plots of the regular solution parametrized by $h_0$ for successively
increasing values of $h_0$. Note that, as $h_0$ increases, the entire
profile of the solution approaches a limiting shape, with a sharp
spike near $\rho=0$.  The spike becomes sharper as we increase $h_0$,
while the solution at larger values of $\rho$ remains almost
unchanged. The limiting solutions indeed appears to be the special
solution $S$ of the previous subsubsection (denoted by the solid line
in Fig. \ref{nc1}). Thus it appears that the solution $S$ forms the
end point of the family of regular supersymmetric solitons.

We will now study in more detail how solutions with large $h_0$ approach the 
special solution $S$. We work in 
the language of the dynamical system \eqref{trunc}. We are given a function 
$f(x)$ that starts out, at large negative values of $x$ (small $\rho$) as 
\begin{equation}\label{incond}
f(x)=h_0 e^{\frac{2x}{3}}\left(1+{\cal O}(e^{2x}) \right).
\end{equation}
We wish to study how $f(x)$ evolves under \eqref{trunc} at later times. 
We are interested in the limit in which $h_0$ is large; as we have 
argued above, we expect $f(x)$ to increase to a value of order unity 
at a time  $x_0 \sim -\frac{3}{2} \ln h_0$, and thereafter stabilize 
exponentially to the fixed point $f=1$. Note that $x_0 \ll -1$ (this follows 
because of our assumption that $h_0$ is large) , so that $f$ should settle down
to very near unity well within the domain of applicability of the dynamical 
system \eqref{trunc}.
\footnote{In this language, the conjecture of the 
previous paragraph is equivalent to the assumption that this fluctuation 
lies within the domain of attraction of the fixed point $f=1$. }

Let us now compute $f(x_1)$ for some fixed ($h_0$ independent) $x_1$ 
that obeys\footnote{Recall that \eqref{trunc} is valid only for large and
negative $x_1$.}
$$x_0 \ll x_1 \ll -1.$$
In order to do this we recall that the equation \eqref{trunc} is invariant 
under translations in $x$. Now the judiciously chosen translation 
$$x'=x+\frac{3 \ln h_0}{2}$$
eliminates the $h_0$ dependence of the 
initial condition \eqref{incond} . Let $\chi(x)$ be the (unique, $h_0
$ independent) solution 
to \eqref{trunc} that reduces at early (large negative) times to 
$\chi(x')= e^{\frac{2x'}{3}}$. It follows that the solution of interest to us  
is 
$$f(x)=\chi(x+\frac{3 \ln h_0}{2}).$$

The key assumption of this section, is that the function $\chi$ lies within 
the domain of attraction of the fixed point $f=1$ (as we have seen above there
is impressive numerical evidence for this assumption). If this is the 
case it follows from \eqref{sp} and
\eqref{linsol} that at large $x'$ (i.e. for $x \gg x_0$;)
$$\chi(x')=1+A e^{-x'}\cos(\frac{x'}{\sqrt{3}}+ \delta)$$
for some unknown, order unity constants $A$ and $\delta$. 
It follows that 
\begin{equation}\label{fthen}
f(x_1)\approx 1+Ae^{-{3\over 2}\log h_0 +x_1} \cos\left({\sqrt{3}\over 2} \log (h_0 +x_1) + \delta\right)
\end{equation}
Of course the $x_1$ dependence of this result may be absorbed into a 
redefinition of $A$ and $\delta$. 

Let $\rho_1=e^{x_1}$. \eqref{fthen} gives us a formula for $h(\rho_1)$ and 
$h'(\rho_1)$ for the solution of interest; these values may be used as an 
`initial conditions' to generate $h(\rho)$ for all 
$\rho > \rho_1$ . The resultant solution will take the form 
$$h(\rho) \approx h_S(\rho) + \delta h(\rho)$$ 
where $h_S(\rho)$ is the special solution $S$ and $\delta h(\rho)$ is a small 
fluctuation (of order $ \sim {\cal O}(\frac{1}{h_0^{\frac{3}{2}}})$ ) about this 
solution. To leading order in this small parameter, the function 
$\delta h(\rho)$ obeys a linear differential equation, and so 
depends linearly on $h(\rho_1)$ and $h'(\rho_1)$. It follows that 
\eqref{fthen} then determines the behaviour of every observable 
(like the charge) of the solution that depends only on the behaviour of 
$h(\rho)$ for $\rho$ of order unity or greater. In particular it follows 
that the dependence of the charge of solutions on $h_0$ is given approximately 
by  
\begin{equation}\label{chargepred}
q(h_0)  \approx  q_c + A e^{-{3\over 2}\log h_0} \cos\left({\sqrt{3}\over 2} \log h_0 + \delta\right)
\end{equation}
for some constants $A,\delta$. While $A$ and $\delta$ can only be 
determined numerically, we have a sharp analytic 
prediction for the form \eqref{chargepred}. A similar formula applies 
for the vacuum expectation value of the operator dual to $\phi$ as a function 
of $h_0$.

We have verified the prediction \eqref{chargepred} numerically; in Fig. 
\ref{roscq} we present a plot of the rescaled oscillations of $q$ about $q_c$.
This graph displays precisely the damping (reflected in the $h_0$ dependent
renormalization of the $y$ axis in Fig. \ref{roscq}) and the oscillations 
predicted by \eqref{chargepred}. We will give more details below.

\subsubsection{$h(\rho) \approx  \frac{a}{\rho}$}
\label{sssec:matchb}

Next we consider solutions with $\alpha=1$, i.e solutions that behave
near $\rho=0$ like $\frac{a}{\rho}$.  The one parameter set of these
solutions may be labeled by $a$.  At small $\rho$ our solution takes
the form ${a\over \rho} P(\rho)$ where $P(\rho)$ is a regular power
series. The first few terms in the power series expansion are given by
\begin{equation}
\label{singularone}
h(\rho) = {1\over \rho}\left(a + {1\over 3 a^2} \rho + {18 a^4 -5 \over 36 a^5} \rho^2 +{-90 a^4 + 31 \over 270 a^8}\rho^3
+{\cal O}(\rho^4)\right)
\end{equation}
The formal procedure that generates the power series
\eqref{singularone} treats the term proportional to unity (on the LHS
of \eqref{solode}) as subleading compared to the term proportional to
$h^3 \rho^2$. When $a \ll 1$, this is justified only when $\rho \ll
a^3$.  Consequently we expect the expansion \eqref{singularone} to
break down at $\rho_{sb} \sim a^3$.

\begin{figure}[h]
\begin{minipage}[t]{0.5\linewidth} 
\centering
\psfrag{a}[t]{ $\rho$}
\psfrag{b}[r]{$h(\rho)$}
\includegraphics[totalheight=0.18\textheight]{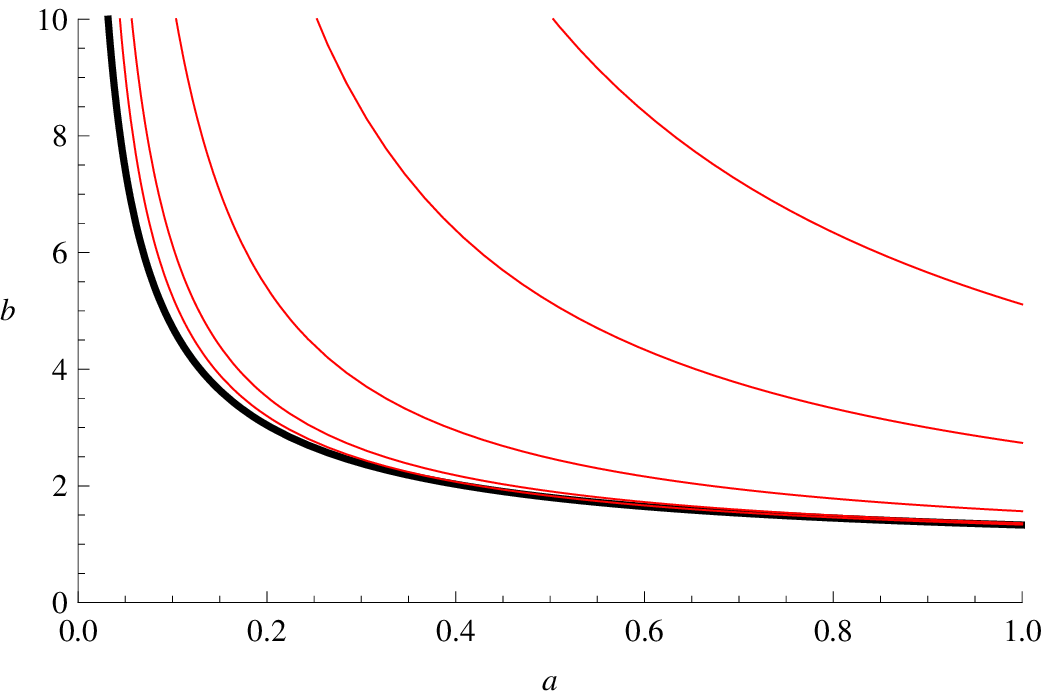}
\caption{Convergence of the numerical solutions for singular solitons
  with an ${a\over \rho}$ singularity to the special singular solution
  $S$ as we decrease $a$.  The black line corresponds to the solution
  $S$ with $\rho^{-2/3}$ behavior near $\rho=0$. The red lines
  correspond to singular solitons of $a = 0.35, 0.5, 1, 2.5,5 $,
  starting from the lowest red curve and going up.}
\label{nc2}
\end{minipage}
\hspace{1cm} 
\begin{minipage}[t]{0.5\linewidth}
\centering
\psfrag{a}[t]{$\log a$}
\psfrag{b}[r]{${q-q_c \over e^{3 \log a}}$}
\includegraphics[totalheight=0.18\textheight]{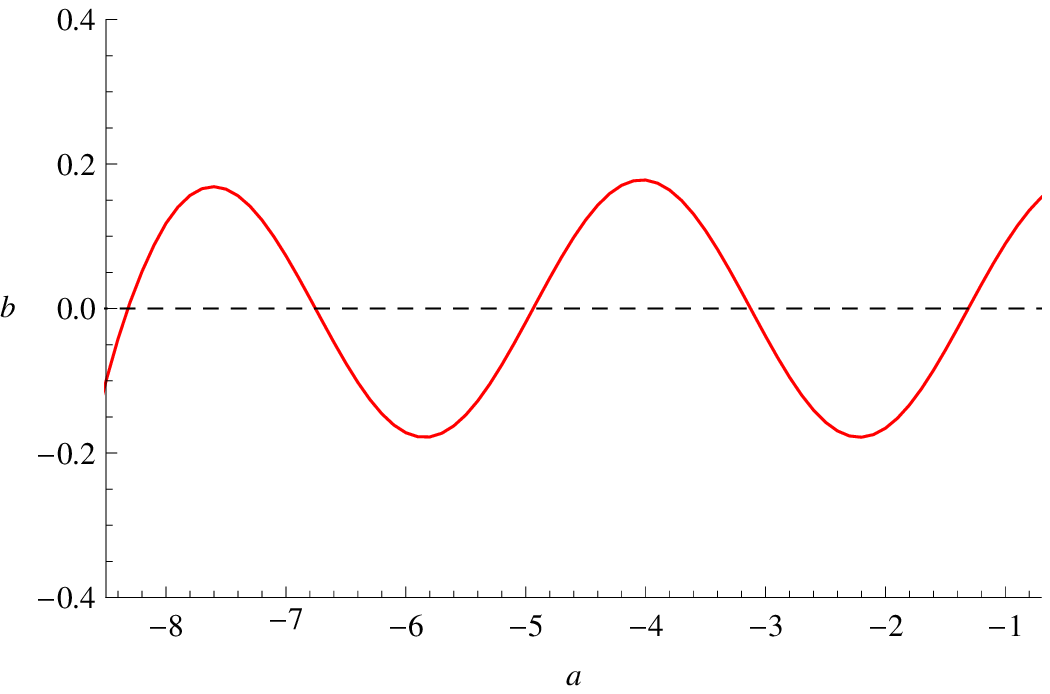}
\caption{The damped oscillations of $q$
  around the critical value $q$ for small $a$.}
\label{sosc}
\end{minipage}
\end{figure}

As in the previous subsubsection, the solution presented here must 
reduce to one of the other solutions of this section when $ a^3 \ll  
\rho \ll 1$.  Noting that, at the point of breakdown of \eqref{singularone}, 
 the function $h$ may be estimated by $h\sim \frac{1}{a^2} 
\sim \rho_{sb}^{-\frac{2}{3}}$, it is natural to guess that the solution 
of this subsubsection tends to the special solution $S$ in the limit of 
small $a$. We now present strong numerical evidence in support of this guess.
In figure \eqref{nc2} we present numerical plots of the solution of this 
subsection for a range of decreasing values of $a$. Note that the solution 
converges to the solution $S$ (denoted by the solid line in Fig. \eqref{nc2})
at small $a$. 

For $\rho\ll 1$ the metric, gauge field and scalar corresponding to this
solution take the following form 
\begin{equation}
\label{metricS}
\begin{split}
& ds^2 \approx - r^2 dt^2 + {4 r^2 \over a^4} dr^2 + r^2 d\Omega_3^2 \cr
& A(r) \approx {r^2\over a^2} dt\cr
&\phi(r) \approx {a^2\over r^2}
\end{split}
\end{equation}

Precisely as in the previous subsubsection, we can analytically study the 
approach of the solution with small $a$ to the solution $S$. Repeating 
an analysis very similar to that of the previous subsubsection, we 
conclude that the dependence of, for instance, the charge of the solution 
on $a$ is given by the formula 
$$
q(h_0) \approx q_c + A e^{3 \log a} \cos\left(\sqrt{3} \log a + \delta\right)
$$ for some constants $A,\delta$ which cannot be determined
analytically. We have verified this prediction numerically (see
Fig. \ref{sosc}).  A similar formula applies for the vacuum
expectation value of the operator dual to the field $\phi$, as a
function of $h_0$.

\subsubsection{The generic solution, $\alpha=2$}

Finally we move to the case $m=2$. Now we find a two parameter set of
solutions, labeled by two arbitrary constants $a,b$, which have the
form
\begin{equation}
\label{singulartwo}
h(\rho) = {1\over \rho^2}\left[a + {1-b^2 \over 2a}\, \rho^4 - {(b^2-1)(3a(5b^2-1)-2) \over 
24 a^4}\,\rho^8 + {\cal O}(\rho^{12})\right]
\end{equation}

The value $b=1$ is special; as we have already remarked $h=1+ 
\frac{a}{\rho^2}$ is an exact solution; hints of this fact are already visible 
in the expansion \eqref{singulartwo}. It follows in particular that the values 
$b=1$ lies outside the basin of attraction of the fixed point $S$ at least 
when $b=1$. A very rudimentary numerical investigation suggests that this 
is also true for all values of $a$ at (for example) $b=2$. Although we 
have not carefully investigated this question, it seems possible that 
the solutions with $\alpha=2$ are completely disconnected from all the other 
solutions studied above. 

For the reasons outlined in the previous paragraph, the 
 `generic' solutions of this subsubsection will make no further appearance 
in our paper. We suspect that all - or at least most - of these solutions are 
genuinely singular, in the sense that they cannot be regarded as the limit of 
smooth solutions. We leave a fuller study of these solutions, and their 
possible physical significance, to future work.

\subsection{'Regular' supersymmetric Solutions}
\label{ssec:regsolnum}

In this section we present the results of a numerical analysis of the
space of 'regular' supersymmetric solutions. Let us first describe what
we believe the space of these solutions to be. The smooth
supersymmetric solitons (with $\alpha=0$) of the previous subsection
are clearly regular.  However this space of solutions ends at finite
charge (see below) as $h_0 \to \infty$. As we have described above,
this line of solutions spirals into (and ends in ) the special solution $S$.
 We have also seen above that another line
of solutions - those with $\alpha=1$ and small values of $a$ - spiral 
out of the solution $S$.  As we will see below, the charge of this new
line of solutions increases without bound at large $a$. It thus
seems that the solutions with $\alpha=0$ and $\alpha=1$ may be
regarded as two different segments of a single line of supersymmetric
solutions.  The two segments are joint together (by a very intricate non 
intersecting double spiral structure) at the special solution $S$.
At least one member of this line of solutions exists at every value of
the charge, and so constitutes a candidate end point of the phase
diagram Fig. \ref{two}.  We conjecture that it is indeed the case that 
hairy black holes at every value of the charge exist for all energies 
above the BPS bound. In the BPS limit, these solutions reduce to some 
configuration on this special line of solutions; either 
to the smooth supersymmetric soliton (at small charges) or the $\alpha=1$ 
solutions (at large charges). In particular we conjecture that all solutions 
on the special line described in this are `regular', where this word is 
used in the sense specified in the previous subsection. We now proceed to 
study these conjecturally regular solutions in more detail.

\subsubsection{Solitons}

\begin{figure}[h]
\begin{minipage}[t]{0.5\linewidth} 
\centering
\psfrag{a}[t]{$h_0$}
\psfrag{b}[r]{$q$}
\includegraphics[totalheight=0.18\textheight]{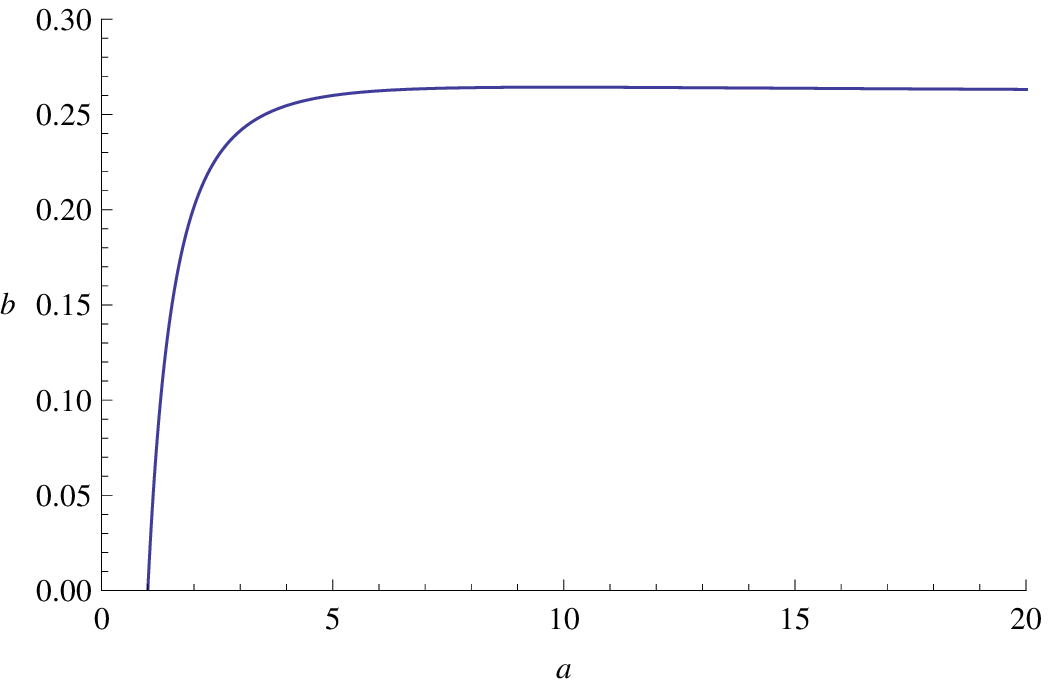}
\caption{Charge $q$ of spherically symmetric supersymmetric regular solitons as a function
of the value $h_0\equiv h(0)$.}
\label{graph1}
\end{minipage}
\hspace{0.5cm} 
\begin{minipage}[t]{0.5\linewidth}
\centering
\psfrag{a}[t]{$h_0$}
\psfrag{b}[r]{$q$}
\includegraphics[totalheight=0.18\textheight]{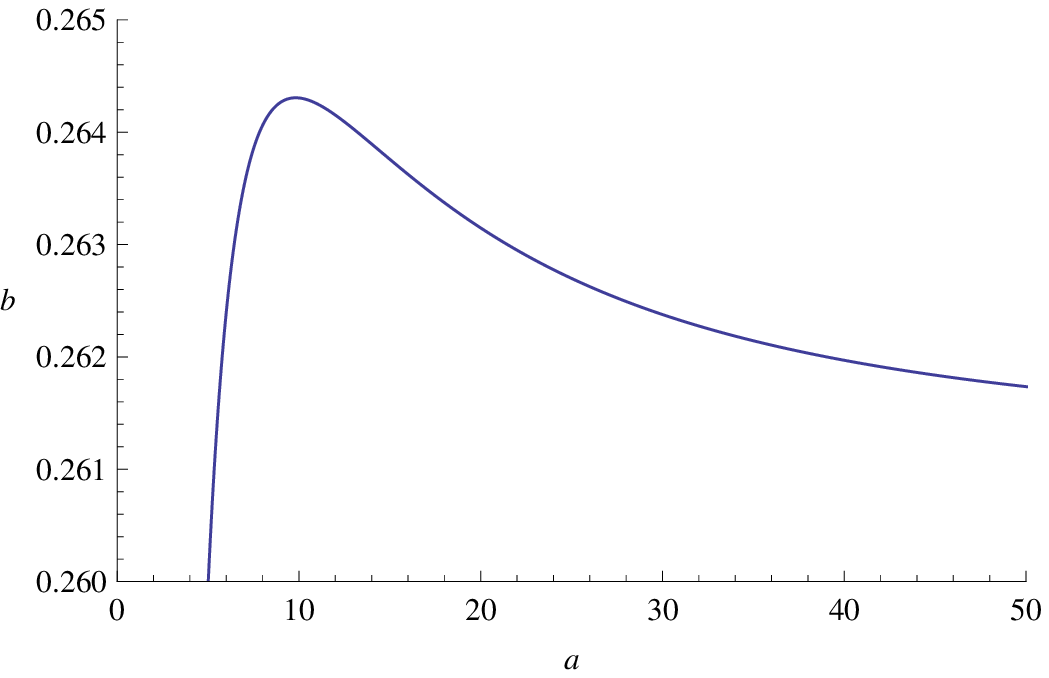}
\caption{The same graph with different scales on the axes, where we can see
the maximum charge.}
\label{graph2}
\end{minipage}
\end{figure}

\begin{figure}[h]
\centering
\psfrag{a}[t]{$\log h_0$}
\psfrag{b}[r]{$q-q_c$}
\includegraphics[totalheight=0.18\textheight]{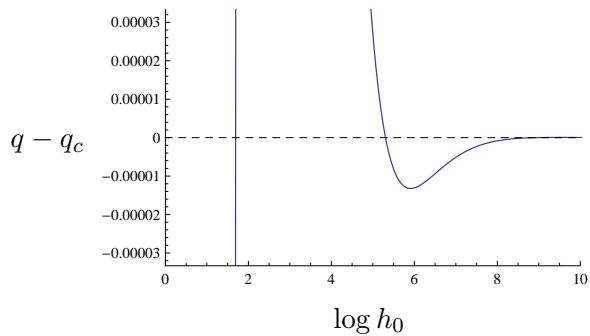}
\caption{Magnification of the previous graph. We wee the next oscillation around $q_c$.}
\label{rdq}
\end{figure}

We first present the numerical analysis of regular solutions of
\eqref{solode}. For this we fix the initial conditions $h(0) = h_0\geq
1$ \footnote{The condition $h_0\geq 1$ is necessary since the scalar
  field at $\rho=0$ is given by $\phi(0) = 2 \sqrt{h_0^2 -1}$ and has
  to be real by assumption \eqref{ansatz}.} and $h'(0)=0$ and
integrate the equation outwards. For each value of $h_0$ we compute
the solution $h(\rho)$ numerically and we evaluate the charge
$q(h_0)$. The results can be seen in Figs \ref{graph1}, \ref{graph2}
and \ref{rdq}, in various magnifications.  
\footnote{To partly check the validity of the numerics it is easy to perform a
perturbative analysis of equation \eqref{solode}, similar to that of
the previous sections i.e. in a small amplitude of $h(\rho)-1$. One
finds agreement between numerics and perturbation theory
(i.e. convergence of the perturbative solution to the numerical one,
for small enough values of $h_0-1$). Notice that the case $h_0=1$ is
precisely empty AdS.}

The most striking feature of the numerical analysis is the existence
of a maximum value\footnote{We have solved the equations numerically
  using Mathematica.}
\begin{equation}
\label{maxcharge}
q_{max} \approx 0.2643
\end{equation}
for the charge of regular supersymmetric solitons. This charge is
obtained for the value $h_{q_{max}}\approx 9.821$ of the initial
conditions at the center. The existence of a maximum charge for these
solitons was also noticed in \cite{Liu:2007xj}\footnote{Notice that we
  are working in slightly different conventions from
  \cite{Liu:2007xj}, in which $q_{there}= 2 q_{here}$. This is
  consistent with the maximum charge $(q_m)_{there} = 0.529$ reported
  in that paper.}.  For higher values of $h_0$ the charge of the
solution starts to decrease and asymptotically it approaches the
limiting value
\begin{equation}
\label{critcharge}
q_{c}\approx 0.2613
\end{equation}
as $h_0 \rightarrow \infty$. 

A more careful analysis reveals that the convergence of the function
$q(h_0)$ towards the critical value $q_c$ is not monotonic. Instead,
the function $q(h_0)$ undergoes slow oscillations around the critical
value, as can be seen in Fig. \ref{rdq} and in more detail in
Fig. \ref{roscq}. These oscillations are periodic, with damped
amplitude, if expressed in terms of $x=\log h_0$. As we explained in
section \ref{sssec:matcha} the asymptotic form of these oscillations
can be determined analytically by matching the regular solution for
large $h_0$ to the special $\rho^{-2/3}$ solution $S$ and we have the
following asymptotic formula for large $h_0$
\begin{equation}
\label{qosc}
q(h_0) \approx q_c  + A e^{-\gamma \log h_0 } \cos(\omega \log h_0 + \delta)
\end{equation}
In section \ref{sssec:matcha} we saw that the period of the oscillations and the damping constant can
 be determined analytically from
the matching procedure to be
\begin{equation}
\gamma = 3/2,\quad \omega = \sqrt{3} /2.
\end{equation}
while $A$ and $\delta$ cannot be fixed in this way. If one tries to
fit this formula to the numerical data one finds
\begin{equation}
\gamma \approx 1.50,\qquad \omega \approx 0.87,\qquad  A \approx -0.19,\qquad
\delta\approx 0.12
\end{equation}
which are in very good agreement with the exact values.

The number of regular solitonic solutions as a function of the charge
are as follows: for small enough charge there is only one solution. As
we increase the charge, at some point we hit the first oscillation
around $q_c$, which increases the number of solutions to
three. Increasing $q$ further we encounter the second oscillation and we have five solutions, and so on. As we approach the critical value
$q_c$ from below the number of solutions is always an odd integer
which goes to infinity. Now let us consider the large charge behavior.
For $q>q_{max}$ we have no regular solitonic solution. As we decrease
the charge and we go below $q_{max}$ we first find two solution.  As
we decrease further we encounter the first oscillation above $q_c$,
giving us four solutions, then the second oscillation to six solution
and so on. Hence for $q>q_c$ we always have an even number of
solutions (possibly zero) which tends to infinity as we approach $q_c$
from above.

Notice that since the BPS bound $m=3q$ is satisfied for all of these
solutions, the figures \ref{graph1},\ref{graph2} also show the
dependence of the mass of the solution on the value of the field at
the center. This qualitative behavior, i.e. the existence of a maximum
mass, and of a slightly lower critical value of the mass which is
approached asymptotically for large central density via a function
which undergoes damped oscillations, is typical in related problems in
general relativity \cite{Sorkin:1981wd, Page:1985em, Hubeny:2006yu}. 
\footnote{\label{foot:chandra}Generally,
  when a family of gravitational solutions has the property that their
  mass has a local maximum for some value of the initial conditions at
  the center, it is the sign that one of the two branches (to the left
  or right of the local maximum) is unstable (under radial
  perturbations) and thus unphysical.  This is the analogue of the
  ``Chandrasekhar instability'': if we expand the equations of motion
  around the solution at the local maximum of the mass they have a
  zero mode, since the total mass does not change to first order in
  the perturbation. Generically this zero mode will be stable on one
  side and unstable (i.e. tachyonic) on the other side of the local
  maximum.  This is what happens for example in the case of boson
  stars \cite{Schunck:2003kk}.  In our case the solutions are
  supersymmetric for all values of $h_0$. It would be very interesting
to check what this implies about their stability.}
To our knowledge, however, this is the first time such behaviour has been 
observed in family of supersymmetric solutions.

Let us now study the expectation value of the scalar operator dual
to $\phi$, which we denote by $\langle {\cal O}_\phi \rangle$, as a
function of $h_0$. Since ${\cal O}_\phi$ is an operator of dimension
$\Delta=2$, its expectation value can be determined from the large $r$
expansion of $\phi$ as
$$
\phi(r) = {\langle {\cal O}_\phi \rangle \over r^2} + ...
$$ We plot the results in figures
\ref{vevreg1},\ref{vevreg2},{\ref{vevreg3} in appendix \ref{app:numresults}. The qualitative behavior
  is similar to that of the charge $q$: the expectation value $\langle
  {\cal O}_\phi \rangle$ is an increasing function of $h_0$ up to a
  maximum value
\begin{equation}
\label{maxvev}
\langle {\cal O}_\phi \rangle_{max} \approx 1.8906
\end{equation}
which is realized for $h_0 \approx 6.580$ and then decreases 
and approaches the asymptotic value
\begin{equation}
\label{critvev}
\langle {\cal O}_\phi \rangle_c \approx 1.8710
\end{equation}
as $h_0 \rightarrow \infty$, while performing small oscillations around it. 
Again the oscillations can be determined following the logic of section \ref{sssec:matcha} 
and are captured by the formula
\begin{equation}
\label{vevosc}
\langle {\cal O}_\phi\rangle (h_0) \approx \langle {\cal O}_\phi \rangle_c 
+ A e^{-\gamma \log h_0} \cos\left(\omega \log h_0  + \delta\right)
\end{equation}
The analytic prediction is $\gamma = {3\over 2}, \,\,\omega={\sqrt{3}
  \over 2}$. If one tries to fit this formula to the numerical data
one finds
\begin{equation}
\gamma \approx 1.50,\qquad \omega \approx 0.87,\qquad  A \approx-0.66 ,\qquad
\delta\approx 0.48
\end{equation}
which are in good agreement with the exact values.

Before we proceed let us point out that the value of $h_0$ at which we
have the maximum charge ($h_{q_{\max}}\approx 9.821$) differs from the
one at which we have the largest expectation value $\langle {\cal
  O}_\phi\rangle_{max}$, which turns out to be $h_{\langle {\cal
    O}_\phi\rangle_{max}} \approx 6.580$. More generally, and as we
will see more clearly in subsection \ref{subsec:spacesolutions}, while
there are pairs of regular solitonic solutions with the same charge
$q$ or the same expectation value $\langle {\cal O}_\phi \rangle$,
there are no such pairs which have the same $q$ and same $\langle
{\cal O}_\phi \rangle$ simultaneously.

\subsubsection{Solutions with $\alpha=1$}
 \label{subsec:singsol}

 We now present the results of a numerical investigation of the other 
segment of the line of (conjecturally) 'regular' supersymmetric solutions: 
those whose small $\rho$ behaviour is given by $a/\rho$ for
small $\rho$. We compute the entire solution numerically and calculate
the charge $q$ as a function of $a$. The results are shown in Figs
\ref{graph3},\ref{graph4},\ref{sdq}. As we see in the figures the charge of this
family starts (at small $a$) precisely at the point $q=q_c$
\eqref{critcharge} where the family of regular solitons ended, then as
we increase $a$ the charge seems to decreases, down to a minimum value
\begin{equation}
\label{chargemin}
q_{min} = 0.2605
\end{equation}
and then increases all the way to arbitrarily large values.
\begin{figure}[h]
\begin{minipage}[t]{0.5\linewidth} 
\centering
\psfrag{a}[t]{ $a$}
\psfrag{b}[r]{$q$}
\includegraphics[totalheight=0.18\textheight]{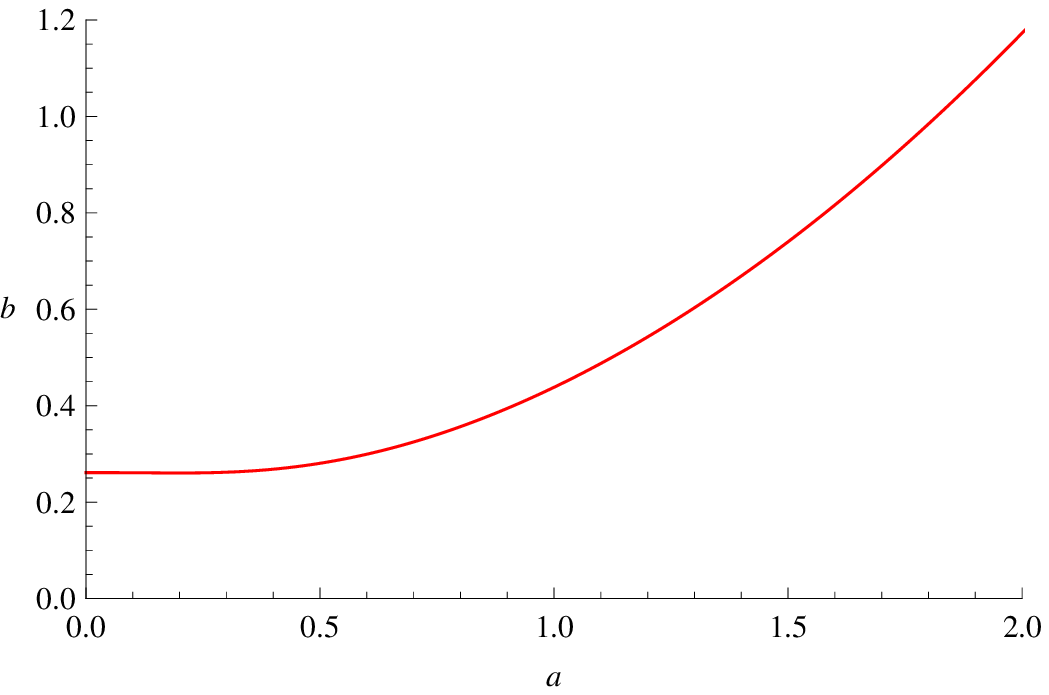}
\caption{ Charge $q$ of spherically symmetric supersymmetric solitons with a singularity
of the form ${a \over \rho}$ at $\rho=0$.}
\label{graph3}
\end{minipage}
\hspace{0.5cm} 
\begin{minipage}[t]{0.5\linewidth}
\centering
\psfrag{a}[t]{$a$}
\psfrag{b}[r]{$q$}
\includegraphics[totalheight=0.18\textheight]{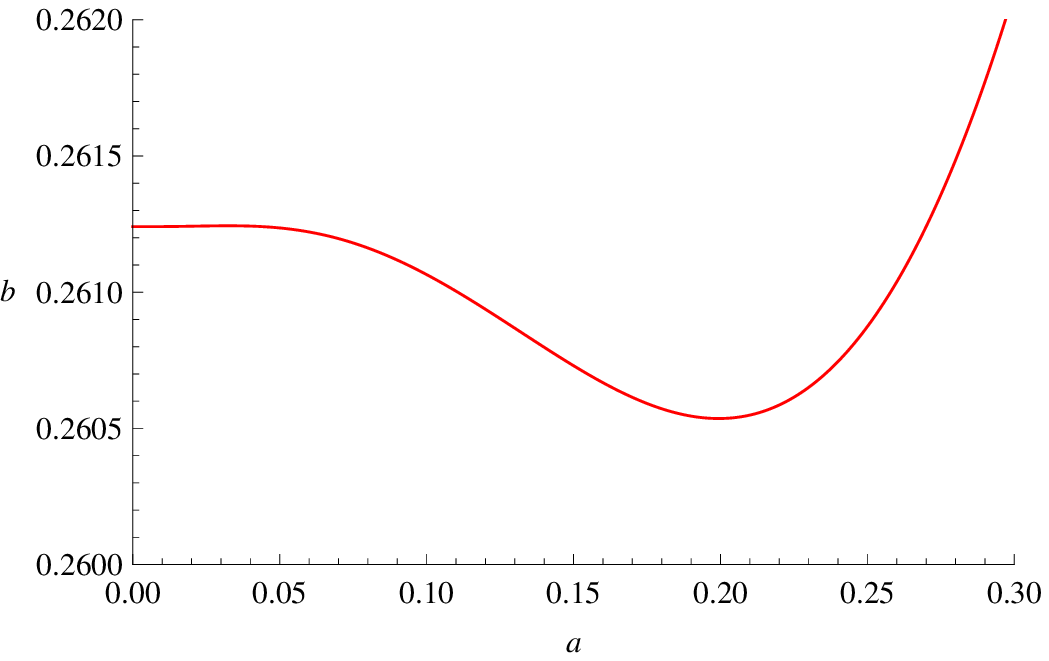}
\caption{Detail of the previous graph with different scales on the axes, where we can see
the minimum of the charge near $a=0$.}
\label{graph4}
\end{minipage}
\end{figure}

\begin{figure}[h]
\centering
\psfrag{a}[t]{$\log a$}
\psfrag{b}[r]{$q-q_c$}
\includegraphics[totalheight=0.18\textheight]{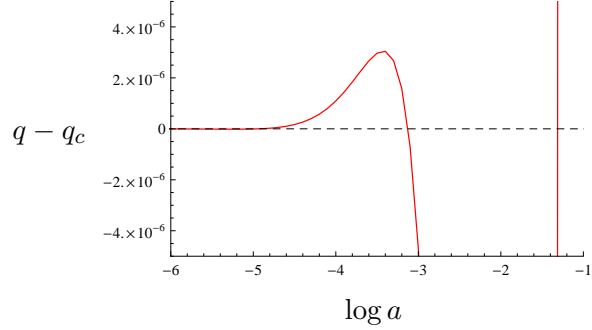}
\caption{Magnification of the previous graph. We wee the next oscillation around $q_c$.}
\label{sdq}
\end{figure}
As before, a more careful analysis shows that the entire radial
profile of solutions with ${a\over \rho}$ singularities converges to
the special solution $S$ in the limit $a\rightarrow 0$, as shown in
Fig. \ref{nc2}. Again, a closer inspection shows that in the regime
between $a=0$ and the point where $q=q_{min}$ the function $q(a)$ is
not monotonically decreasing, but rather is undergoing small damped
oscillations around the value $q_c$ as a function of $\log a$. This is
shown in Fig. \ref{sosc}. For small values of $a$ the form of these
oscillations can be determined by the matching procedure discussed in
section \ref{sssec:matchb} and we find the following formula
\begin{equation}
\label{qoscb}
q(a) \approx q_c  + A e^{\gamma \log a } \cos(\omega \log a  + \delta)
\end{equation}
with the analytically determined values (see \ref{sssec:matchb})
$\gamma=3,\,\, \omega=\sqrt{3}$.  From the numerics we find
\begin{equation}
\gamma \approx 3.00,\qquad \omega \approx 1.73,\qquad  A \approx 0.18,\qquad
\delta\approx 0.70
\end{equation}
which are in good agreement with the exact values.

We find similar behavior for the expectation value $\langle {\cal
  O}_\phi\rangle$ as shown in Figs \ref{vev3},\ref{vev4} in appendix
\ref{app:numresults}: the expectation value starts at the point
\eqref{critvev} where the regular soliton family ended, it then goes
down to
\begin{equation}
\label{vevmin}
\langle {\cal O}_\phi\rangle_{min} \approx 1.8658
\end{equation}
and then increases indefinitely. 

Finally we have the following oscillatory behavior for small $a$ which is shown in Fig. \ref{soscb}
\begin{equation}
\label{vevoscb}
\langle {\cal O}_\phi\rangle (a) \approx \langle {\cal O}_\phi \rangle_c 
+ A e^{\gamma \log a} \cos\left(\omega \log a + \delta\right)
\end{equation}
with the exact values $\gamma = 3,\,\,\omega = \sqrt{3}$. From the
numerics we find
\begin{equation}
\gamma \approx 2.97,\qquad \omega \approx 1.74,\qquad  A \approx 0.59,\qquad
\delta\approx -0.34
\end{equation}

Let us mention that the numerical results depicted in Fig. \ref{par1}
agree with the perturbative analysis of section \ref{sec:soliton} in the
regime of small $q$. According to the results of that section we expect that
for small $q$ the expectation value $\langle {\cal O}_\phi\rangle$ goes like
$$
\langle {\cal O}_\phi\rangle = 4 \sqrt{q} + ...
$$ One can indeed verify that the small $q$ behavior of the curve in
Fig. \ref{par1} agrees with this result.

\subsubsection{An analytic solution at large charge}
\label{sub:largecharge}

While we have no analytic control of $\alpha=1$ solutions in general,
we can see from the numerical analysis that for large $a$ the charge
of the solution can be well approximated by the formula $ q =
{a^2\over 4} + {\rm subleading}$. In fact, in the limit of large $a$
one can find an analytic form of the solution as follows: let us
consider the first factor $(1+\rho^2 h^3)$ on the LHS of equation
\eqref{solode}. At small value of $\rho$ the term $\rho^2 h^3$
dominates over the $1$ since by assumption $h\sim {a\over \rho}$. At
large values of $\rho$ the same is true since $h\sim 1+ {2q \over
  \rho^2}$. Hence it is not unreasonable to assume that in the limit
of very large $a$ we can make the approximation $1+\rho^2 h^3 \approx
\rho^2 h^3$ for the entire range of $\rho$. Then the differential
equation \eqref{solode} becomes
\begin{equation}
\label{simplerode}
\rho\, h  \left(3\,h' + \rho\, h''\right)= \left[4- \left(2\,h +
 \rho\, h'
\right)^2\right] 
\end{equation}
This equation can be solved exactly\footnote{The general solution
of \eqref{simplerode} is $
h(\rho) = \sqrt{1 + {c_1 \over \rho^2} + {c_2\over \rho^4}} $.} and
if we impose the desired behavior near $\rho=0$ the solution is
\begin{equation}
\label{largeqlimit}
h_{\infty}(\rho) = \sqrt{1 +{a^2 \over \rho^2}}
\end{equation}
It is not hard to check that in the large charge limit the numerical
solutions do indeed converge towards the solution \eqref{largeqlimit}
in the entire range of $\rho$, as shown in figure \ref{nc3}. The
solution \eqref{largeqlimit} goes like ${a\over \rho}$ near $\rho=0$
and like $1+{a^2 \over 2 \rho^2}$ for large $\rho$. As we said this
implies that the charge $q\sim {a^2 \over 4}$. One also finds that in
this limit the expectation value goes like $\langle {\cal
  O}_\phi\rangle \sim a^2$. So in the limit of large $q$ we have
$$
\langle {\cal O}_\phi\rangle = 4q + ...
$$ which describes the behavior of the red curve in Fig. \ref{par1} for
large $q$.

\begin{figure}[h]
\centering
\psfrag{a}[t]{ $\rho$}
\psfrag{b}[r]{${h(\rho)\over h_{\infty}(\rho)}$}
\includegraphics[totalheight=0.20\textheight]{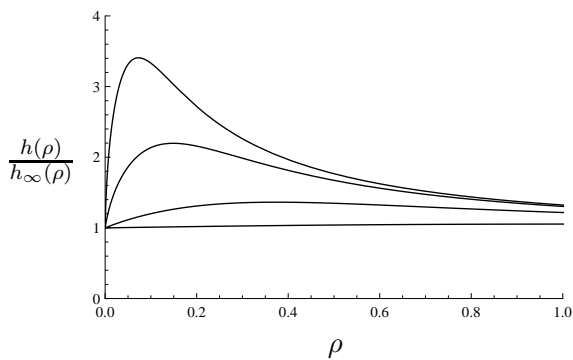}
\caption{Convergence of the numerical solutions for singular solitons
  with an ${a\over \rho}$ singularity to the family of approximate
  solutions $h_{\infty(\rho)}$ as we increase $a$. We plot the ratio
  of the two functions for various values of $a$ and we see that it
  converges to 1 as we raise $a$. The values plotted are $a= 0.1,
  0.2, 0.5, 1.5$, from top to bottom.}
\label{nc3}
\end{figure}

\subsection{Phase Structure of `regular' supersymmetric solutions}
\label{subsec:spacesolutions}

Let us now put everything together and describe the space of
supersymmetric solutions.  In Figs \ref{par1},\ref{par2} we show the
expectation value $\langle {\cal O}_\phi\rangle$ and the charge $q$ of
the family of regular solitons (blue curve) and those with an ${a\over
  \rho}$ singularity (red curve). As we explained above the two
families meet at the solution $S$ with $\rho^{-2/3}$
behavior, which is denoted by a black dot. Near the point $S$ the two
curves develop into two intertwined spirals which are asymptotically
described by equations
\eqref{qosc},\eqref{vevosc},\eqref{qoscb},\eqref{vevoscb}. We zoom into
the point S in Figs. \ref{spiral1},\ref{spiral2} in appendix 
\ref{app:numresults}.

\begin{figure}[h]
\begin{minipage}[t]{0.5\linewidth} 
\centering
\psfrag{a}[t]{ $q$}
\psfrag{b}[r]{$\langle {\cal O}_\phi\rangle$}
\includegraphics[totalheight=0.18\textheight]{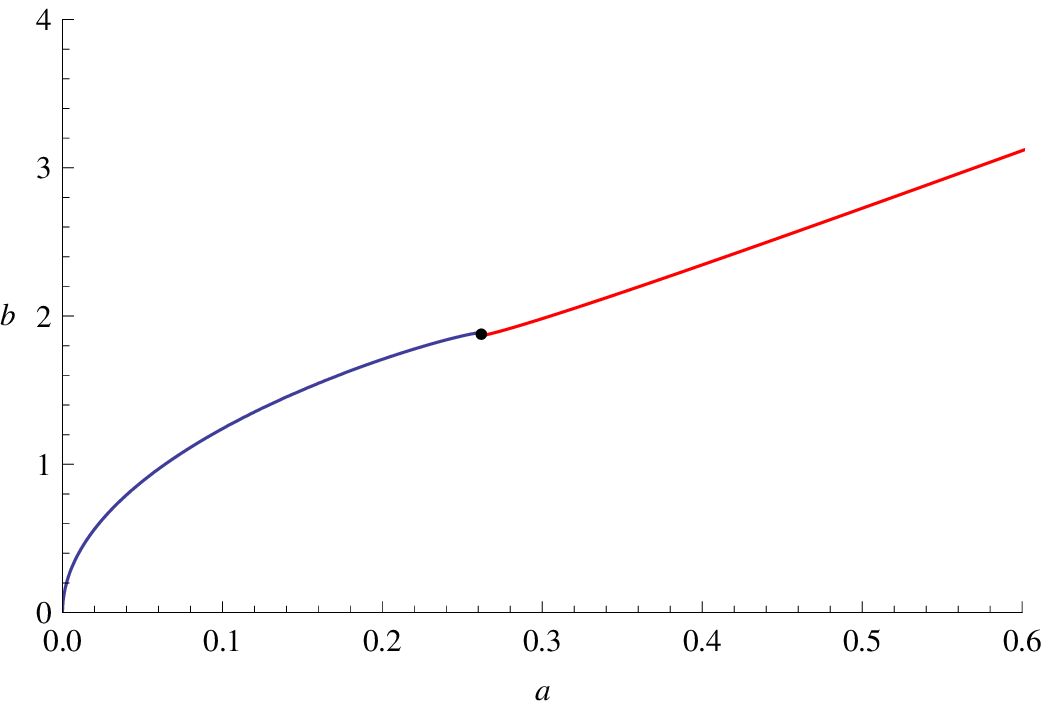}
\caption{ Expectation value $\langle {\cal O}_\phi \rangle$ vs charge $q$ for the
family of regular solitons (blue) and for the family of solitons with an ${a\over \rho}$ singularity 
(red). The two curves meet at the point denoted by the black dot which corresponds to the special solution
$S$ with $\rho^{-2/3}$ behavior.}
\label{par1}
\end{minipage}
\hspace{0.5cm} 
\begin{minipage}[t]{0.5\linewidth}
\centering
\psfrag{a}[t]{$q$}
\psfrag{b}[r]{$\langle {\cal O}_\phi\rangle$}
\includegraphics[totalheight=0.18\textheight]{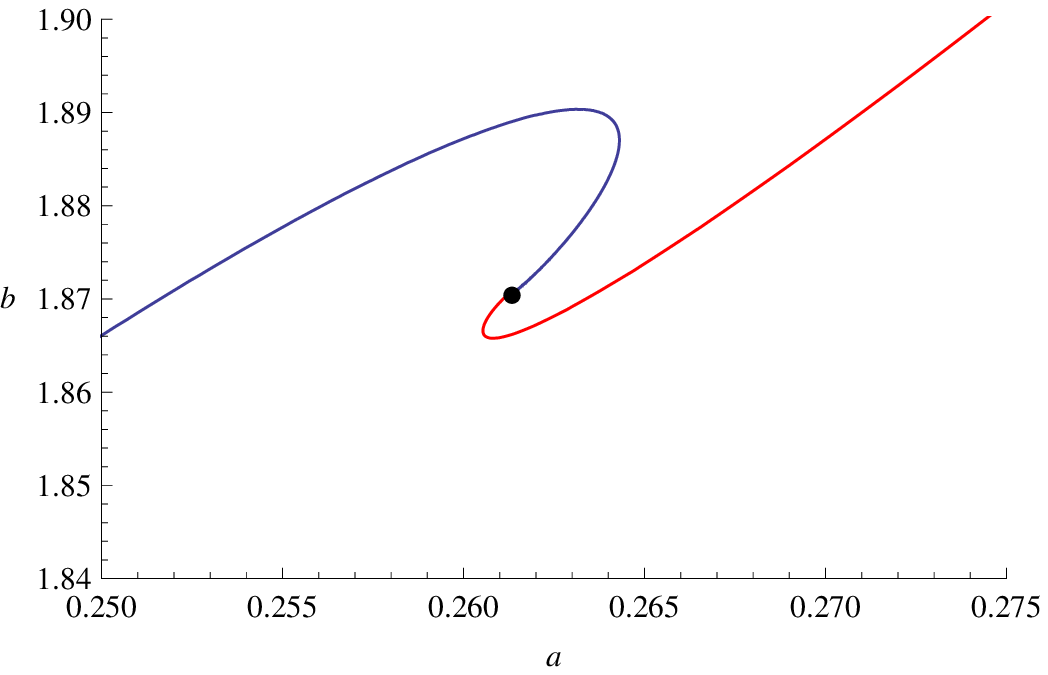}
\caption{Detail of the previous graph around the point $S$ where the
  two families meet. The blue curve is the regular soliton and the
  red curve is the soliton with the ${a\over \rho}$ singuality.}
\label{par2}
\end{minipage}
\end{figure}

From these figures we see that the curves are non intersecting, which
means that if we fix the charge $q$ and the expectation value $\langle
{\cal O}_\phi\rangle$ there is at most one solution. If we consider
the number of solutions as a function of the charge $q$ leaving
$\langle {\cal O}_\phi\rangle$ arbitrary we have the following
pattern: for small enough $q$ we have only one regular solitonic
solution.  As we increase $q$ we first encounter the point $q_{min}$
where two new solutions appear, bringing the total number of
solutions to 3.  Increasing $q$ we cross a point where two more
regular solutions are added and the total number of solutions becomes
5. This pattern continues as we approach the point $q_c$ with an
ever-increasing number of total solutions (at each step we add,
alternatively, either two regular or two singular solutions). Notice
that this number is always odd. After we cross the point $q_c$ the
pattern is reversed, we successively lose pairs of solutions until we
end up with a single singular solution for $q>q_{max}$.  Similar
statements hold if we look for solutions with given expectation
$\langle {\cal O}_\phi\rangle$ instead of given charge.  However, the
fact that the curves in these figures are non-intersecting, means that
if we specify both $q$ and $\langle {\cal O}_\phi\rangle$ there is
always at most one solitonic solution with these values.

The phase diagram we have proposed for our gravitational system is
depicted in Fig. \ref{two}. In this diagram we have included a phase
transition curve that meets the BPS line near $q=q_c\approx 0.2613$;
we will now explain our rational for doing so.  In this paragraph we
assume that both the regular and the singular supersymmetric solutions
found before, may each be obtained as a limit of non singular non
extremal hairy black hole solutions. As we saw above for any given
value of $q$ we may have either one, or a larger odd number of
supersymmetric configurations depending on whether $q$ lies outside or
inside the interval $(q_{min},q_{max}) = (0.2605,2643)$. It follows
that there may exist more than one near supersymmetric regular hairy
black hole solutions in the charge range $q \in (q_{min}, q_{max})$
approximately centered around $q_c$, on which we now focus. These
configurations differ by the expectation value $\langle {\cal
  O}_\phi\rangle$ of the operator dual to the field $\phi$.  Let
$S_R(q, \delta e )$ denote the entropy of the hairy black hole that
reduces to the regular soliton with the largest value of $\langle {\cal
  O}_\phi\rangle$ when $\delta e \to 0$ (here $\delta e$ denotes the
energy above BPS).  Let $S_{S}(q, \delta e)$ denote the entropy of the
hairy black hole that reduces to the singular supersymmetric solutions
with the smallest value of $\langle {\cal O}_\phi\rangle$. We suspect 
that 
\begin{equation} \label{phaset}\begin{split}
&S_R(q, \delta e) > S_S(q, \delta e)  ~~~~{\rm when}~~ 
q<q_P(\delta e)\\
&S_S(q, \delta e) > S_R(q, \delta e) ~~~~{\rm when}~~
q>q_P(\delta e)\\
\end{split}
\end{equation}
for some $q_P(\delta e)$ such that $\lim_{\delta e \to 0} q_P(\delta
e) \in (q_{min}, q_{max})$. Moreover we suspect that the entropies 
of the plethora of intermediate phases that appear at charges near to $q_c$ 
are always smaller than either $S_S(q, \delta e)$ or  $S_R(q, \delta e)$. 
In other words we suspect that our system undergoes the micro canonical 
analogue of a a single first order phase
transition at $q=q_P(\delta e)$; this is the black curve we have
depicted in Fig. \ref{two}. The phase transition curve, which
originates at the BPS line, could either extend all the way to the
phase transition curve between RNAdS and hairy black holes, or could
terminate somewhere in the bulk of the hairy black hole phase, at a
triple point analogous to the water steam system. Of course the considerations 
of this paragraph have been highly speculative. It would be very
interesting to investigate this further.

Before we continue we would like to mention that it would be
important to clarify the stability of these solutions under
linearized perturbations.  As mentioned in footnote \ref{foot:chandra}
on general grounds one would expect regular solutions past $q_{max}$
to be unstable. On the other hand our solutions are supersymmetric and
from this point of view it would seem more natural to believe that
they are stable. This is a confusing issue that deserves further study.

\section{Thermodynamics in the Micro Canonical Ensemble}\label{sec:thermoMicro}

In this section we present thermodynamical formulae for RNAdS black holes, 
the supersymmetric solitons, and hairy black holes, in a small charge and 
near extremal limit. We also demonstrate that the leading order 
thermodynamical formulae for hairy black holes are reproduced by modeling
them by a non interacting mix of a soliton and an RNAdS black hole with $
\mu=1$. 

\subsection{RNAdS Black Hole}\label{ssec:RNAdSBH}

The basic thermodynamics for an RNAdS black hole is summarised by the following
formulae
\begin{equation} \label{basictherm} \begin{split}
m \equiv \frac{M}{N^2}&= \frac{3}{4} R^2\left[1+R^2+ \mu^2 \right]\\
q \equiv \frac{Q}{N^2}&=\frac{\mu}{2} R^2\\
s \equiv \frac{S}{N^2}&= \pi R^3\\
T&=\frac{1}{2\pi R}\left[1+2R^2 - \mu^2\right]\\
\end{split}
\end{equation}
were $Q$ is the charge, $M$ is the mass of the black hole, $S$ is its 
entropy, $T$ its temperature and $\mu$ its chemical potential. Note that 
\begin{equation}\label{extboundA}
\mu^2\leq (1+2 R^2).
\end{equation}
(this follows from the requirement that $R$ is the outer rather than the inner 
event horizon of the black hole).

In this paper we are interested in small RNAdS black holes - i.e. black holes 
with $m \ll 1$ and $q \ll 1$ that are also very near extremality. The mass of RNAdS black holes at fixed charge is bounded from below by the mass of the extremal black hole 
of the same charge; at small $q$ we have
\begin{equation}\label{ext}
m \geq 3 \left(q + q^2 - 
2 q^3 \right) + {\cal O}(q^4)
\end{equation}\
For every pair $(m, q)$ that obeys this inequality, there exists a 
unique black hole solution. In this paper we are interested in black holes whose 
mass above extremality of of order ${\cal O}(q^3)$. \footnote{Note that the mass 
of an extremal black hole, at charge $q$, exceeds the mass of a BPS black hole at the 
same charge by $3 q^2 -6 q^3+{\cal O}(q^4)$. Consequently the deviation of the mass of 
our black holes from the BPS bound is given by $3q^2 +(\delta-6)q^3$, and in particular 
is ${\cal O}(q^2)$ rather than ${\cal O}(q^3)$.} For this reason we define 
the shifted and rescaled mass variable $\delta^2$ by 
\begin{equation}\label{dsq}
\delta^2 q^3 = m-3 \left(q + q^2 - 
2 q^3 \right)
\end{equation}
We are interested in $q \ll 1$ but $\delta$ of order unity. In this regime the  
entropy, chemical potential and temperature
of this black hole is given by 
\begin{equation}\label{thbh2}
\begin{split}
s &=\pi q^\frac{3}{2} \bigg[2 \sqrt{2}+\left(2 \sqrt{3} \delta -6 \sqrt{2}\right) q+\left(\frac{3
   \delta ^2}{\sqrt{2}}-10 \sqrt{3} \delta +33 \sqrt{2}-\frac{24
   \sqrt{3}}{\delta }\right) q^2+O\left(q^3\right)\bigg]\\
\mu &= 1+\left(2-\sqrt{\frac{2}{3}} \delta \right) q+\left(\frac{\delta
   ^2}{3}-6+\frac{4 \sqrt{6}}{\delta }\right) q^2+O\left(q^3\right)\\
\\
 \pi T &= q^\frac{1}{2}\bigg[ \frac{\delta }{\sqrt{3} }-\frac{\left(3 \sqrt{2} \delta ^3-10
   \sqrt{3} \delta ^2+24 \sqrt{3}\right) q}{6  \delta}+O\left(q^2\right)\bigg]
\end{split}
\end{equation}
\footnote{The reader may worry that the appearance of inverse powers of 
$\delta$ in   \eqref{thbh2} signify that black hole thermodynamics 
degenerate in the extremal limit; this, however, is not the case. At 
extremality $\delta=\delta_{ext}(q)$. The function $\delta_{ext}(q)$ starts
out at ${\cal O}(q)$ and so is small at small charge, but does not identically 
vanish. Infact $\delta_{ext}(q)$ may be determined as a power expansion in $q$ 
by equating the temperature in \eqref{thbh2} to zero. Plugging this function 
into the remaining expressions in \eqref{thbh2}, yields nonsingular, analytic
expressions as a function of $q$.}

Although the black holes we study are very small, their temperature is very small (it scales 
like $\sqrt{q}$) because we focus on the near extremal limit \footnote{In contrast 
small black holes in \cite{our} all had a very high temperatures.}. Moreover the black hole 
temperature decreases as we decrease $\delta$, reaching zero at $\delta =0 +{\cal O}(q)$.  

Note also that the chemical potential $\mu$ of these black holes is unity at leading order. 
The first correction to this leading order result is of order ${\cal O}(q)$ and is positive
when $\delta^2<6$ but negative otherwise. This already suggests that RN-AdS black holes 
with $\delta^2 <6$ are unstable to super radiant decay; we will see below that this is 
indeed the case. As we will see below, the end point of the resultant tachyon condensation 
process is a hairy black hole. 

Finally note that the radius of the black holes we study is ${\cal O}(\sqrt{q})$ so that 
the entropy is of order ${\cal O}(q^{\frac{3}{2}})$.

\subsection{Supersymmetric Soliton}\label{ssec:GSSoliton}
The mass and charge of the supersymmetric soliton are given by 
\begin{equation}\label{bhartarit}
\begin{split}
m=& \frac{3}{4} \left(  
\frac{\epsilon ^2}{4}+\frac{\epsilon ^4}{192}+\frac{\epsilon
   ^6}{1920}+\frac{169 \epsilon ^8}{2211840}+{\cal O}\left(\epsilon ^{10}
\right) \right)\\
q = & \frac{1}{2} \left(
  \frac{\epsilon ^2}{8}+\frac{\epsilon ^4}{384}+\frac{\epsilon
   ^6}{3840}+\frac{169 \epsilon ^8}{4423680}+{\cal O}\left(\epsilon ^{10}
\right) \right)
\end{split}  
\end{equation}
Note that 
\begin{equation}\label{ssh} \begin{split}
m&=3q\\
\mu&=1\\
\end{split}
\end{equation}
Of course the soliton is dynamically stable as it is supersymmetric. 
It carries no entropy.

\subsection{A non interacting mix of the black hole and soliton}\label{ssec:mix}

In this subsection we will determine the thermodynamics of a 
hypothetical non interacting mixture of the small black holes and the supersymmetric 
solitons of the previous subsection. 

Of the net mass $m$ and charge $q$ of the system, let mass $3 q_s$ and and charge $q_s$ lie 
in condensate so that the  mass and charge of the black hole are given by 
\begin{equation}\label{masschargesplit} \begin{split}
m_b&=m-3q_s \\
q_b&=q-q_s \\
\end{split}
\end{equation}
The charge $q_s$ is determined by maximising the entropy of the system, which 
determines the black hole chemical potential to be unity. This condition 
gives 
\begin{equation}\label{machb}
m_b = 3 q_b + 3q_b^2 
\end{equation}
(this exact formula may also be verified to ${\cal O}(q^4)$ by setting $\delta^2=6$ in 
\eqref{thbh2})
plugging \eqref{masschargesplit} into \eqref{machb} yields a quadratic 
equation for $q_s$. Solving this equation we find

\begin{equation}\label{msl}
\begin{split}
q_s &= \left(q - 
\sqrt{\frac{m-3q}{3}}\right)\\
q_b &= \sqrt{\frac{m -3q}{3}}
\end{split}
\end{equation}
The squared radius of the black hole is given by $R_b^2 = {2 q_b}$. At leading order 
the entropy and temperature of the mixture are given by  
\begin{equation}\label{ent} \begin{split}
 s&= \pi R^3 = \pi \left( 2 \sqrt{\frac{m-3q}{3}} 
\right)^\frac{3}{2}\\
T&=\frac{\sqrt{2}}{\pi} \left({\frac{m-3q}{3}} \right)^{\frac{1}{4}}
\end{split}
\end{equation}
As the $m-3q$ for the mixture is of order $q^2$, it is convenient to define a shifted 
and rescaled mass variable 
$$q^2 \rho = \frac{4}{3} \left(m-3q \right)$$
in terms of which 

\begin{equation}\label{enttemp} \begin{split}
 s&= \pi q^{\frac{3}{2}} \rho^{\frac{3}{4}}\\
T&=\sqrt{q} \frac{\rho^{\frac{1}{4}}}{\pi} \\
\end{split}
\end{equation}

We will see below that these results correctly reproduce the leading
order thermodynamics of hairy black holes.

\subsection{Hairy Black Hole}\label{ssec:GSHairyBH}

Once we have our solutions for hairy black holes from Appendix \ref{app:PertExp}, 
the evaluation of their thermodynamic charges and potentials is a 
straight forward exercise. At low orders in the perturbative expansion we 
find
\begin{equation}\label{keshtaap}
 \begin{split}
 &\left(\frac{8 G_5}{3\pi}\right)M = \frac{4 M}{3 N^2} =
\frac{4m}{3}=\left[ 2 R^2+R^4+R^6+{\cal O}(R^8)\right] \\
+&\epsilon^2\left[\frac{1}{4}+\frac{R^2}{4}+\left(-2 \log (R)-\frac{\log
   (2)}{2}-\frac{3}{2}\right) R^4+{\cal O}(R^6)\right] \\
+&\epsilon^4\left[\frac{1}{192} + \frac{29 R^2}{576} + {\cal O}(R^4)\right]+ \epsilon^6\left[\frac{1}{1920} + {\cal O}(R^2)\right]+ {\cal O}(\epsilon^8)\\
\\
 & \left(\frac{4 G_5}{\pi}\right)Q = \frac{2 Q}{N^2}=2q=
  \left[ R^2 + \frac{R^6}{2} + {\cal O}(R^8)\right]\\
+&\epsilon^2\left[\frac{1}{8}+\frac{R^2}{8}+\left(-\log (R)-\frac{\log
   (2)}{4}-\frac{13}{16}\right) R^4+{\cal O}(R^6)\right] \\
+&\epsilon^4\left[\frac{1}{384} + \frac{29 R^2}{1152} + {\cal O}(R^4)\right]+ \epsilon^6\left[\frac{1}{3840} + {\cal O}(R^2)\right]+ {\cal O}(\epsilon^8)\\
\\
&\mu = \left[1 + \frac{R^4}{2} + {\cal O}(R^6)\right] + \epsilon^2\left[ \frac{R^4}{6} + {\cal O}(R^6)\right]\\
& +\epsilon^4\left[{\cal O}(R^4)\right] + \epsilon^6\left[{\cal O}(R^2)\right]+{\cal O}(\epsilon^8)\\
\\
&4\pi T = \left[4 R - 2 R^3 + {\cal O}(R^5)\right]\\
 +& \epsilon^2\left[\frac{R}{2} + R^3\left(-12 \log (R)-\log (8)-\frac{89}{12}\right)+ {\cal O}(R^5)\right]\\
&+ \epsilon^4\left[\frac{3 R}{32} + {\cal O}(R^3)\right] +\epsilon^6\left[{\cal O}(R)\right] +{\cal O}(\epsilon^8)
\end{split}
\end{equation}
These quantities obey the second law
$$dM - 3\mu~ dQ - T dS = 0$$
where $S = \text{Entropy} = \frac{\pi^2}{2 G_5}R^3 = \pi N^2 R^3 $.

Upon setting $\epsilon=0$ in \eqref{keshtaap} we find we find a formula 
for the instability curve of the RNAdS black hole. 
Eliminating $R$ we find that this curve lies along the curve  
\begin{equation}\label{instab}
m= 3 q + 3 q^2 +{\cal O}(q^4), 
\end{equation} 
i.e. along the curve $\delta^2=6$ in the notation of subsection 
\ref{ssec:RNAdSBH}. Note that, along this curve, 
 $\mu$ deviates from the value unity only at ${\cal O}(R^4)$.

We will now compute the entropy of the hairy black hole as a function of 
its mass and charge. Let us define a rescaled energy above BPS 
$$\rho q^2  = \frac{4}{3}\left(m - 3 q\right)$$

 It may be verified that 
\begin{equation}\label{hairent}
 \begin{split}
  R^2 &= q \sqrt{\rho}\left[1 + \frac{q}{2}\left(- 2  + \sqrt{\rho}\right) +{\cal O}(q^2)\right]\\
\epsilon^2 &= 8 q \left(2  - \sqrt{\rho}\right ) - \frac{8q^2}{3}\left(2 +  \sqrt{\rho}-\rho\right) + {\cal O}\left( q^3 \right)
\end{split}
\end{equation}

The temperature and chemical potential of the black hole are are given by
\begin{equation} \label{te}
 \begin{split}
 T &= \sqrt{q} \frac{\rho^{\frac{1}{4}}}{4\pi}\left[4 + q (6 - 5\sqrt{\rho}) + {\cal O}(q^2)\right]\\
\mu &= 1 + q \frac{\rho}{2} + {\cal O}(q^2)
 \end{split}
\end{equation}
while its entropy is given by
\begin{equation}\label{enthair}
\begin{split}
 S = \pi R^3 &= q^{\frac{3}{2}} \pi\rho^{\frac{3}{4}}
\bigg[1 + \frac{3q}{4}\left(-2  + \sqrt{\rho}\right) + q^2 \frac{100 - 76 \sqrt{\rho} + 13\rho}{32}\\
&+{\cal O}\left(q^3\right)\bigg]
\end{split}
\end{equation}
Note that \eqref{te} and \eqref{enthair} agree with \eqref{enttemp} at leading order. 
It follows that, atleast for thermodynamical purposes, the hairy black hole solution may 
be regarded as a non interacting mix of a small RNAdS black hole and the soliton, at leading order.

\section{Hairy Rotating Black Holes}\label{rot}

\subsection{Thermodynamics of small Kerr RNAdS Black Holes}

Large classes of explicit Kerr RNAdS black hole solutions have been 
presented (and their thermodynamics worked out) in 
\cite{Gibbons:2004uw, Gibbons:2004ai, Cvetic:2004ny, 
Chong:2005hr, Chong:2005da, Chong:2006zx, Cvetic:2005zi}. 
In this subsection we focus on the special case of small near extremal 
black hole with self dual angular momenta and all three charges equal. 
Concretely, we study black holes with mass $M= N^2 m$, charge $Q=N^2 q$ and 
angular momentum $J=N^2 q^2 j$ \footnote{Concretely, $J$ is the value of 
$J_z$ in one of the two $SU(2)$ factors of $SO(4)$. As in the earlier part 
of this paper, the mass $M$ is normalized to agree with the scaling dimension
of dual operators, while the charge $Q$ is normalized to be unity for a 
complex chiral field $Z$.}.  We will take $q$ to be small, but allow $j$ to 
be arbitrary. All our formulae below are presented in a power series expansion 
in $q$ but are exact in $j$.  

The extremality curve for the black holes we study is given by
\begin{equation}\label{extcurve}
 m_{ext}(q, j) = 3 q + \left(3 + \frac{j^2}{3}\right)q^2 + \left(-6 + \frac{10}{9}j^2 - \frac{4}{81} j^4\right)q^3+ {\cal O}(q^4)
\end{equation}
In contrast the BPS bound for the theory is 
\begin{equation}\label{extcurve}
 m_{BPS}(q, j) = 3 q + 2 q^2 j
\end{equation}
Note that
\begin{equation} \label{bhext}
m_{ext}-m_{BPS}=q^2 \frac{(j-3)^2}{3}+{\cal O}(q^3).
\end{equation}
In particular the mass of the extremal black hole always $\geq$ that of the 
BPS black hole. The extremal black hole is also BPS only when 
\begin{equation}\label{condBPS}
 j=3 - 2 q + 3 q^2  + {\cal O}(q^3)
\end{equation}

As in previous sections, we will be interested in black holes whose energy 
deviates from extremality only at order ${\cal O}(q^3)$. To focus in on these 
energies, it is convenient, as in previous sections, to define a shifted and 
rescaled energy variable $\delta^2$ by 
\begin{equation}\label{del}
\begin{split}
 \delta^2 q^3 &= m - \left[3 q + \left(3 + \frac{j^2}{3}\right)q^2 + \left(-6 + \frac{10}{9}j^2 - \frac{4}{81} j^4\right)q^3 \right]\\
\end{split}
\end{equation}
As above we work in a power series expansion in $q$ but our formulae are all 
exact in $\delta$. With this notation, the thermodynamical potentials 
of small near extremal RNAdS
black holes is given by
\begin{equation}\label{mut}
 \begin{split}
  \mu &= 1 +\left(2 - \frac{2}{9}j^2 - \sqrt{\frac{2}{3}}\delta\right)q + {\cal O}(q^2)\\
\Omega &= \frac{j}{3} - \left(\frac{10}{9} j - \frac{j\delta}{\sqrt{6}} -\frac{8 j^3}{81}\right)q + {\cal O}(q^2)\\
\pi T &= q^\frac{1}{2}\left[\left(\frac{\delta}{\sqrt{3}}\right) + {\cal O}(q^2)\right]\\
s &= \pi q^\frac{3}{2}\left[2\sqrt{2} + \left(\frac{-12 + j^2 + 2\sqrt{6}\delta}{\sqrt{2}}\right)q + {\cal O}(q^2)\right]\\
 \end{split}
\end{equation}

We are particularly interested in two different two parameter surfaces in this 3 dimensional 
space of black holes. The first of these surfaces is the subspace of extremal black holes
given by \eqref{extcurve}. Specializing to this surface we find
\begin{equation}\label{exmut}
 \begin{split}
  \mu =& 1 + \left[2 - \frac{2j^2}{9}\right]q + \left[\frac{10}{243} j^2 \left(2 j^2-9\right)-6\right]q^2 + {\cal O}(q^3)\\
\Omega =& \frac{j}{3} + \left[\frac{2}{81} j \left(45-4 j^2\right)\right]q + \left[\frac{j}{729}  \left(28 j^4-150 j^2-1917\right)\right]q^2 + {\cal O}(q^3)\\
s = & \pi q^\frac{3}{2}\left[2\sqrt{2} + \left(\frac{-12 + j^2}{\sqrt{2}}\right)q +\left(\frac{3 j^4+104 j^2-1584}{24 \sqrt{2}}\right)q^2+ {\cal O}(q^3)\right]\\
 \end{split}
\end{equation}

The second surface of interest, as in previous sections, is the subspace of black holes 
with $\mu=1$. We may solve for $\delta$ in terms of $q$ in order to set $\mu=1$; 
the solution is given by
\begin{equation}\label{muone}
 \delta = -\sqrt{\frac{2}{27}}(j^2 - 9) + \left[\frac{j^2\left( -162 + 45 j^2 + j^4\right)}{486\sqrt{6}(j^2 - 9)}\right] q + {\cal O}(q^4)
\end{equation}
Substituting this value of $\delta$ in \eqref{mut} we find
\begin{equation}\label{mutst}
 \begin{split}
\Omega &=  \frac{j}{3} + \left(\frac{j(9 + j^2)}{81}\right)q + \left(\frac{j(-18 + j^2)}{243}\right) q^2 + {\cal O}(q^3)\\
\pi T &= q^\frac{1}{2}\left[\frac{\sqrt{2}}{9}( 9 - j^2) - \left(\frac{j^2(45 + 11 j^2)}{486\sqrt{2}}\right)q + {\cal O}(q^2)\right]\\
s &= \pi q^\frac{3}{2}\left[2\sqrt{2} - \left(\frac{j^2}{3\sqrt{2}}\right)q  +\left(\frac{7 j^4}{648\sqrt{2}}\right) q^2 + {\cal O}(q^3)\right]
 \end{split}
\end{equation}

Notice that the $\mu=1$ surface intersects the extremality surface along the line of 
supersymmetric black holes. Indeed, plugging the relation \eqref{condBPS} into 
\eqref{exmut} we find 
\begin{equation}\label{exmutsusy}
 \begin{split}
  \mu &=1 +{\cal O}(q^3) \\ 
\Omega&= 1+{\cal O}(q^2)
\end{split}
\end{equation}
In fact it may be shown that the equation $\mu=\Omega=1$ is exact for 
supersymmetric black holes \footnote{We thank Seok Kim for explaining this 
to us.}

\subsection{Hairy Rotating Black Holes as a non interacting mix}

We have seen in the previous section that the thermodynamics of Hairy 
$AdS$ black holes is very simply reproduced, at leading order, by a simple 
physical picture of the black hole as a non interacting mix of the RNAdS 
black hole and the soliton. In this section we will simply assume the same 
is true of charged rotating hairy black holes. In other words we assume 
that there exist charged rotating
black holes whose thermodynamics, at leading order, is reproduced by an arbitrarily weakly interacting mix of the spherically symmetric soliton
of previous sections \footnote{It should also be possible to study rotating 
black holes in equilibrium with solitons made out of the condensates of 
other supergravity modes, e.g. modes of the graviton. At small total angular 
momentum, however, the only modes of the chiral ring satisfy the 
thermodynamical requirement of non interacting equilibrium. This changes
at angular momenta exceeding unity at which point equilibrium between spinning
black holes and the solitons of this paper becomes impossible (see below), but 
rotating black holes could presumably equilibrate with other solitons. We 
thank H. Reall for a discussion on this point, and leave the discussion of 
black holes immersed in other solitonic backgrounds to future work.}
, and 
the small spinning RNAdS Kerr black hole. As in the previous section, the condition 
for thermodynamical equilibrium of this mix is simply the requirement that the 
Kerr RNAdS black hole, that participates in this mix, has $\mu=1$. 

The charge, angular momentum and energy of such an equilibrated mix is given by 
\begin{equation} \begin{split} \label{splitq}
 q&=q_b+q_s\\
jq^2 &=j_b q_b^2\\
m&=\left[3q_b+\frac{1}{3}(j_b^2+9)q_b^2  +{\cal O}(q_b^3)\right] + 3 q_s
\end{split}
\end{equation}
The three equations \eqref{splitq} may be used to solve for $q_b$, $q_s$ and $j_b$ in 
terms of $q$, $j$ and $m$. At leading order we find
\begin{equation}\label{splitsol}
 \begin{split}
  q_b &= \left[\frac{(m-3q)+\sqrt{(m-3q)^2 - 4 j^2 q^4}}{6}\right]^\frac{1}{2}\\
  j_b&= \frac{6 j q^2}{(m-3q)+\sqrt{(m-3q)^2 - 4 j^2 q^4}}\\
  q_s&= q-\left[\frac{(m-3q)+\sqrt{(m-3q)^2 - 4 j^2 q^4}}{6}\right]^\frac{1}{2}\\
 \end{split}
\end{equation}

It then follows that the entropy, temperature and angular chemical potential of the mix, 
are given at leading order by 
\begin{equation}\label{mixeta}
 \begin{split}
  s &= 2\sqrt{2}\pi\left[\frac{(m-3q)+\sqrt{(m-3q)^2 - 4 j^2 q^4}}{6}\right]^\frac{3}{4}\\
\pi T&=\sqrt{2}\left(\frac{(m-3q)+\sqrt{(m-3q)^2 - 4 j^2 q^4}}{6} \right)^{\frac{1}{4}}\left(1-\frac{4j^2q^4}{\left((m-3q)+\sqrt{(m-3q)^2 - 4 j^2 q^4}\right)^2} \right)\\
\Omega& = \frac{j}{3}\\
 \end{split}
\end{equation}

Note that the hairy black holes of this subsection exist only for $j<3$ (this follows 
because $j_b \geq j$ whenever $q_s>0$ (as we have assumed), but black holes with $j_b>3$ have negative temperature and so a naked singularity). 
When this condition is satisfied, they exist provided 
\begin{equation}\label{range}
 3q+2jq^2\leq m\leq 3q+\frac{1}{3}(j^2+9)q^2 -\frac{2}{81}j^2 (j^2+9)q^3
\end{equation}
The range of this existence interval shrinks to zero as $j$ approaches $3$.

At the upper end of this mass range we find, from \eqref{splitsol}, that 
$q_b=q$, $j_b=j$ and $q_s=0$; i.e.  the hairy black hole reduces to a pure 
RNAdS black hole. At the lower end of this range (i.e. when the system charges 
satisfy the BPS bound  $m-3q=2jq^2$) on the 
other hand \eqref{splitsolbps} reduces to 
\begin{equation}\label{splitsolbps}
 \begin{split}
  q_b &= q \left[\frac{j}{3}\right]^\frac{1}{2}\\
  j_b&= 3\\
  q_s&= q\left(1-\sqrt{\frac{j}{3}}\right)\\
 \end{split}
\end{equation}
At this lower end, then, the temperature of the hairy black vanishes. At this end 
the hairy black hole is neither pure soliton (unless $j=0$) 
or pure black hole (unless $j=3$; at which value the upper and lower end of \eqref{range}
coincide) but a mix of soliton and black hole. The special feature of this mix is that 
$j_b=3$, so that the participating black hole is supersymmetric. At the lower edge, then, 
the system is a weakly interacting mix of a supersymmetric black hole and the 
supersymmetric condensate, and is itself supersymmetric. This is of course intuitive. An extremal hairy black hole is  given by 
a non interacting mix of an extremal Kerr RNAdS black hole with $\mu=1$ and a soliton.
But, as we have seen above, extremal Kerr RNAdS black holes are BPS at $\mu=1$. It follows
that (within the approximations of this subsection), that the extremal hairy black hole is also BPS. 

Note that the approximations of this subsection predict a two parameter family of BPS hairy black holes (the entire lower edge of \eqref{range}) in comparison with the one parameter set of BPS Kerr RNAdS black holes. 
These black holes constitute the lower end of the mass range \eqref{range}.

To end this section we should emphasize that all the formulae of this 
section are predicated on the assumption that hairy black hole thermodynamics
may be reproduced, at leading order, by a non interacting mixture of 
a Kerr RNAdS black hole and a scalar condensate. While we have checked by 
explicit calculation that this guess is true in the absence of rotation, 
we have not yet performed any such check for rotating black holes. The reader
should, consequently, regard the formulae of this section as conjectural. 
As we have remarked in the introduction, it may well prove technically possible
to verify (or diversify) the predictions of this subsection by an explicit 
perturbative calculation of the sort presented earlier in this paper. We leave 
this effort for future work.

\section{Discussion}\label{sec:discussion}

In this paper we have studied charged black holes in global $AdS$
spaces.  We have focused mainly on spherically symmetric black holes
with equal diagonal $SO(6)$ charges. At small values of the charge we
have demonstrated that the spectrum of $AdS$ hairy black holes extends
all the way down to the BPS bound. We have also conjectured that this
result applies at all values of the charge. The evidence for this last
conjecture is not yet overwhelming, it would
certainly be worthwhile to gather other evidence (e.g. from numerical
solutions of the relevant differential equations) to support or refute
this claim, as also to verify the precise nature of the large charge
singular supersymmetric solutions proposed in subsection
\ref{sub:largecharge}.

The special singular solution $S$, discussed in section
\ref{sec:nsoliton}, appears to play a very special role in the space
of supersymmetric solutions of the theory. It would be interesting to
study the near singularity geometry of this solution in more detail;
in particular it would be very interesting to determine the 10
dimensional lift of this solution, as well those of the $\alpha=1$
singular solutions of section \ref{sec:nsoliton}. It is conceivable
that these singular five dimensional solutions have a regular ten
dimensional lift. If this is not the case the ten dimensional
perspective should at least yield valuable insight into the nature of
the singularities in these solutions. It may also be worthwhile to re
investigate the nature of $\alpha=2$ supersymmetric solutions from the
ten dimensional viewpoint.

Let us momentarily turn to the consideration of black branes in
asymptotically Poincare AdS spaces. These solutions may be obtained
from the large charge limit of the black hole solutions studied in
this paper. RNAdS black branes solutions exist at all values of
$\rho_e \geq c_{e} \rho_q^{\frac{4}{3}}$ with $c_e={9 \over
  2}\pi^{2/3}$. These solutions are presumably unstable to the
condensation of the $\phi$ field for energy densities $\rho_e \leq
c_{s} \rho_q^{\frac{4}{3}}$ for some $c_{s}>c_{e}$. If hairy black
branes at a given charge density have a lower bound on their energy
density, the equation for this lower bound must also take the form
$\rho_e \geq c_{m} \rho_q^{\frac{4}{3}}$ with $c_{m}<c_{e}$. It is
clearly of interest to know the value of $c_{m}$. Our conjecture that
hairy black holes descend down to their BPS bound even at large charge
amounts to the prediction that $c_{m}=0$, i.e. there is no lower bound
for the mass density of hairy black branes, at any fixed finite charge
density. This phenomenon is not without known precedent; as we explain in
appendix \ref{app:bps}, a similar result is true of one charge (rather
than 3 equal charge, as studied in this paper) RNAdS black branes.

On a related note, the recent paper \cite{Herzog:2010vz} has presented 
an analytic determination of the constant $c_s$ (and the hairy black brane 
solution in the neighborhood of this critical density) in a system with some 
similarities to the one studied in this paper. In particular the scalar field 
in \cite{Herzog:2010vz} has the same value of the mass as in the current 
work, but carries infinite charge (this is the so called the probe 
approximation that ignores the backreaction of gauge dynamics on the 
metric). It would be interesting to investigate whether the results of 
\cite{Herzog:2010vz} can be generalized to the study of our Lagrangian 
\eqref{spclag}.

It would be of interest to extend the results of this paper to the
case of charged black holes with three unequal $SO(6)$ charges, and
particularly to study special limits in which one or two of these
charges are turned off.  In particular, in the special case where only
one $SO(6)$ charge is nonzero, ordinary RNAdS black holes extend all
the way down to the BPS bound, so there is no pressing reason for
these solutions to exhibit a superradiant instability, or for one
charge hairy black brane solutions to exist. This question deserves
further investigation.

It it would be fascinating to verify the correctness or otherwise of
the tentative predictions of section \ref{rot} for the spectrum of
small rotating black holes. In particular, if new physically
acceptable supersymmetric hairy black holes really do exist, it would
be fascinating to determine them and to analyse their properties.

As we have mentioned in the introduction, it is possible that the black holes
we have constructed suffer from a superradiant instability towards the emission
of the gravity modes dual to $TrX^n +Tr Y^n + Tr Z^n$. It should in principle 
be straightforward (though perhaps tedious in practice) to check 
whether or not this is the case, by computing the imaginary part of the 
relevant quasinormal modes, along the lines of Appendix A of \cite{our}.

In this paper we have focused only on a very particular kind of hairy
black hole; one whose condensate is the zero mode of a scalar field in
$AdS_5$ space.  It is likely that there exist many different such
hairy solutions with various different gravitational field
condensates. In the current paper we have likely illuminated only a
very small part of an intricate and fascinating structure of hairy
black hole solutions in $AdS_5 \times S^5$. It would be fascinating to
more fully uncover the structure of new black hole solutions in $AdS_5
\times S^5$, and perhaps most importantly, to understand their
properties directly from the dual Yang Mills point of view.

\subsection*{Acknowledgments}

We would like to thank J. Bhattacharya, J. de Boer, R. Loganayagam,
M. Shigemori, A. Strominger, N. Suryanarayana and S. Trivedi for
useful discussions. We would like especially to thank S. Kim for
collaboration over the initial stages of this project, and several
very useful discussions over every stage of this project.  We would
also like to especially thank V. Hubeny and M. Rangamani for
suggesting we look for oscillatory behavior of the supersymmetric
solutions near the critical charge. We would also like to thank
J. Bhattacharya, C. Herzog, V. Hubeny, M. Rangamani, H. Reall for
useful comments on a draft version of this manuscript. SM would like
to thank the Weizmann Institute and the organizers of the SERC school
at Punjab University, Chandigarh and the Institute of Mathematical
Sciences, Chennai, for hospitality while this work was being
completed. The work of SM was supported in part by a Swarnajayanti
Fellowship.  S.B. and S.M. would also like to acknowledge their debt
to the steady and generous support of the people of India for research
in basic science.

\appendix

\section{Results for the  Perturbative Expansion of Hairy 
Black Holes}\label{app:PertExp}

\subsection{Far Field Solution}
{\it Scalar Field:}
\begin{equation}\label{phi1out}
\begin{split}
 \phi^{out}_{(1,0)}(r) =&\frac{1}{1 + r^2}\\
\phi^{out}_{(1,2)}(r) = & \frac{2 \left[\left(r^2+1\right) \log
   \left(1+\frac{1}{r^2}\right)-1\right]}{\left(r^2+1\right)^2}\\
\phi^{out}_{(1,4)}(r) = & \frac{8 r^4+14 r^2+2}{r^2
   \left(r^2+1\right)^3}+\frac{2 \left(r^2+1\right) \log
   ^2\left(1+\frac{1}{r^2}\right)-\left(9 r^2+13\right) \log
   \left(1+\frac{1}{r^2}\right)}{
   \left(r^2+1\right)^2}\\
\phi^{out}_{(3,0)}(r) =& \frac{1}{8(1 + r^2)^3}\\
\phi^{out}_{(3,2)}(r) = & \frac{r^2+6 \left(r^2+1\right) \log
   \left(1+\frac{1}{r^2}\right)-5}{8
   \left(r^2+1\right)^4}  \\
\phi^{out}_{(5,0)}(r) = & \frac{6 r^4+4 r^2+55}{2304 \left(r^2+1\right)^5}\\
\phi^{out}_{(5,2)}(r)=& \frac{1}{6912 \left(r^2+1\right)^6}\bigg[-1091 + 923 r^2 + 562 r^4 + 282 r^6 + 24 r^8\\
&+6 \left(r^2+1\right) \left(4 r^8+16 r^6-6 r^4-4 r^2-271\right) \log
   \left(\frac{r^2}{r^2+1}\right)\bigg]
\end{split}
\end{equation}

{\it Metric and Gauge Field:}
\begin{equation}\label{metgout}
 \begin{split}
  f^{out}_{(2,0)}(r) =& -\frac{1}{4(1 + r^2)},~~g^{out}_{(2,0)}(r) = 0,~~A^{out}_{(2,0)}(r) =  -\frac{1}{8(1 + r^2)}\\
f^{out}_{(2,2)}(r) = & -\frac{\log \left(1+\frac{1}{r^2}\right)}{r^2+1} -\frac{r^4-3 r^2-2}{4r^2 \left(r^2+1\right)^2}\\
g^{out}_{(2,2)}(r) = &\frac{1}{4 \left(r^2+1\right)^3}\\
A^{out}_{(2,2)}(r) =& -\frac{\log \left(1+\frac{1}{r^2}\right)}{2 \left(r^2+1\right)} + \frac{-r^4+2 r^2+1}{8 r^2 \left(r^2+1\right)^2}\\
f^{out}_{(4,0)}(r) = &-\frac{r^4}{192\left(1 + r^2\right)^3},~~g_{(4,0)}(r) = \frac{r^4}{192\left(1 + r^2\right)^3}\\
A^{out}_{(4,0)}(r) =& -\frac{3 + 3 r^2 + r^4}{384\left(1 + r^2\right)^3}\\
f^{out}_{(4,2)}(r) = & -\frac{29 r^6+71 r^4+96 r^2-24 \left(r^6+r^4\right) \log
   \left(\frac{r^2}{r^2+1}\right)+36}{576 \left(r^2+1\right)^4}\\
g^{out}_{(4,2)}(r) = &\frac{29 r^6+71 r^4+108 r^2+24 \left(r^6+r^4\right) \log
   \left(1+\frac{1}{r^2}\right)+36}{576 \left(r^2+1\right)^6}\\
A^{out}_{(4,2)}(r) = & -\frac{-9 + 90 r^4 + 92 r^6 + 29 r^8}{1152 r^2\left(1 + r^2\right)^4}- \frac{r^4+3 r^2+3}{48\left(1 + r^2\right)^3} \log
   \left(\frac{r^2}{r^2+1}\right)
\end{split}
\end{equation}

\subsection{Intermediate Solution}
\begin{equation}\label{phimid}
\begin{split}
 \phi^{mid}_{(1,0)}(y) =&1\\
\phi^{mid}_{(1,2)}(y) = & -y^2-4 \log (R)-2 \log \left(y^2-1\right)-2\\
\phi^{mid}_{(1,4)}(y) = & 8 \log ^2(R)+\left(4 y^2+8 \log
   \left(y^2-1\right)+\frac{4}{y^2-1}+26\right) \log (R)
+\frac{y^6+5 y^4+4 y^2}{y^2-1}\\
&+\frac{\log \left(y^2-1\right) \left(2 y^4+11 y^2+2
   \left(y^2-1\right) \log \left(y^2-1\right)-15\right)+\log
   (4)-16}{y^2-1}\\
\phi^{mid}_{(3,0)}(y) =& \frac{1}{8}\\
\phi^{mid}_{(3,2)}(y) = & \frac{1}{8} \left(-3 y^2-12 \log (R)-6 \log \left(y^2-1\right)-5\right) \\
\phi^{mid}_{(5,0)}(y) =& \frac{55}{2304}\\
\phi^{mid}_{(5,2)}(y)=&\frac{-813 y^2-3252 \log (R)-1626 \log \left(y^2-1\right)-1091}{6912}
 \end{split}
\end{equation}

{\it Metric and Gauge Field:}
\begin{equation}\label{metgmid}
 \begin{split}
  f^{mid}_{(2,0)}(y) =& -\frac{(y^2 -1)^2}{4 y^4}\\
g^{mid}_{(2,0)}(y) =& 0\\
A^{mid}_{(2,0)}(y) = & -\frac{y^2 -1}{8y^2}\\
f^{mid}_{(2,2)}(y) = & \frac{\left(y^2-1\right) \left[y^4+\left(8 y^2-4\right) \log (R)+\left(4
   y^2-6\right) \log \left(y^2-1\right)+\log (4)-1\right]}{4 y^4}\\
g^{mid}_{(2,2)}(y) = &\frac{y^4 \left[y^2-4 \log (R)+2 \log \left(y^2-1\right)-\log
   (4)-5\right]}{4 \left(y^2-1\right)^3}\\
A^{mid}_{(2,2)}(y) =& \frac{y^4+\left(8 y^2-4\right) \log (R)+\left(4 y^2-6\right) \log
   \left(y^2-1\right)+\log (4)-1}{8 y^2}\\
f^{mid}_{(4,0)}(y) =& 0\\
g^{mid}_{(4,0)}(y) =& 0\\
A^{mid}_{(4,0)}(y) =& -\frac{1}{128}\left(1- \frac{1}{y^2}\right)\\
f^{mid}_{(4,2)}(y) =& -\frac{\left(y^2-3\right) \left(y^2-1\right)}{16 y^4}\\
g^{mid}_{(4,2)}(y) =& \frac{y^4 \left(y^2-4 \log (R)+2 \log \left(y^2-1\right)-\log
   (4)-7\right)}{16 \left(y^2-1\right)^3}\\
A^{mid}_{(4,2)}(y)=& \frac{y^4-2 y^2+\left(8 y^2-4\right) \log (R)+2 \left(2 y^2-3\right) \log
   \left(y^2-1\right)+\log (4)+5}{64 y^2}
 \end{split}
\end{equation}

\subsection{Near Field Solution}
{\it Scalar field}
\begin{equation}\label{phiin}
\begin{split}
 \phi^{in}_{(1,0)}(z) =&1\\
\phi^{in}_{(1,2)}(z) = & -8 \log (R)-\frac{1}{2} \log (z+1) (\log (z+1)+4)-\text{Li}_2(-z)-2 \log
   (2)-\frac{\pi ^2}{6}-3\\
\phi^{in}_{(3,0)}(z) =& \frac{1}{8}\\
\phi^{in}_{(3,2)}(z) = &\frac{2 (4+3 \log (2)) z+\pi ^2 (z+1)+6 \log (2)+12}{8 (z+1)}\\
&+\frac{3}{8} (\log (z)+i \pi -2) \log (z+1)-3 \log (R)
+\frac{3}{8} \text{Li}_2(z+1)\\
&-\frac{(3 z (z+1)+2) \log ^2(z+1)}{16 z (z+1)}\\
\phi^{in}_{(5,0)}(z) =& \frac{55}{2304}
 \end{split}
\end{equation}

{\it Metric and Gauge Field:}
\begin{equation}\label{metgin}
 \begin{split}
 f^{in}_{(2,4)}(z) =& -z^2-\frac{1}{2} (2 z+1) \log (z+1)\\
g^{in}_{(2,-2)}(z) =& \frac{(2 z+1) \log (z+1)-4 z}{32 z^2 (z+1)^2}\\
A^{in}_{(2,2)}(z) = & \frac{1}{4} (-z-\log (z+1))\\
f^{in}_{(4,4)}(z) =&\frac{z (4 z+2)+\log (z+1) ((z+1) \log (z+1)-1)}{16 (z+1)}\\
g^{in}_{(4,-2)}(z) = & \frac{(5-12 z) z^2+(4 (z-2) z-3) \log (z+1) z+(3 z (z+1)+1) \log
   ^2(z+1)}{256 z^3 (z+1)^3}\\
A^{in}_{(4,2)}(z) =& -\frac{z (z+2 \log (z+1)-3)}{64 (z+1)}
 \end{split}
\end{equation}

\section{Supersymmetric solitons in $AdS_5\times S^5$ and the planar limit}
\label{app:bps}
\subsection{One-charge solitons}

In this paper we studied supersymmetric configurations in $AdS_5\times
S^5$ by considering a truncation of ${\cal N}=8$ supergravity where
all 3 charges corresponding to the three orthogonal planes of $SO(6)$
are taken to be equal. The spherically symmetric supersymmetric
solutions of this truncation are 1/8 BPS from the ten dimensional
point of view. It is instructive to briefly review solutions of ${\cal
  N}=8$ supergravity where we turn on only one of these charges.
Spherically symmetric supersymmetric configurations of this type are
1/2 BPS and belong to the LLM family \cite{Lin:2004nb}. We look for
static, spherically symmetric 1/2 BPS solitonic solutions with the
topology of $AdS_5\times S^5$ (so that in the LLM language they have
the topology of a disk). These can be studied in a truncation of
${\cal N}=8$ to $U(1)^3$ coupled to 3 hypermultiplets, by turning on
only one of the $U(1)$ charges and can be expressed in the language of
\cite{Chong:2004ce, Liu:2007xj} as\footnote{We use somewhat different
  notation with $ r_{there}= \rho_{here},\quad
  (H_1)_{there}=(H_2)_{there}=1, \quad (H_3)_{there}= h_{here},\quad
  (A_i)_{there}=-(A_i)_{here},\quad
  2\sinh \phi_{i,there}=\phi_{i,here}$.}
\begin{equation}
\label{halfbps}
\begin{split}
& ds^2 = -{1 + \rho^2 h \over h^{-2/3}} dt^2 +{h^{1/3}\over1 + \rho^2
    h } d\rho^2 + h^{1/3} \rho^2 d\Omega_3^2 \cr & A_1=A_2=0,\qquad A_3 = h^{-1}
  dt,\qquad X_1 = X_2 = h^{1/3},\qquad X_3 = h^{-2/3}\cr
& \phi_1=\phi_2=0, \qquad \phi_3 =2\sqrt{(h + \rho h'/2)^2-1}
\end{split}
\end{equation}
where $A_i$ are the 3 $U(1)$ gauge fields, $X_i$ the scalars in the vector multiplets
constrained to satisfy $X_1 X_2 X_3=1$ and $\phi_i$ scalars in the hypermultiplets. This is a solution if 
$h(\rho)$ satisfies the equation
\begin{equation}
\label{halfode}
(1+\rho^2 h)(3h' + \rho h'') = \rho[4-(2h+\rho h')^2]
\end{equation}
Notice the different powers of $h$ in the metric \eqref{halfbps}
relative to the 3-equal-charge case \eqref{metricbps}. Also, unlike equation
\eqref{solode}, now \eqref{halfode} can be solved analytically and we
find the following set of smooth solutions \cite{Chong:2004ce, Liu:2007xj}
$$
h(\rho) = \sqrt{1 + 2{1+2q \over \rho^2} + {1\over \rho^4}} - {1\over \rho^2}
$$ The parameter $q$ corresponds to the charge. In contrast to the
3-charge case which was studied in the rest of the paper, now $q$ is
unbounded from above i.e. there is no maximum charge of smooth
spherically symmetric 1/2 BPS solitons similar to \eqref{critcharge}.

In particular we can try to recover a spacetime which is
asymptotically the Poincare Patch by taking the large charge limit. We
consider the following scaling
\begin{equation}
\begin{split}
& t=k^{-1}\,\tau,\quad \rho= k\, u,\quad \Omega_i = k^{-1}\, x_i,\quad
  q = q_0\, k^2\cr
& \tau\,, u \,, x_i\,,q_0\, = {\rm const}\qquad ,\qquad  k\rightarrow \infty
\end{split}
\end{equation}
Then we find the following solution
\begin{equation}
\begin{split}
& ds^2 = -u^2 \left(1 + {4 q_0 \over u^2} \right)^{1/6} d\tau^2
+ {1\over u^2 \left(1 + {4 q_0 \over u^2}\right)^{1/3} }d u^2
+u^2 \left(1+ {4 q_0 \over u^2}\right)^{1/6} dx_i^2 \cr
& A_1=A_2=A_3 = 0 \cr
& X_1=X_2 = \left( 1+ {4 q_0 \over u^2}\right)^{1/6},\qquad
X_3  =  \left( 1+ {4 q_0 \over u^2}\right)^{-1/3}\cr
&\phi_1=\phi_2 =0,\qquad \qquad\qquad\qquad\quad\phi_3  =  
{4q_0 \over u \sqrt{4 q_0 +u^2}} \cr
\end{split}
\end{equation}
which is asymptotically the Poincare Patch and has scalar fields turned on in the interior.

Let us finally consider the non-extremal one-charge RNAdS black branes. We start
with a non-extremal one-charge RNAdS black hole in global AdS. We have the following solution
\begin{equation}
\begin{split}
& ds^2 = - H^{-2/3} f dt^2 + H^{1/3} (f^{-1} d\rho^2 + \rho^2 d\Omega_3^2) \cr
&A_1=A_2 = 0,\quad A_3 = {\sqrt{b(\mu+b)} \over \rho^2 +b}dt\cr
& X^1 = X^2 = H^{1/3},\qquad X^3 = H^{-2/3}\cr
& \phi_1=\phi_2=\phi_3 =0\cr
& H = 1 + {b\over \rho^2},\qquad f = 1- {\mu \over \rho^2} + \rho^2 H
\end{split}
\end{equation}
whose charge is $q=\sqrt{b(\mu+b)}$ and mass $m= {3\over 2}\mu + q$. The black hole
is extremal (and singular, i.e. the superstar) when $\mu=0$. We now take the scaling
\begin{equation}
\begin{split}
& t=k^{-1}\,\tau,\quad \rho= k\, u,\quad \Omega_i = k^{-1}\, x_i,\quad
  b = b'\, k^2,\quad \mu =\mu' k^4 \cr
& \tau\,, u \,, x_i,\,b',\,\mu' = {\rm const}\qquad ,\qquad  k\rightarrow \infty
\end{split}
\end{equation}
Then we find the following non-extremal charged brane solution 
\begin{equation}
\begin{split}
& ds^2 = - \left(1+ {b' \over u^2}\right)^{-2/3} \left(u^2 + b' -{\mu' \over u^2}\right) d\tau^2 +\left(1+ {b' \over u^2}\right)^{1/3} \left( \left(u^2 + b' -{\mu' \over u^2}\right)^{-1} du^2 + u^2 dx_i^2\right) \cr
& A_1=A_2 = 0 ,\quad A_3 = {\sqrt{b' \mu'} \over u^2 + b'^2} dt\cr
& X^1 = X^2 = \left(1+ {b' \over u^2}\right)^{1/3},\qquad X^3 = \left(1+ {b' \over u^2}\right)^{-2/3}
,\quad \phi_1=\phi_2=\phi_3 =0
\end{split}
\end{equation}
The energy density and charge density of this solution is (up to numerical factors of
order one)
$$
\rho_e \approx \mu',\qquad \rho_q \approx \sqrt{b' \mu'}
$$
We notice that
$$
{\rho_e \over \rho_q^{4/3}} \approx(\mu')^{1/3} (b')^{-2/3}
$$ we notice that by taking $\mu'$ small enough we can make this ratio
as small as we like. Hence for any finite charge density, the energy
density of one-charge RNAdS black branes can be made arbitrarily small
i.e. the constant $c_e$ mentioned in section \ref{sec:discussion} is
$c_e=0$ for the one-charge case.

\subsection{Three-charge solitons}

A similar scaling can be performed for the 1/8 BPS (3-equal charge) soliton. For this one
takes the following scaling of solution of the ${a\over \rho}$ form
\begin{equation}
\begin{split}
& t=k^{-1}\,\tau,\quad \rho= k\, u,\quad \Omega_i = k^{-1}\, x_i,\quad
  a = 2\sqrt{q_0}\, k\cr
& \tau\,, u \,, x_i\,,q_0\, = {\rm const}\qquad ,\qquad  k\rightarrow \infty
\end{split}
\end{equation}
For large $a$ the solution \eqref{largeqlimit} becomes a good
approximation and from \eqref{metricbps} we end up with the following hairy,
asymptotically Poincare Patch solution of \eqref{restlag}
\begin{equation}
\label{planarthree}
\begin{split}
& ds^2 = -u^2 \left(1 + {4 q_0 \over u^2} \right)^{1/2} d\tau^2
+ {1\over u^2 \left(1 + {4 q_0 \over u^2}\right) }d u^2
+u^2 \left(1+ {4 q_0 \over u^2}\right)^{1/2} dx_i^2 \cr
& A = 0 \qquad ,\qquad \phi  = 
{4q_0 \over u \sqrt{4 q_0 +u^2}} \cr
\end{split}
\end{equation}
It would be interesting to explore whether the solution \eqref{planarthree} has any
relation to hairy black branes near the BPS limit.

\section{Some numerical results}
\label{app:numresults}

\begin{figure}[h]
\begin{minipage}[t]{0.5\linewidth} 
\centering
\psfrag{a}[t]{$h_0$}
\psfrag{b}[r]{$\langle {\cal O}_\phi\rangle$}
\includegraphics[totalheight=0.18\textheight]{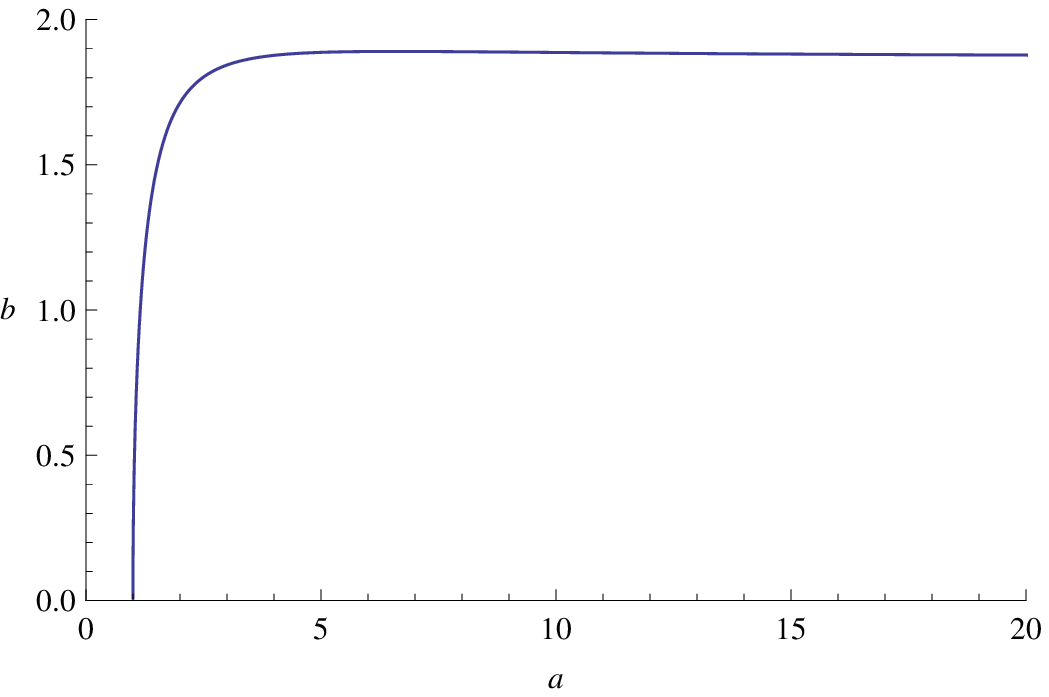}
\caption{Expectation value $\langle {\cal O}_\phi \rangle$ of the operator dual to the field $\phi$ as a function
of the value of $h_0$ at $r=0$.}
\label{vevreg1}
\end{minipage}
\hspace{0.5cm} 
\begin{minipage}[t]{0.5\linewidth}
\centering
\psfrag{a}[t]{$h_0$}
\psfrag{b}[r]{$\langle {\cal O}_\phi\rangle$}
\includegraphics[totalheight=0.18\textheight]{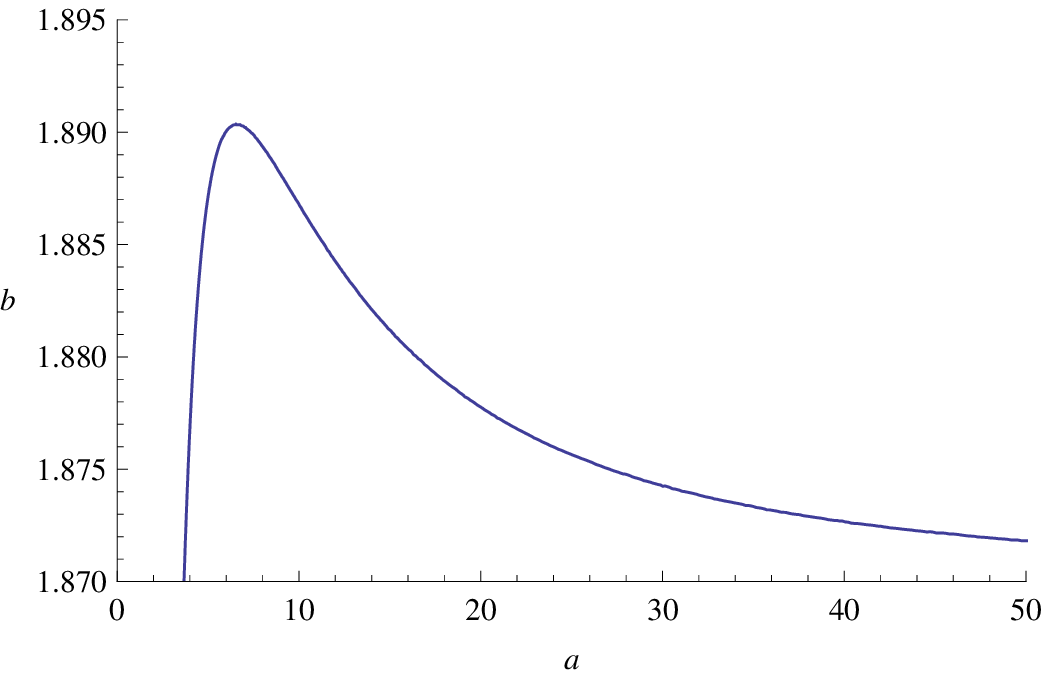}
\caption{Detail of the previous graph with different scales on the axes, where we can see
the local maximum.}
\label{vevreg2}
\end{minipage}
\end{figure}

\begin{figure}[h]
\begin{minipage}[t]{0.5\linewidth} 
\centering
\psfrag{a}[t]{$\log h_0$}
\psfrag{b}[r]{$\langle {\cal O}_\phi\rangle-\langle {\cal O}_\phi \rangle_c$}
\includegraphics[totalheight=0.18\textheight]{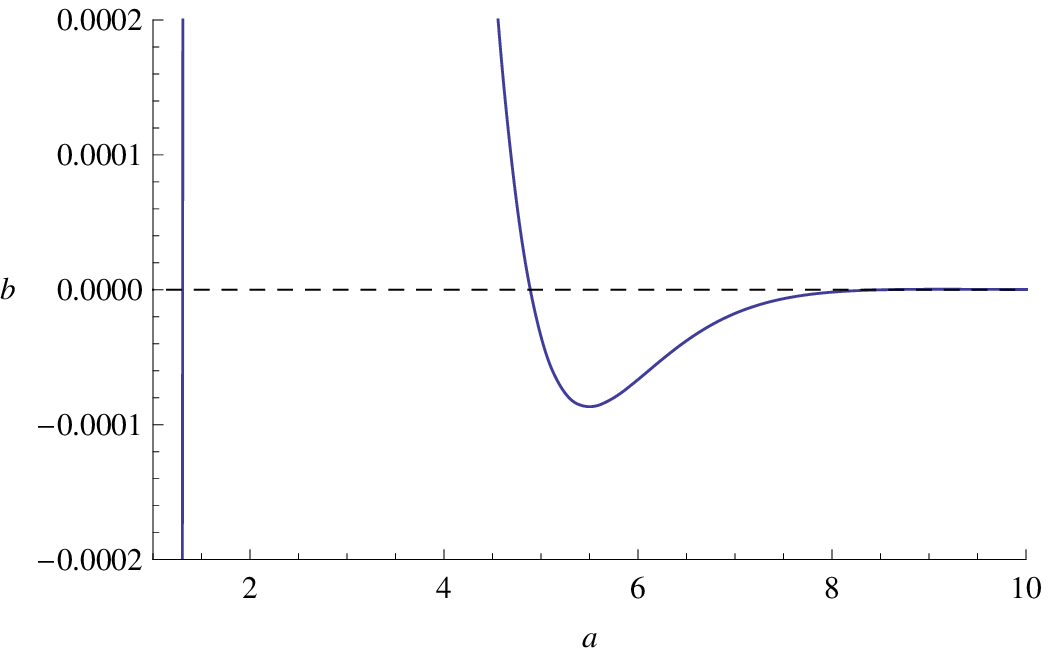}
\caption{Magnification of the previous plot.. We see the next oscillation around $\langle {\cal O}_\phi \rangle$.}
\label{vevreg3}
\end{minipage}
\hspace{1.0cm} 
\begin{minipage}[t]{0.5\linewidth}
\centering
\psfrag{a}[t]{$\log h_0$}
\psfrag{b}[r]{${\langle {\cal O}_\phi\rangle-\langle {\cal O}_\phi\rangle_c \over e^{-{3\over 2} \log h_0}}$}
\includegraphics[totalheight=0.18\textheight]{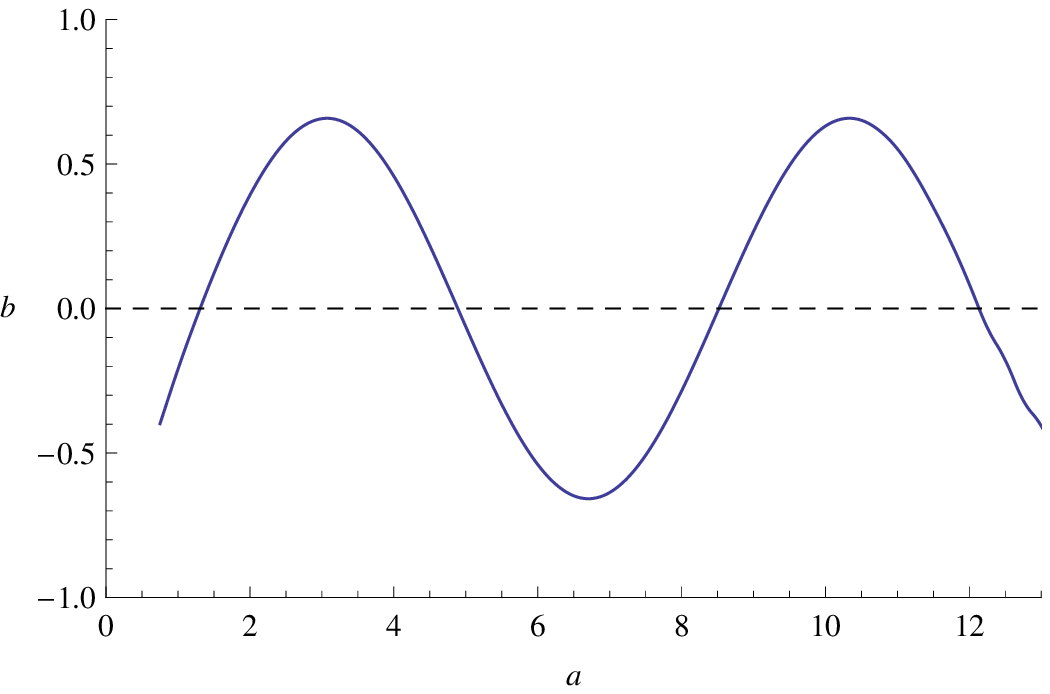}
\caption{The damped oscillations of $\langle {\cal O}_\phi\rangle$
  around the critical value $\langle {\cal O}_\phi\rangle_c$ for large
  $h_0$.}
\label{vevreg4}
\end{minipage}
\end{figure}

\begin{figure}[h]
\begin{minipage}[t]{0.5\linewidth} 
\centering
\psfrag{a}[t]{ $a$}
\psfrag{b}[r]{$\langle {\cal O}_\phi\rangle$}
\includegraphics[totalheight=0.18\textheight]{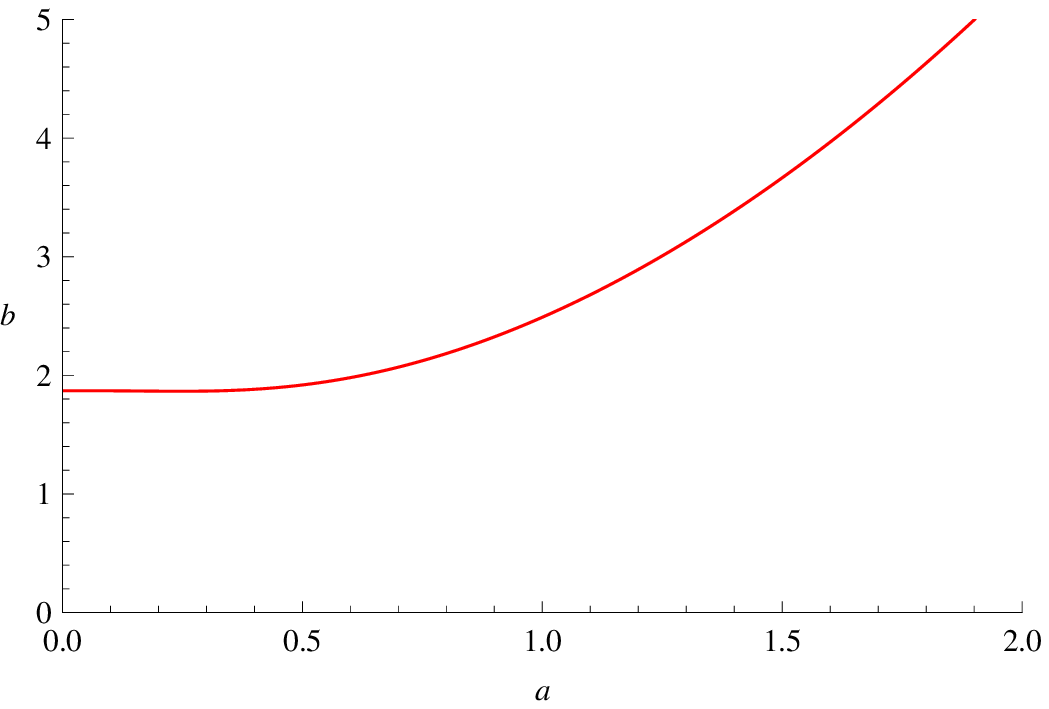}
\caption{Expectation value $\langle {\cal O}_\phi\rangle$ for
  spherically symmetric supersymmetric solitons with a singularity of
  the form ${a \over \rho}$ at $\rho=0$.}
\label{vev3}
\end{minipage}
\hspace{0.5cm} 
\begin{minipage}[t]{0.5\linewidth}
\centering
\psfrag{a}[t]{$a$}
\psfrag{b}[r]{$\langle {\cal O}_\phi\rangle$}
\includegraphics[totalheight=0.18\textheight]{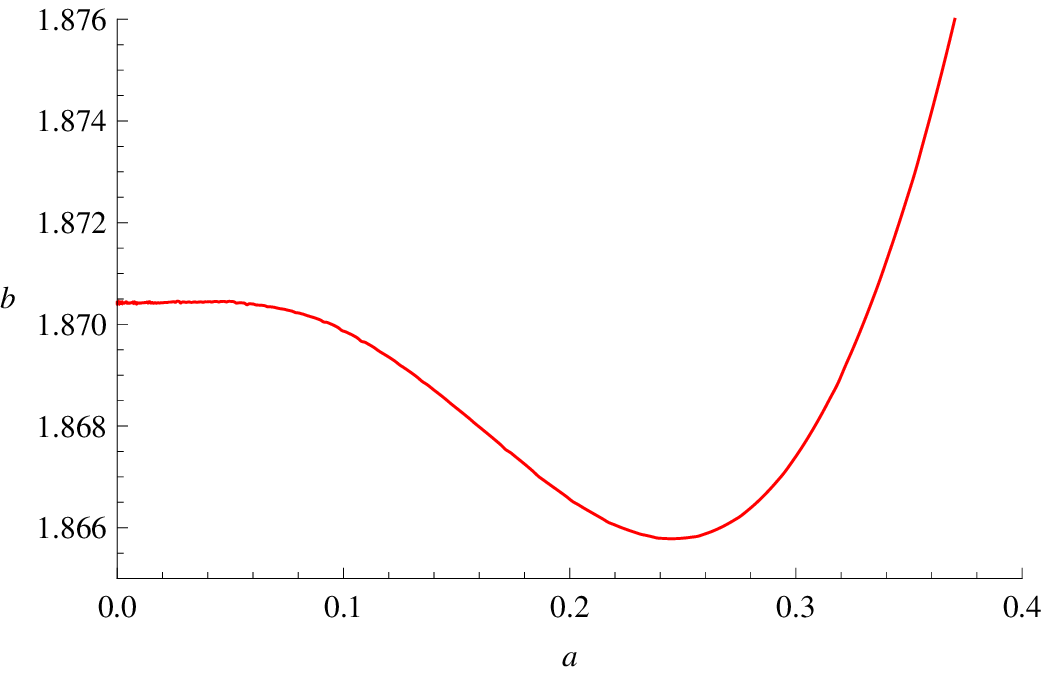}
\caption{Detail of the previous graph with different scales on the axes, where we can see
the minimum of $\langle {\cal O}_\phi\rangle$.}
\label{vev4}
\end{minipage}
\end{figure}

\begin{figure}[h]
\begin{minipage}[t]{0.5\linewidth} 
\centering
\psfrag{a}[t]{$\log a$}
\psfrag{b}[r]{$\langle {\cal O}_\phi\rangle-\langle {\cal O}_\phi\rangle_c$}
\includegraphics[totalheight=0.18\textheight]{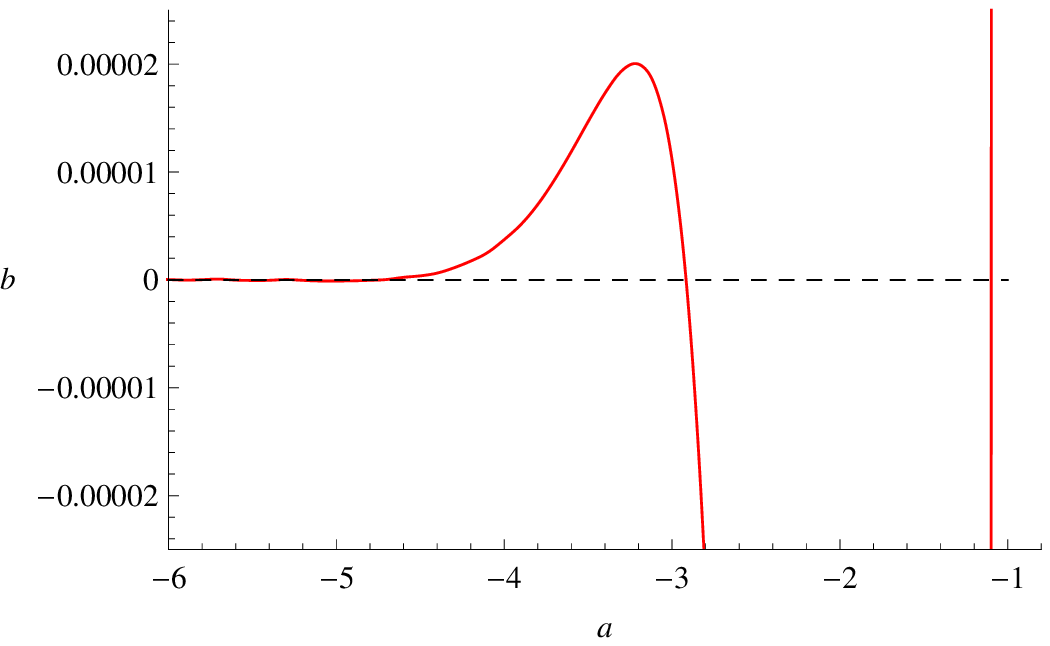}
\caption{Magnification of the previous graph. We wee the next oscillation around $\langle {\cal O}_\phi\rangle$.}
\label{sdqb}
\end{minipage}
\hspace{1.cm} 
\begin{minipage}[t]{0.5\linewidth}
\centering
\psfrag{a}[t]{$\log a$}
\psfrag{b}[r]{${\langle {\cal O}_\phi\rangle-\langle {\cal O}_\phi\rangle_c \over e^{3 \log a}}$}
\includegraphics[totalheight=0.18\textheight]{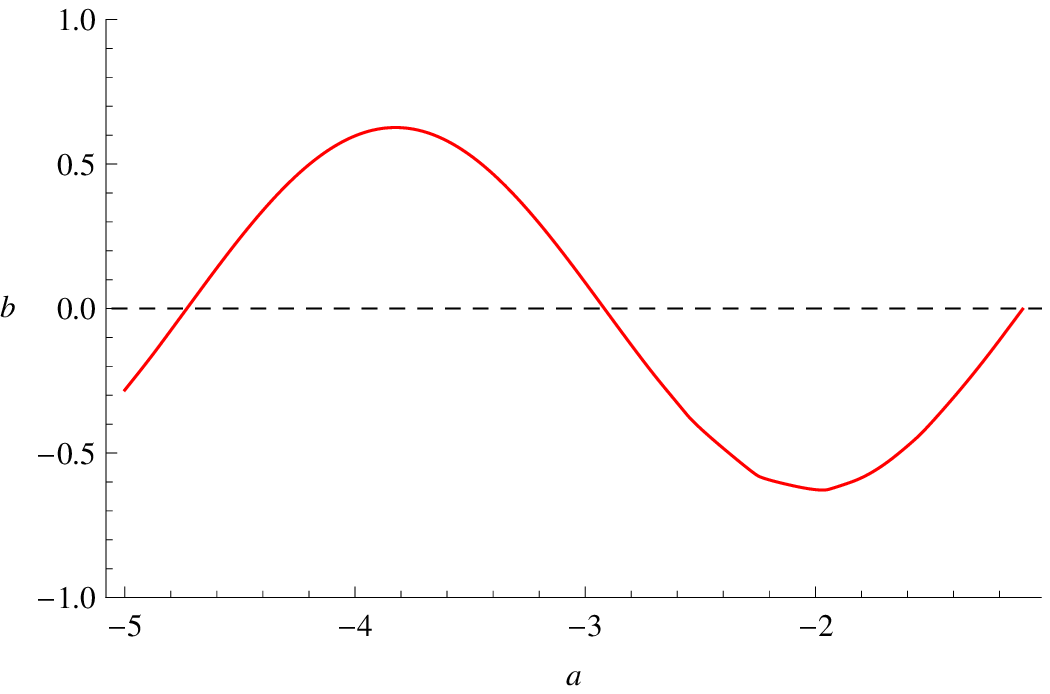}
\caption{The damped oscillations of $\langle {\cal O}_\phi\rangle$
  around the critical value $\langle {\cal O}_\phi\rangle_c$ for small $a$.}
\label{soscb}
\end{minipage}
\end{figure}

\begin{figure}[h]
\begin{minipage}[t]{0.5\linewidth} 
\centering
\psfrag{a}[t]{ $q-q_c$}
\psfrag{b}[r]{$\langle {\cal O}_\phi\rangle-\langle {\cal O}_\phi\rangle_c$}
\includegraphics[totalheight=0.18\textheight]{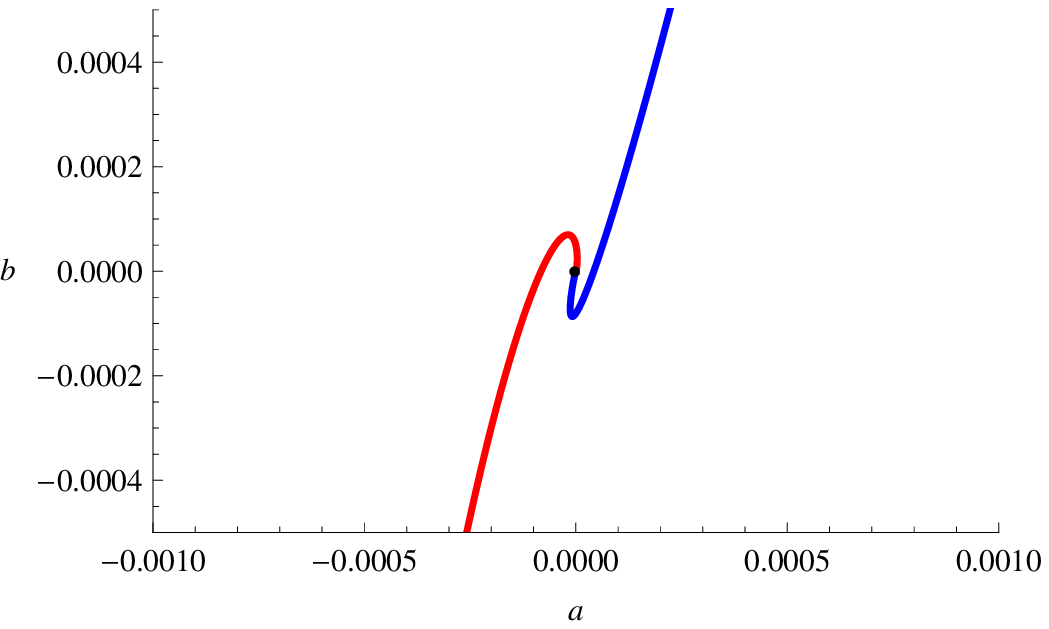}
\caption{Zooming in around the point $S$. The blue curve is the regular soliton and the
  red curve is the soliton with the ${a\over \rho}$ singuality.} 
\label{spiral1}
\end{minipage}
\hspace{1.0cm} 
\begin{minipage}[t]{0.5\linewidth}
\centering
\psfrag{a}[t]{$q-q_c$}
\psfrag{b}[r]{$\langle {\cal O}_\phi\rangle-\langle {\cal O}_\phi\rangle_c$}
\includegraphics[totalheight=0.18\textheight]{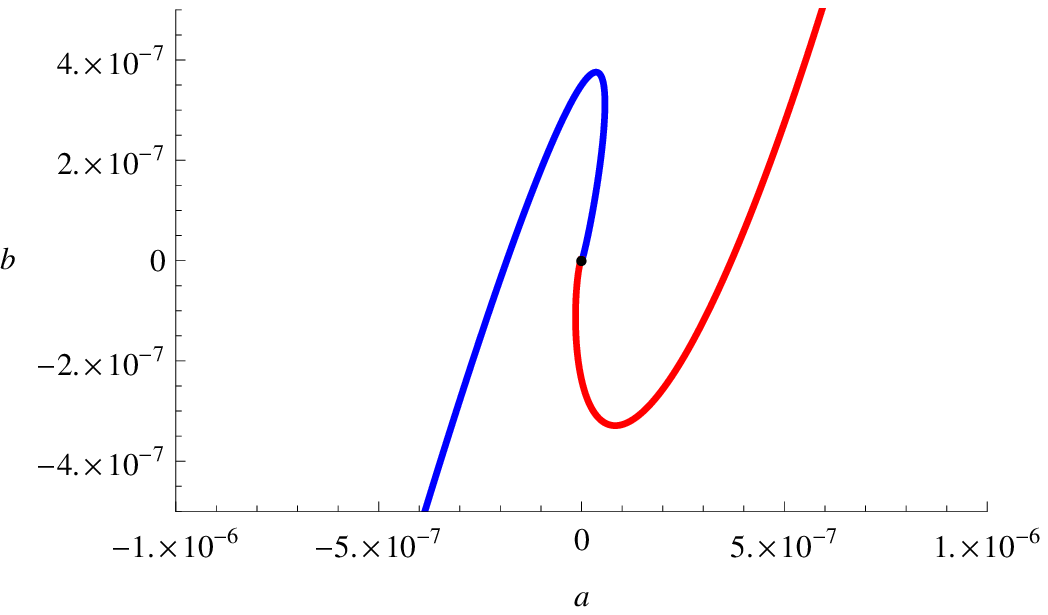}
\caption{Further magnification around the point $S$. The blue curve is the regular soliton and the
  red curve is the soliton with the ${a\over \rho}$ singuality.} 
\label{spiral2}
\end{minipage}
\end{figure}

\nocite{*}
\bibliographystyle{JHEP}
\bibliography{hairybh}

\end{document}